\documentclass[a4paper]{esannV2}
\usepackage[dvips]{graphicx}
\usepackage[latin1]{inputenc}
\usepackage{amssymb,amsmath,array}
\usepackage{mathtools}
\usepackage{changepage} 
\usepackage{float}

\usepackage[breaklinks]{hyperref} 

\usepackage{longtable}
\usepackage{makecell}
\usepackage{pdfpages}
\usepackage{subcaption}
\usepackage{amsthm}
\usepackage{sidecap}
\usepackage[acronym]{glossaries}
\usepackage{enumitem}
\usepackage{multicol}
\usepackage{caption}
\usepackage{lipsum}

\hypersetup{%
  colorlinks=true,
  linkcolor=.,
  linkbordercolor=white,
  citecolor=.,
  urlcolor=blue,
  pdfborder={0 0 0},
  urlbordercolor={1 1 1},
  citebordercolor={1 1 1},
  filecolor=.
}

\makeatletter
\Hy@AtBeginDocument{%
  \def\@pdfcitebordercolor{0 0 0}
  \def\@pdfurlbordercolor={0 0 0}
  \def\@pdfborder{0 0 0}
  \def\@pdfborderstyle{/S/U/W 1}
}
\makeatother
\usepackage{listings}
\usepackage{color}

\definecolor{dkgreen}{rgb}{0,0.6,0}
\definecolor{gray}{rgb}{0.5,0.5,0.5}
\definecolor{mauve}{rgb}{0.58,0,0.82}

\theoremstyle{definition}
\newtheorem{exmp}{Example}

\makeglossaries

\newglossaryentry{mainnet}
{
        name=Mainnet,
        description={is the alias for the live Ripple network where real money are being traded daily, and is meant for production.}
}
\newglossaryentry{testnet}
{
        name=Testnet,
        description={is an alternate network meant for testing; usually the public connects to the "amundsen" connector for testing.}
}
\newglossaryentry{ethereum}
{
        name=Ethereum \acrshort{poa},
        description={is the Ethereum test network using the proof of authority to validate transactions. At the time of writing there is proof of authority network Kovan, Rinkeby, Solokl and G{\"o}rli. However none of them is in production. They are all deployed as test networks}
}
\newglossaryentry{moneyd}
{
        name=Moneyd,
        description={An ILP provider, allowing all applications on an end-user computer to use funds on the live ILP network}
}
\newglossaryentry{switchapi}
{
        name=Switch \acrshort{api},
        description={A SDK for cross-chain trading between BTC, ETH, DAI and XRP with Interledger Streaming}
}
\newglossaryentry{glspow}
{
        name= Proof of Work (PoW),
        description={A consensus protocol introduced by Bitcoin, known as mining, which involves answering to a mathematical problem that requires considerable work to solve, but is easily verified once given the answer}
}
\newglossaryentry{xrp}
{
        name= XRP,
        description={Ripple's digital payment asset which is used for Interledger payments}
}

\newacronym{ilp}{ILP}{Interledger Protocol}
\newacronym{api}{API}{Abstract Programming Interface}
\newacronym{dlt}{DLT}{Distributed Ledger Technology}
\newacronym{poa}{PoA}{Proof of Authority}
\newacronym{pow}{PoW}{Proof of Work}
\newacronym{gui}{GUI}{Graphical User Interface}
\newacronym{fsm}{FSM}{Finite State Machine}
\newacronym{btp}{BTP}{Bilateral Transfer Protocol}
\newacronym{ilsp}{ILSP}{Interledger Service Provider}
\newacronym{spsp}{SPSP}{Simple Payment Setup Protocol}

\begin{document}
\title{Deep dive into Interledger: \\Understanding the Interledger ecosystem}

\author{Lucian Trestioreanu, Cyril Cassagnes, and Radu State
%
%
\vspace{.3cm}\\
%
Ripple UBRI @ Interdisciplinary Centre for Security, Reliability and Trust, \\ University of Luxembourg \\
29, Avenue JF Kennedy, 1855 Luxembourg, Luxembourg
}
\setlength{\tabcolsep}{0pt}

\maketitle

\vspace{\fill}
\begin{abstract}
At the technical level, the goal of Interledger is to provide an architecture and a minimal set of protocols to enable interoperability between any value transfer systems.  The Interledger protocol is a protocol for inter-blockchain payments which can also accommodate FIAT currencies. To understand how it is possible to achieve this goal, several aspects of the technology require a deeper analysis. For this reason, in our journey to become knowledgeable and active contributors we decided to create our own test-bed on our premises. By doing so, we noticed that some aspects are well documented but we found that others might need more attention and clarification. Despite a large community effort, the task to keep information on a fast evolving software ecosystem up-to-date is tedious and not always the main priority for such a project. The purpose of this tutorial is to guide, through several examples and hands-on activities, community members who want to engage at different levels. The tutorial consolidates all the relevant information from generating a simple payment to ultimately creating a test-bed with the Interledger protocol suite between Ripple and other distributed ledger technology.
\end{abstract}
\vspace{\fill}

\newpage
    \tableofcontents
    \listoffigures
    \listoftables
\newpage

\section{What this document covers}
\label{section:intro}  

The scope of this document is to provide a walk-through the Interledger ecosystem starting with the \acrfull{ilp}. This ecosystem encompasses Ripple validating servers, \acrshort{ilp} connectors, \gls{moneyd}, \gls{switchapi} and more. The main goal of the document is to create a panoramic understanding of the ecosystem, how the main project systems and tools are interconnected together with practical insights into the how-tos of using the various associated systems and tools. 

In our Ripple journey, we needed a comprehensive practical understanding of the general picture, but the required information was rather sparse, making it difficult to join the different bits and pieces together in order to form a complete test-bed (private network). Indeed, a positive indicator is the level of activity of the community to push the vision proposed by Ripple. Consequently, tutorials and various other resources can become outdated after a few weeks. In order to overcome this problem, this document consolidates the required documentation to setup and configure all the different parts. We will provide as much detail as possible in order to build and deploy a private Ripple network and so an Ethereum network.

The rest of this document is organized as follows. 
In Section~\ref{section:community}, we review the state of the online documentation and the different community communication channels. In Sections \ref{section:environment}, \ref{section: custapps}, \ref{sec:connectors} and \ref{section: ledgers} we remind all the important aspects of Ripple, which we subsequently evaluate in Section~\ref{section:eval}. 
Finally, we wrap-up the document in Section~\ref{section:conclusions}.


\section{Who this document is for}

No prerequisites regarding the Interledger ecosystem are expected from the reader. However, developers, computer science students or people used to deal with computer programming challenges should be able to reproduce our setup without struggle.

\section{The Interledger community}
\label{section:community}

The community defines different communication channels\footnote{\href{https://interledger.org/community.html}{https://interledger.org/community.html}, accessed June 2019} for different purposes in order to interact, get support and discuss the evolution of the protocol. The main channels are:
    \begin{itemize}
        \item The company website of Ripple\footnote{\href{https://ripple.com/}{https://ripple.com/}, accessed June 2019} itself, established in 2012. The website provides all information related to business activities with RippleNet:
        \begin{itemize}
            \item The Interledger Forum\footnote{\href{https://forum.interledger.org/}{https://forum.interledger.org/}, accessed April 2020}
            \item The Ripple Dev Blog\footnote{\href{https://ripple.com/dev-blog/}{https://ripple.com/dev-blog/}, accessed June 2019}
            \item The \gls{xrp} Chat Forum\footnote{\href{http://www.xrpchat.com/}{http://www.xrpchat.com/}, accessed June 2019}
            \item The \gls{xrp} Ledger Dev Portal\footnote{\href{https://developers.ripple.com/}{https://developers.ripple.com/}, accessed June 2019} for the {\em rippled} servers (i.e. validators and trackers) and general concepts\footnote{\href{https://developers.ripple.com/docs.html}{https://developers.ripple.com/docs.html}, accessed June 2019}. This is the development portal of the \gls{xrp} Ledger, built on the Ripple open-source platform called {\em rippled}, which is the reference implementation. Thanks to their open source code-base they also have a Bug Bounty program\footnote{\href{https://ripple.com/bug-bounty/}{https://ripple.com/bug-bounty/}, accessed June 2019}.
        \end{itemize}
        \item The Interledger website\footnote{\href{https://interledger.org/}{https://interledger.org/}, accessed June 2019}. This website aggregates pointers towards all resources\footnote{\href{https://interledger.org/community.html}{https://interledger.org/community.html}, accessed June 2019}. Here is a sample of the resources accessible from the website:
        \begin{itemize}
            \item The Interledger Forum\footnote{\href{https://forum.interledger.org}{https://forum.interledger.org}, accessed June 2019}
            \item The reference implementation\footnote{\href{https://github.com/interledgerjs}{https://github.com/interledgerjs}, accessed June 2019} for \acrshort{ilp} connector
            \item The Rafiki\footnote{\href{https://github.com/interledgerjs/rafiki}{https://github.com/interledgerjs/rafiki}, accessed June 2019} \acrshort{ilp} connector,
            which generally has the same purpose as the reference implementation but a newer, different architecture
            \item The Interledger\footnote{\href{https://interledger.org/calls.html}{https://interledger.org/calls.html}, accessed June 2019} community calls.
        \end{itemize}
        \item The Interledger Whitepaper publication - \textit{A Protocol for Interledger Payments} - Stefan Thomas and Evan Schwartz -- self-published online\footnote{\href{https://interledger.org/interledger.pdf}{https://interledger.org/interledger.pdf}, accessed June 2019}
        \item Academic publication: \textit{Interledger: creating a standard for payments} - Hope-Bailie, Adrian and Thomas, Stefan - Proceedings of the 25th International Conference Companion on World Wide Web, p. 281--282
        \item Not peer-reviewed research paper: \textit{The Ripple protocol consensus Algorithm and Analysis of the \gls{xrp} Ledger Consensus Protocol}\footnote{\href{https://arxiv.org/abs/1802.07242}{https://arxiv.org/abs/1802.07242}, accessed June 2019}
        
        \item The Interledger Payments Community Group\footnote{\href{https://www.w3.org/community/interledger}{https://www.w3.org/community/interledger}, accessed June 2019} whose scope is broader than \acrfull{ilp} and aims to create an open, universal payment scheme built on Web standards.
    \end{itemize}

Regarding practical experimentation with Ripple, Ethereum and Interledger, at the time of writing, the sources of documentation are sparse, disseminated over different channels like websites, forums, conference slides. Therefore, we collect and connect a relevant selection of these sources.
A few tutorials on the Medium Interledger blog\footnote{\href{https://medium.com/interledger-blog}{https://medium.com/interledger-blog}, accessed June 2019} concern setting up and starting a JavaScript connector, which is the reference implementation of the Interledger specifications and related protocols. Recently, for the same connector, Strata Labs has provided a quick-start bundle.


\section{The Interledger ecosystem}
\label{section:environment}

RippleNet\footnote{\href{https://ripple.com/ripplenet/}{https://ripple.com/ripplenet/}, accessed July 2019} aims to create a friction-less experience for sending and receiving money globally. The company targets institutions (e.g. Banks) of the financial sector. The key benefit of the solution is modernizing the traditional systems, which are many times expensive and take days to settle. However, one type of infrastructure won't fulfill all the requirements. Therefore, they strongly support the \acrfull{ilp} initiative in order to realize the vision of an international friction-less payments routing system. In other words, a standard for bridging diverse financial systems. 

\textit{"Ripple has no direct competitors in crypto space, as it is fundamentally different from the most cryptocurrencies: it's more centralized, \textbf{totally currency agnostic} and uses probabilistic voting (and not \gls{glspow}) to confirm transactions."} \cite{agnostic}. This is possible thanks to the \acrfull{ilp}, and because the transaction verification process of RippleNet is not coupled to a mint process, which in the case of some others cryptocurrencies generates a direct income.

Nonetheless, besides lower cost and faster settlement than classic banking transactions, one of the most interesting aspects regarding the \acrfull{ilp} is that it will seamlessly manage payments when the sender's currency is different from the receiver's currency, or when the sender's payments network is different from the receiver's payments network. For instance, a payment is issued on the fiat currency network using MasterCard, VISA, wire-transfer and the receiver receives it on an account (also known as a {\em Wallet}) created within a payment system using a crypto-currency. The \acrfull{ilp} offers a means to bridge the crypto-currencies and fiat to enable interoperable and fast value exchange. Therefore, it is paramount to underline that \acrfull{ilp} is not a blockchain, a token, nor a central service.

In \acrshort{ilp}, money is actually not moved meaning that \acrshort{ilp} doesn't decrease or increase of the total amount of electronic money in circulation. A connector swapping both currency has an account for each payment system it supports. Account balances are open and closed between parties involved in a particular transaction according to transaction instructions of each payment system involved. The parties are the sender, intermediaries (connectors) and the receiver. This statement is derived from the original Ripple explanation regarding connectors: \newline
\indent \textit{"Interledger connectors do not actually move the money, they rely on plugins for settlement. Plugins may settle by making a payment on an external payment system like (automated clearing house) ACH or they may use payments channels over a digital asset ledger like \gls{xrp} Ledger or Bitcoin"} \cite{conngit}.

In other words, when the receiver's currency is different from the sender's currency, also, no money is leaving the sender's network and no money enters the receiver's network. What happens is that at some point along the chain, some connector with accounts on both payment systems, keeps the sender's currency in one wallet (belonging to same ledger as the sender) and forwards the money towards the receiver, now denominated in the receiver's currency, from its other wallet holding that currency on the second ledger - the same ledger with the receiver's. The main difference with the classic system running today is that with \acrshort{ilp}, the end-to-end payment becomes completely seamless thanks to the automation of many parts provided by the \acrfull{ilp} Suite. For example, whereas in the classical banking systems there is no direct way for the sender to know when the recipient received the money, with \acrshort{ilp} this information is available immediately.\newline
\indent In this section, we are going to quickly go through the payment infrastructure where all elements belong to one of the following three categories: distributed Ledger, user-level apps for payment, and connectors. Then, we will discuss the protocols stack and we will go through each of these components, providing details on each of them.

\subsection{Main components of a unified payment infrastructure}
\label{section:infractructure}
\begin{figure}[h!]
    \includegraphics[width=0.8\textwidth]{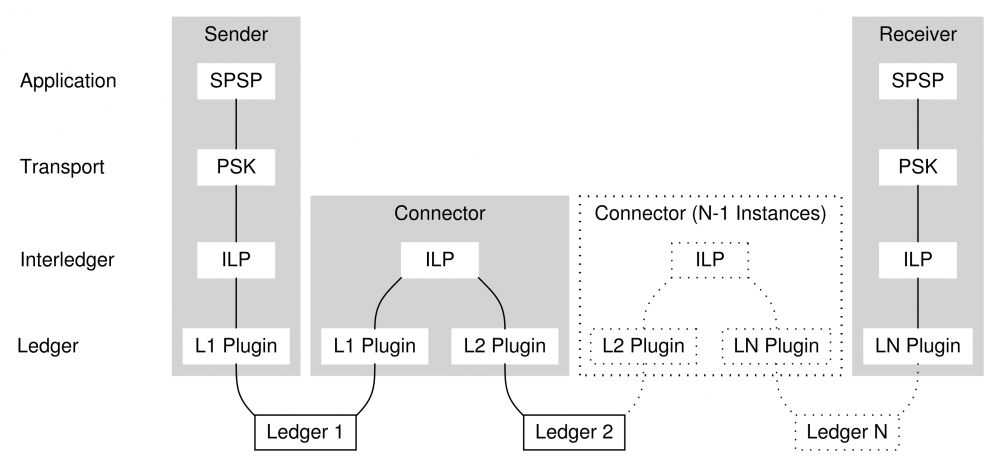}
    \caption[Payment chain]{Payment chain.}
    \label{fig:paymentchain}
\end{figure}

From a high level point of view, the infrastructure comprises three main components:  

\paragraph{\textbf{Ledgers.}} In the context of Interledger, a \textit{Ledger} is any accounting system that holds user accounts and balances. It can be linked to cryptocurrencies like Bitcoin, Ethereum, \gls{xrp} or classic banks, PayPal and more. Here we are going to use the \gls{xrp} ledger and the ETH ledger which are distributed ledgers.\\
\indent The \gls{xrp} ledger comprises of a network of servers running the {\em rippled} software in "validator" or "tracking" mode, and the connectors can plug into it. The servers run the blockchain, record the user accounts and validate the transactions as they arise from connectors. Figure \ref{fig:paymentchain} illustrates the Ledger layer designed by the early adopters of the Interledger ecosystem. Each Connector is linked to at least two Ledgers (e.g. Ledger 1 = \gls{xrp} Ledger and Ledger 2 = Ligthning) with dedicated plugins. \newline
\indent The user accounts are opened and stored on-ledger. In the case of the \gls{xrp} ledger, this is how the related account info looks like at the time of opening the account:

\begin{lstlisting}
{
   "result" : {
      "account_id" : "rMqUT7uGs6Sz1m9vFr7o85XJ3WDAvgzWmj",
      "key_type" : "secp256k1",
      "master_key" : "NIP SELF EDGY AQUA COME BAWD RING NEAL HINT HACK HEAT ADEN",
      "master_seed" : "shjZQ2E3mYzxHf1VzYBJCQHqLvt7Y",
      "master_seed_hex" : "925A2949624EF8CFA8A6C6A6E9211B2C",
      "public_key" : "aB4XmhndLn6C3sbsp5qK4Cy3GG9mU4KVe3wqWHatuudZX7CMhsvC",
      "public_key_hex" : "024B8511437A9A20E57C21A42A463DEEFE49D1DBE48ECA7FEEDE50048D02D92152",
      "status" : "success"
   }
}
\end{lstlisting}
- \textit{"account\_id"} is the public address of the account, used to identify the account and perform transactions. \\
- \textit{"master\_seed"} is the private key of the account. It is "the lock" of the account and should be kept safe and private.\\
- \textit{"master\_key"} can be used to regenerate the account details if needed.\\
- \textit{"public\_key"} can be used by third parties for verification.

\paragraph{\textbf{\acrshort{ilp} connectors.}} The reference implementation of the connector specification is in JavaScript and so is the new Rafiki connector. However, other implementations are also written, in Java and Rust. Figure \ref{fig:paymentchain} shows that connectors are bridging all ledgers and their end-users, represented there by the \textit{"sender"} and \textit{"receiver"}.

Connectors are run by different entities and offer payment inter-operability across the payment platform to the "customers" running a "customer app". The connectors are the "service providers", or "market makers", or "liquidity providers", because they provide end-users access to other payment networks, provide payment routing, exchange and liquidity. In order to do this, the Connectors make use, among others, of the \acrfull{ilp}, and \acrshort{ilp} addresses.\\

\begin{adjustwidth}{0.5cm}{}
\small{
\textit{\acrshort{ilp} address.} In order to be able to identify themselves and their users, and route the payments in a global network, the need for unique identification in the internet's IP style has arisen. As such, unique \acrshort{ilp} addresses are being assigned to each Interledger node. In ILP a node can be a \textit{sender}, a \textit{connector} or a \textit{receiver}. All nodes implement \acrshort{ilp}. 

On the production network, any \acrshort{ilp} address starts with "g". Other prefixes can be "private" , "example" , "peer" , "self" ..
We provide below some \acrshort{ilp} address examples. These are thoroughly explained on the Interledger website \cite{ILPaddr}. \\

\noindent \textit{g.scylla} - \textit{"scylla"} is a connector and \textit{"g.scylla"} is its \acrshort{ilp} address \\
\textit{g.acme.bob} - is the address held by Bob on the \textit{"acme"} connector \\
\textit{g.us-fed.ach.0.acmebank.swx0a0.acmecorp.sales.199.~ipr.cdfa5e16-e759-4ba3-88f6-8b9dc83c1868.2} - is a complex, real-life address.\\}
\end{adjustwidth}

The connectors accept "dial-up connections" from the "customer apps", like an internet provider would provide internet services by accepting a dial-up connection from a home internet dial-up modem. As such, they function as Interledger Service Providers (\acrshort{ilsp}s). \textit{'An \acrshort{ilsp},} [depicted in Figure \ref{fig: netovw}], \textit{is a connector that accepts unsolicited incoming peering requests. It will have multiple child nodes connect to it and will assign them each an address and then route \acrshort{ilp} packets back and forth for them onto the network. (It's modelled on the idea of an ISP that provides customers access to the Internet)'} \cite{bailieILSP}. When opening payment channels and routing payments for their customers, depending on the number of customers, a connector could bound significant amounts of money. They assume some risks and expect to make small profits in exchange for providing the service.

\paragraph{\textbf{Customer apps}} are the third component of the infrastructure, and they provide end-users access to the Interledger payment system. Examples of customer apps are \gls{moneyd} or \gls{switchapi}. Figure \ref{fig:paymentchain} shows the different layers for end-users, i.e. participants willing to generate a payment.

\subsection[The Money transfer system]{The Money transfer system.} 
\label{sec: mts}

The system of money transfer over \acrshort{ilp} involves recording and manipulating money at different levels:
\begin{itemize}
    \item \textit{The Bilateral Balance} kept between two peers
    \item \textit{Settlement} on the \textit{payment channel (paychan)}, which involves signing claims that are recorded on the payment channel opened between the two peers. The claims concern the transactions between the two peers resulting from adjusting the Bilateral Balance above.
    \item \textit{On-Ledger recorded transactions}, resulted, for example, from redeeming the previous claims submitted on the payment channel. On-Ledger transactions can also be submitted, for example, by using the Ripple \acrshort{api}.
\end{itemize}

\subsubsection[The Bilateral Balance]{The Bilateral Balance.}
\label{subsec: bilbal}
Two directly connected peers hold a balance between them, for the \acrshort{ilp} packets of value they exchange. The balance is maintained real-time and it can increase or decrease, according to the value they exchange. 

\begin{figure}[h!]
   			\begin{minipage}{\textwidth}
   			\centering
       		\includegraphics[width=0.8\textwidth]{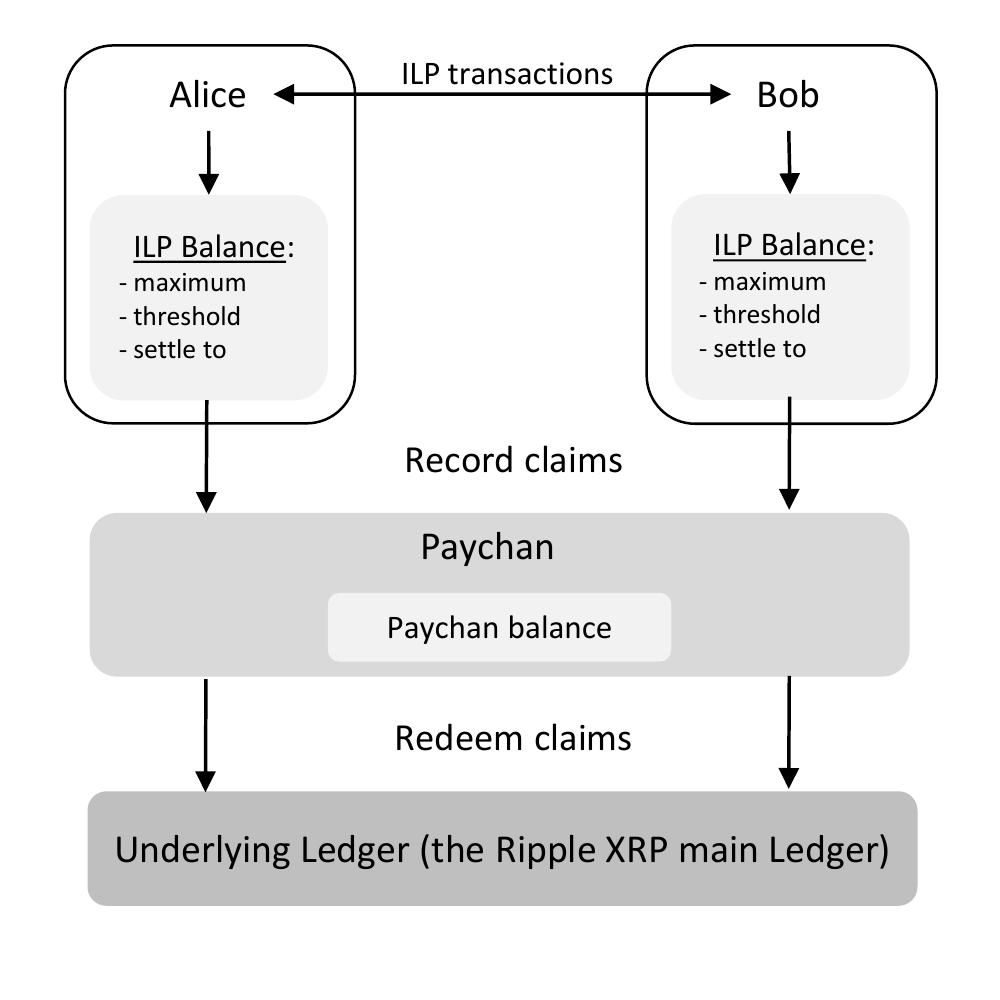}
			\caption[Money transfer]{The money transfer system.}
            \label{fig: money_transfer}
        	\end{minipage}
		\end{figure}

The balance parameters are:

\begin{lstlisting}
balance: {
    maximum: '20000000000', //maximum amount that the other party can owe
    settleThreshold: '-50000000', //the balance value that triggers an automatic settlement
    settleTo: '0' //balance value after settlement
}
\end{lstlisting}

Concerning Figure \ref{fig: money_transfer}, if for example at one given moment, according to the total balance, Alice owes Bob 20 \gls{xrp}, Alice's balance with Bob will be -20 \gls{xrp}, while Bob's balance with Alice will be 20. Their balances will offset each other. This balance can be seen in the Moneyd-GUI (Graphical User Interface). Concerning Figure \ref{fig: money_transfer}, this is the "\acrshort{ilp} Balance", and it is a not-yet-settled balance. \\

\indent \textit{Maximum Balance.} Peers can set the \textit{maximum balance} they are willing to trust (or risk) in relation to each other in the plugins configuration (for the reference js connectors and \gls{moneyd}). \newline

\indent \textit{Settle Threshold.} The \acrshort{ilp} peer balance can be settled manually or automatically. The connector plugins can be set to automatically settle this balance on the paychan by using the "\textit{settleThreshold}" flag. If, for example, Alice has configured a settlement threshold of -15, this means that she will settle with Bob as soon as she owes him 15 \gls{xrp} or more. \newline
\indent \textit{"In a correctly configured peering the additive inverse (negation) of the settlement threshold of one peer will be less than the maximum balance of the other peer."} \cite{peecleset}\newline

\indent \textit{settleTo.} The automatic \textit{Settlement} process should attempt to re-establish the \acrshort{ilp} balance to this amount. This is done through a paychan claim. \newline

\subsubsection[Payment channels]{Payment channels.}
\label{subsec: paychans}
Payment channels, or in short \textit{"paychans"}, are an important feature of nowadays Interledger. 
Some distributed ledgers are defining their own payment channel concept. Therefore, it is important to keep in mind the definition of the Payment Channel agreed in the documentation of Interledger\footnote{\href{https://github.com/interledger/rfcs/blob/master/0027-interledger-protocol-4/0027-interledger-protocol-4.md}{https://github.com/interledger/rfcs/blob/master/0027-interledger-protocol-4/0027-interledger-protocol-4.md}, accessed June 2019}.

Payment channels are opened only between direct peers. \textit{Settlement}, presented in Section \ref{subsec: settlement}, also occurs only between direct Interledger peers. When two peers connect over \acrshort{ilp}, they open a payment channel. Their bilateral transactions will afterwards be carried on to the paychan. Paychans are a solution for faster, cheaper and more secure transactions, especially when the ledger involved is slow or expensive. 

Below is an explained example of an Interledger paychan details \cite{paychans}:
\begin{lstlisting}
{
account: 'rLR52VSZG3wqSrkcpfkSnaKnYoYyPoJJgy', //the payer's XRP account address
amount: '100000', //size of the payment channel
balance: '0', //the amount the payee expects to have already received from the channel
destination: 'rMqUT7uGs6Sz1m9vFr7o85XJ3WDAvgzWmj', //the payee's XRP address
publicKey: 'ED6AF48DB11D68CDF37B22D594DC18B7C1AF2D4157A7F9A487481469A7A7C91AE2', //public key used for the channel. Can be the payer's master key pair. Necessary to verify and redeem claims.
settleDelay: 3600, //(in seconds) provides time for payee to redeem outstanding claims
sourceTag: 2100406056,
previousAffectingTransactionID: '8D8FF4F2AD33FAB476CFBD7256D6138419BAAC6EFDAF16ECEEFAC704752B330A',
previousAffectingTransactionLedgerVersion: 201538 
}
\end{lstlisting}

When two nodes connect over \acrshort{ilp} they negotiate the paychan details according to their business needs. The main characteristic of a paychan is its size, which is the largest claim one can sign before they need to add more money. The paychan and its corresponding details, including size, is recorded on the ledger. This is a guarantee that  obligations will be eventually settled, according to the channel size, with the help of the underlying ledger. The trust invested in the ledger regarding the paychan is implicit, because the ledger was already trusted when opening the main accounts.

When transacting on a paychan, the two parties hold a Bilateral Ledger, which records the transactions performed in-between the two, and the balance. Most of the transactions are performed off-ledger, thus improving the speed and transaction costs also. Only when the peers redeem their recorded paychan claims, the specific transaction is recorded on-ledger. On the payment channel each claim is recorded individually, but they can be later redeemed individually or in bulk on the ledger \cite{paychans}.

\subsubsection[Settlement]{Settlement.} 
\label{subsec: settlement}
Settlement is a core concept used in \acrshort{ilp}, which is part of the larger system of money transfer over \acrshort{ilp}. In practice, \textit{settling} is encountered for example while setting up plugins or in relation to payment channels. The main concept, illustrated in Figure \ref{fig: money_transfer} \cite{settlement}, in practice usually involves a system of Interledger balances and paychan claims. 

 \indent In relation to paychans, \textit{settlement} involves signing a claim for the money owed. Claims do not need to be directly submitted to the ledger, but for the case of the Ripple ledger which is fast and cheap, they can \cite{mdpaychanexpl, paychans}. The process will be reflected in the paychan balance in Figure \ref{fig: money_transfer}. Multiple claims can be signed on the paychan, and the paychan balance will update accordingly. Note that at this point, no amount or transaction has been yet recorded on-ledger (except the initial channel creation and funding), so the ledger accounts for Alice and Bob still show the same balances as before (except for cheap  and fast ledgers like the Ripple ledger which makes it possible to submit claims individually if desired, to make the money available faster). \\
\indent Claims can be redeemed out of the paychan and into the user ledger account in bulk or individually. The paychan can be closed or reused. \newline

\subsubsection[On-Ledger transfers]{On-Ledger transfers.}
\label{subsec: ledger_transfers}
On-ledger transfers can be initiated in different ways. The most relevant in \acrshort{ilp} is redeeming the claims submitted on the paychan. Only at this point, the money will show up in the user wallet. Another possible way to initiate an on-ledger transaction is for example directly using the Ripple-\acrshort{api}. \newline

		 \begin{figure}[h!]
		 \hspace{-1.5cm}
   			\begin{minipage}{1.2\textwidth}
   			\centering
   			\vspace{-0.5cm}
       		\includegraphics[width=1\textwidth]{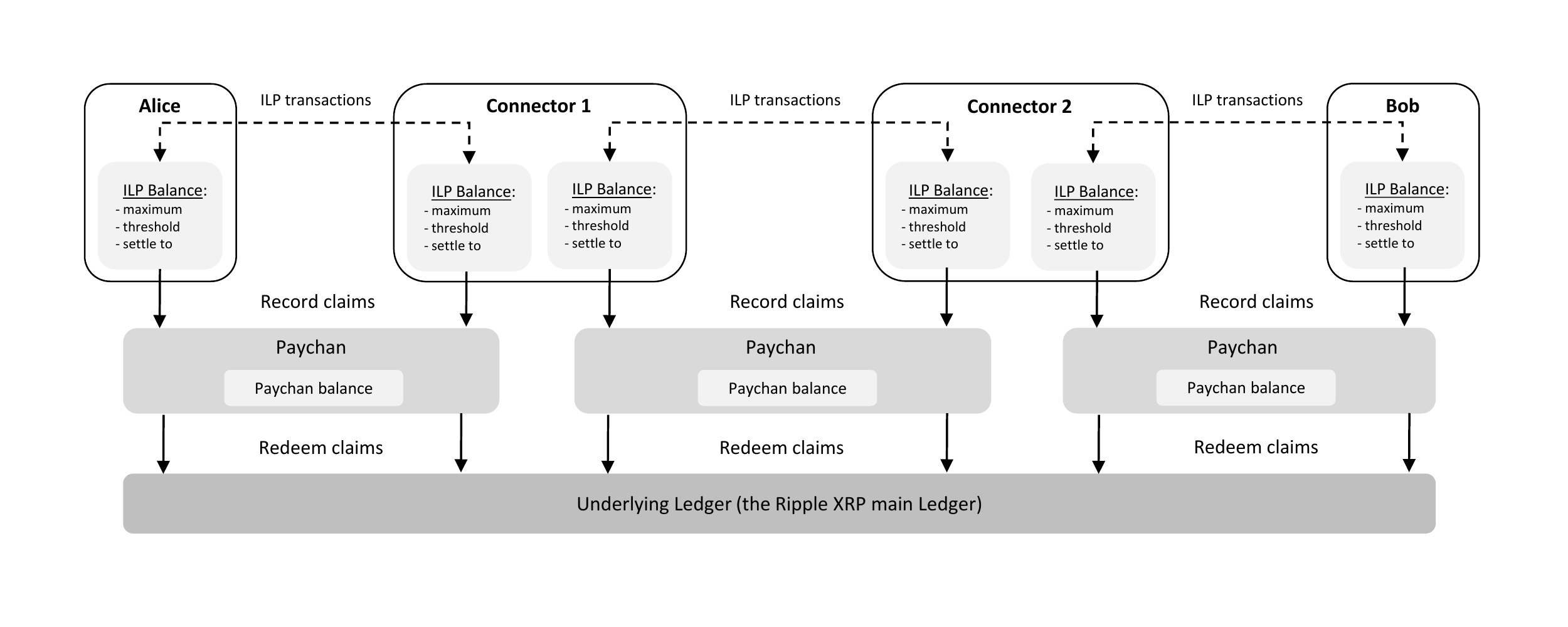}
       		\vspace{-1cm}
			\caption[Money transfer in practice]{The money transfer system in practice.}
            \label{fig: money_transfer2}
        	\end{minipage}
		\end{figure}

\begin{adjustwidth}{0.5cm}{}
\small{
Note: If Alice and Bob are end-users, or customers, running an \acrshort{ilp} customer module to connect to \acrshort{ilsp}s (connectors), the situation presented in Figure \ref{fig: money_transfer} will never happen, because Alice and Bob can never have a direct peering and settlement relationship. Their peering and settlement relationships are with their direct peers, respectively their parent connectors. So in this case, "Bob" should be in fact a "Connector" such that Figure \ref{fig: money_transfer} is correct with respect to real-life situations. As such, a real-life scenario would look in fact as shown in Figure \ref{fig: money_transfer2}: If Alice wants to send Bob 10 USD, the money will end up at Connector 1 and she will settle with Connector 1. In its turn, Connector 1 will pay 10 USD to Connector 2. Connectors 1 and 2 will settle between each other. Further, Connector 2 will forward the 10 dollars to Bob, and settle with Bob. By means of this chain, Alice has in fact sent Bob 10 USD. 
}
\end{adjustwidth}

\subsection{The Interledger protocol suite}
\label{section:protocols}

The Interledger architecture is often compared with the Internet architecture, as in Table \ref{tab: parIPILP}. As a matter of fact, they adopt the same layered approach \cite{Ipilparch}. It involves multiple protocols, but the most important are \acrshort{btp}, \acrshort{ilp}, STREAM and \acrshort{spsp}, which are presented below.

\begin{table}[h]
        \centering
		\caption[A parallel between the Internet and Interledger architectures]{A parallel between the Internet and Interledger architectures. \cite{Ipilparch}}
		\label{tab: parIPILP}
        \setlength{\tabcolsep}{0.6em}
			\begin{tabular}{c|c|c|c}
                \multicolumn{2}{c|}{Internet architecture}&\multicolumn{2}{c}{Interledger architecture} \\
                \hline
                \multicolumn{1}{l|}{L5 Application} & HTTP SMTP NTP  &  \multicolumn{1}{l|}{L5 Application} & SPSP HTTP-ILP Paytorrent\\
                 \multicolumn{1}{l|}{L4 Transport} & TCP UDP QUIC &  \multicolumn{1}{l|}{L4 Transport} & IPR PSK STREAM\\
                  \multicolumn{1}{l|}{L3 Internetwork} & IP &  \multicolumn{1}{l|}{L3 Interledger} & \acrshort{ilp}\\
                   \multicolumn{1}{l|}{L2 Network} & PPP Ethernet WiFi &  \multicolumn{1}{l|}{L2 Link} & BTP\\
                    \multicolumn{1}{l|}{L1 Physical} & Copper Fiber Radio &  \multicolumn{1}{l|}{L1 Ledger} & Blockchains, Central Ledgers
                
			\end{tabular}
    \end{table}


\subsubsection{The Simple Payment Setup Protocol} 
The \acrfull{spsp} is a protocol for exchanging the required information in order to set-up an Interledger payment between a payee and a payer. It is the most widely used Interledger Application Layer Protocol today \cite{scalingconnc}. \acrshort{spsp} makes use of the STREAM protocol to generate the ILP condition and for data encoding.
        
Because STREAM does not specify how to exchange the required payment details, some other protocol and app have to implement this.  \acrshort{spsp} is a protocol that uses HTTP for exchanging payment details between the \textit{sender} and the \textit{receiver}, such as the \textit{\acrshort{ilp} address} or \textit{shared secret} \cite{rfcspsp}. In other words, \acrshort{spsp} is a means for exchanging the server details needed for a client to establish a STREAM connection. It is intended for use by end-user applications,  such as a digital wallet with a user interface to initiate payments. The \acrshort{spsp} Clients and Servers make use of the STREAM module in order to further process the \acrshort{ilp} payments. \acrshort{spsp} messages MUST be exchanged over HTTPS.
        
\textbf{\textit{Payment pointers}} can be used as a persistent identifier on Interledger. They are a standardized identifier for accounts that are able to receive payments \cite{paymentpointer}. \textit{Payment pointers} can also be used as a unique identifier for an invoice to be paid or for a pull payment agreement. The main characteristics of the payment pointers are:
        
        \begin{itemize}
            \item \textit{Unique and Easily Recognizable}
            \item \textit{Simple Transcription}: It should be easy to exchange the payment pointer details with a payee
            \item \textit{Flexible}: avoid being strongly connected to a specific protocol
            \item \textit{Widely Usable}: should be easy to implement and use.
        \end{itemize}

The syntax is: \textit{"\textcolor[HTML]{008000}{\$} \textcolor[HTML]{0087BD}{host} path-abempty"} \cite{paymentpointer}. \newline

The \textbf{\textit{\acrshort{spsp} Endpoint}} is a URL used by the \acrshort{spsp} Client to connect to the \acrshort{spsp} Server, obtain information about it, and set up payments. The \textit{payment pointer} automatically resolves to the \textit{SPSP endpoint} by means of a simple algorithm. If \textit{"path-abempty"} is not specified, it is replaced with \textit{"/.well-known/pay"} \cite{paymentpointer}. Table \ref{tab: pptoep} also explains how a Payment pointer is resolved to an \acrshort{spsp} endpoint.

    \begin{table}[h]
        \centering
		\caption[Payment pointer to Endpoint conversion]{Payment pointer to Endpoint conversion. \cite{paymentpointer}}
		\label{tab: pptoep}
        \setlength{\tabcolsep}{0.7em}
			\begin{tabular}{l|l}
                Payment pointer         & \acrshort{spsp} endpoint \\
                \hline
                \textcolor[HTML]{008000}{\$}\textcolor[HTML]{0087BD}{example.com} & \textcolor[HTML]{008000}{https://}\textcolor[HTML]{0087BD}{example.com}/.well-known/pay\\
                \textcolor[HTML]{008000}{\$}\textcolor[HTML]{0087BD}{example.com}/invoices/12345 & \textcolor[HTML]{008000}{https://}\textcolor[HTML]{0087BD}{example.com}/invoices/12345 \\
                \textcolor[HTML]{008000}{\$}\textcolor[HTML]{0087BD}{bob.example.com} & \textcolor[HTML]{008000}{https://}\textcolor[HTML]{0087BD}{bob.example.com}/.well-known/pay\\
                \textcolor[HTML]{008000}{\$}\textcolor[HTML]{0087BD}{example.com}/bob & \textcolor[HTML]{008000}{https://}\textcolor[HTML]{0087BD}{example.com}/bob\\
			\end{tabular}
    \end{table}
    
    Technically, payments could be performed without the use of \textit{payment pointers / \acrshort{spsp} endpoints}. However, for ease of use, for the practical reasons explained above, the \textit{payment pointers} and \acrshort{spsp} have been introduced on top of the existing technology. 
    
    In conclusion, there are multiple forms of identification, associated to different levels of the system:
    \begin{itemize}
        \item \textit{Ripple account address} or \textit{Ripple wallet address}, used at \textit{ledger level} to identify your user account holding the money on the ledger. It is analog to your bank account number
        \item \textit{ILP address}, used at \textit{ILP level} to uniquely identify your ILP node in the global network. Could be compared to an IP address.
        \item \textit{SPSP endpoint}, used by the \textit{SPSP protocol/app} to set-up an ILP payment
        \item \textit{payment pointer} used by \textit{humans} as an easy-to-handle unique payment identifier, in the same fashion as an email account address. It is also used in association with SPSP.
    \end{itemize}
    
        \begin{figure}[h!]
   			\begin{minipage}{\textwidth}
   			\centering
       		\includegraphics[width=1\textwidth]{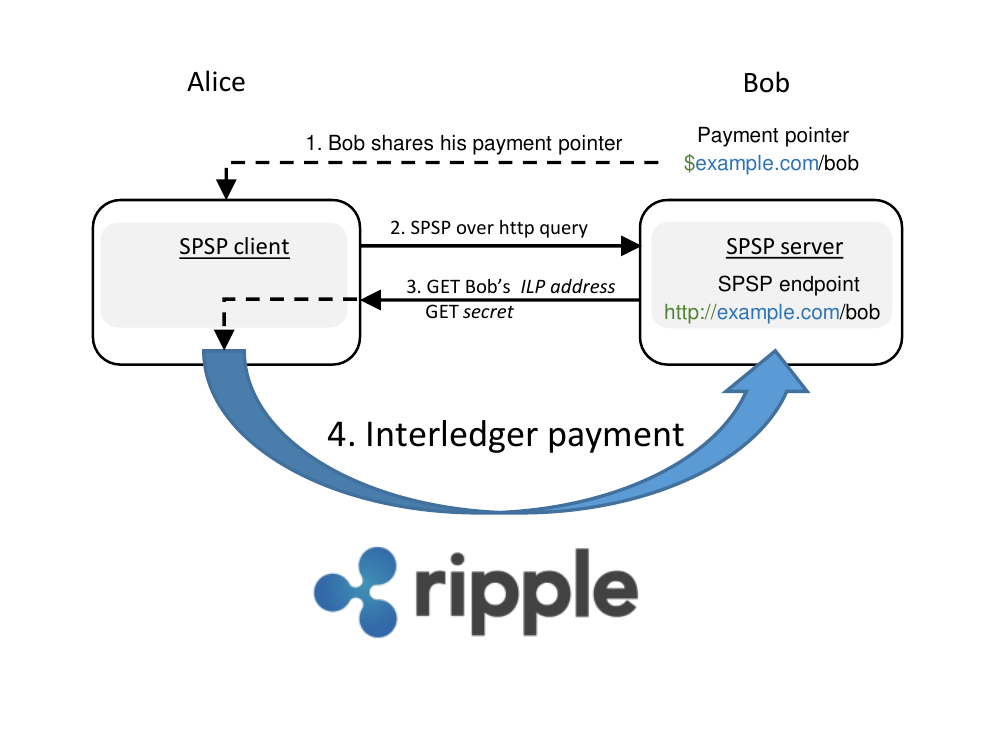}
       		\vspace{-1cm}
			\caption[Example \ref{exmp_spsp}: SPSP payment]{Example \ref{exmp_spsp}: SPSP payment.}
            \label{fig: SPSPex0}
        	\end{minipage}
		\end{figure}
    
    \begin{exmp}
    \label{exmp_spsp}
    Concerning the Figure \ref{fig: SPSPex0}: \\
    
    \noindent - Suppose each Alice and Bob have their own accounts in \gls{xrp} on the \gls{xrp} Ledger. \\
    - Alice wants to pay Bob in \gls{xrp}. \\
    - Bob already has his own\textit{ payment pointer, \textcolor[HTML]{008000}{\$}\textcolor[HTML]{0087BD}{example.com}/bob}, which he easily shares to Alice verbally. This information is easy for Alice to remember and use. (step (1) in Figure \ref{fig: SPSPex0}) \\
    - Supposing the machine already set-up, at her computer, Alice starts up \gls{moneyd}, and is able to pay Bob just by introducing his \textit{payment pointer address} \textit{"\textcolor[HTML]{008000}{\$}\textcolor[HTML]{0087BD}{example.com}/bob"} and the amount paid in Moneyd-\acrshort{gui}, or from a terminal, by using a single line: \\
    
    \textit{"ilp-spsp send --receiver \$example.com/bob --amount 100"}\\
    
    Behind the scenes: \\
    
    \noindent Alice's \acrshort{spsp} client: \\
   \noindent - resolves the \textit{payment pointer "\textcolor[HTML]{008000}{\$}\textcolor[HTML]{0087BD}{example.com}/bob"} to \textcolor[HTML]{008000}{https://}\textcolor[HTML]{0087BD}{example.com}/bob \\
    - connects over HTTPS to Bob's \acrshort{spsp} server, at the address "\textcolor[HTML]{008000}{https://}\textcolor[HTML]{0087BD}{example.com}/bob" (2)\\
    - queries the \acrshort{spsp} server for Bob's \textit{\acrshort{ilp} address} and a \textit{unique secret} (2)\\
    Bob's \acrshort{spsp} server sends Bob's \textit{\acrshort{ilp} address} and a \textit{secret} to Alice's \acrshort{spsp} client (3)\\
    Using this information, Alice's \acrshort{spsp} client starts an \acrshort{ilp} payment to Bob over Interledger (4). \\
    \end{exmp}
        
The usage of public endpoints involves employing HTTP, TLS, DNS, and the Certificate Authority system for the HTTPS request that \acrshort{spsp} makes. \textit{However, the alternatives leave much to be desired. Few people outside of the cryptocurrency world want cryptographic keys as identifiers and, as of today, there is no alternative for establishing an encrypted connection from a human-readable identifier that is anywhere nearly as widely supported as DNS and TLS} \cite{scalingconnc}. \newline

\subsubsection{The Streaming Transport for the Realtime Exchange of Assets and Messages}

The \textbf{Streaming Transport for the Realtime Exchange of Assets and Messages} (STREAM) is a Transport Protocol working with \acrshort{ilp}v4. Application level protocols like \acrshort{spsp} make use of the STREAM protocol to send money. STREAM splits payments into packets, sends them over \acrshort{ilp}, and reassembles them automatically. It can be used to stream micropayments or larger discrete payments and messages. It is a successor of the Pre-Shared Key V2 (PSK2) Transport Protocol and is inspired by the QUIC Internet Transport Protocol.

        \begin{figure}[h!]
   			\begin{minipage}{\textwidth}
   			\centering
       		\includegraphics[width=0.8\textwidth]{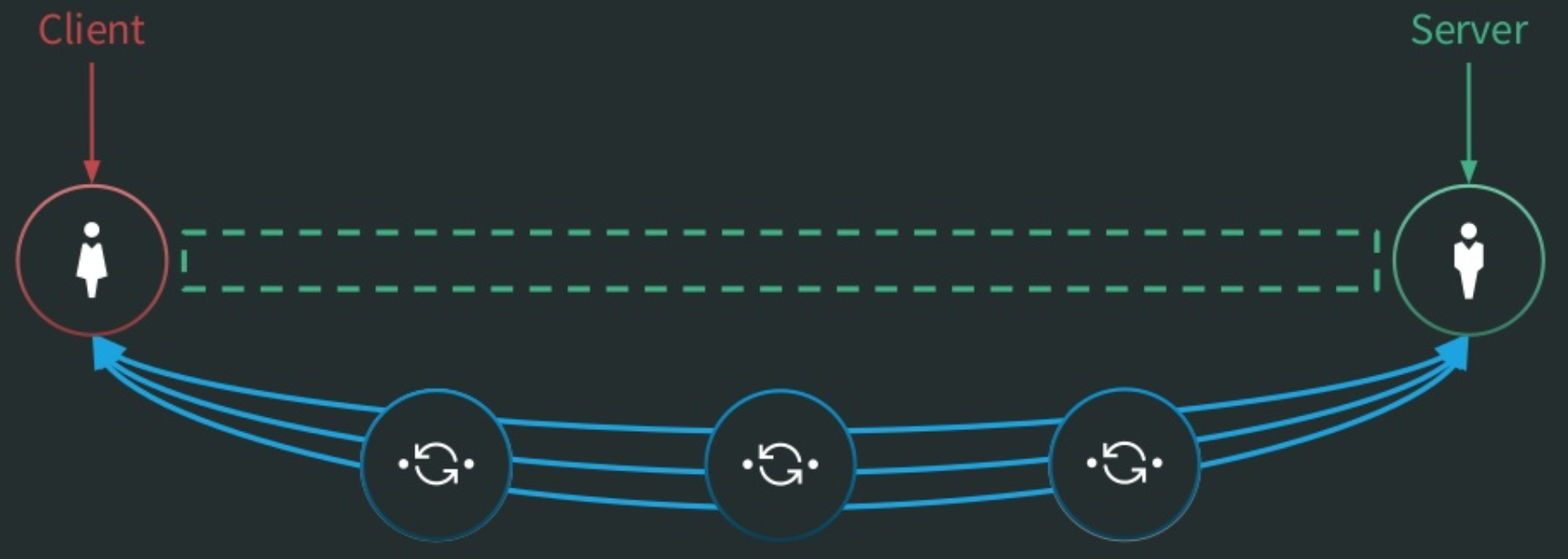}
			\caption[STREAM]{STREAM is a logical, bidirectional channel over ILP. \cite{Ipilparch}}
            \label{fig: STREAMlogic}
        	\end{minipage}
		\end{figure}

As illustrated in Figure \ref{fig: STREAMlogic} with green, a STREAM connection establishes a two-ways, virtual channel of data and money between the payer and payee. STREAM packets are encoded, encrypted, and sent as the data field in \textit{\acrshort{ilp} Prepare} (type 12 \acrshort{ilp} packet), \textit{Fulfill} (type 13 \acrshort{ilp} packet), or \textit{Reject} packets (type 14 \acrshort{ilp} packet). The logical connection is used to send authenticated \acrshort{ilp} packets between the "client" and "server" (the blue connections in Figure \ref{fig: STREAMlogic}). Either the payer or the payee can be the server or the client. STREAM provides authentication, encryption, flow control (ensure one party doesn't send more than the other can process), and congestion control (avoid flooding the network over its processing power).

STREAM servers are waiting for clients to connect over \acrshort{ilp}. The servers connect to a specific plugin on the local machine and wait for the \acrshort{ilp} packets. Usually, \textit{ilp-plugin} is used to connect to \gls{moneyd}. The server generates a unique \textit{\acrshort{ilp} address} and \textit{shared secret}, which will be used to encrypt data and generate fulfillments for \acrshort{ilp} packets in relation to a specific client. The \textit{request} for the address and secret, and the \textit{response}, are not handled by STREAM, but for example by \acrshort{spsp}. After a client has the \acrshort{ilp} address and secret (obtained with \acrshort{spsp} for example), it can connect to the STREAM server by using these credentials \cite{streamedium, streamrfc}.\\

    \begin{exmp}
    \label{exmp_stream}
    We now provide a more advanced explanation regarding the same situation presented in Example \ref{exmp_spsp}. We will refer to Figure \ref{fig: SPSPex01}, and extend the explanation from Example \ref{exmp_spsp}:
    
    	\begin{figure}[h!]
   			\begin{minipage}{\textwidth}
   			\centering
       		\includegraphics[width=1\textwidth]{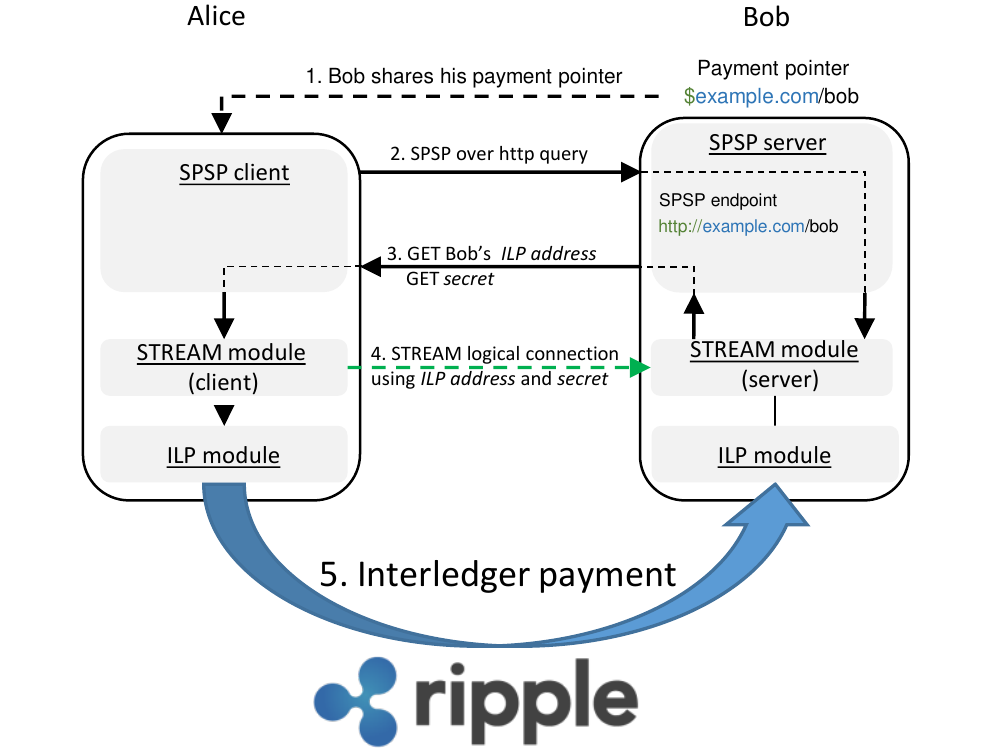}
			\caption[Example \ref{exmp_stream}: STREAM payment]{Example \ref{exmp_stream}: STREAM payment.}
            \label{fig: SPSPex01}
        	\end{minipage}
		\end{figure}

\begin{itemize}
    \item Alice's \acrshort{spsp} client:
        \begin{itemize}
            \item resolves the \textit{payment pointer "\textcolor[HTML]{008000}{\$}\textcolor[HTML]{0087BD}{example.com}/bob"} to \textcolor[HTML]{008000}{https://}\textcolor[HTML]{0087BD}{example.com}/bob
            \item connects over HTTP to Bob's \acrshort{spsp} server at address \textcolor[HTML]{008000}{https://}\textcolor[HTML]{0087BD}{example.com}/bob (2)
            \item queries the \acrshort{spsp} server for Bob's \textit{\acrshort{ilp} address} and a \textit{unique secret} (2). The \acrshort{spsp} server forwards the request to the STREAM server module and fetches the answer
        \end{itemize}
    \item Bob's \acrshort{spsp} server sends Bob's \textit{\acrshort{ilp} address} and the \textit{secret} to Alice's \acrshort{spsp} client (3)
    \item Alice's \acrshort{spsp} client passes the credentials to the STREAM client module which initiates a logical STREAM connection over \acrshort{ilp}, using the \acrshort{ilp} modules (4)
    \item Bob receives his payment over the Interledger (5).
\end{itemize}
\end{exmp}

\begin{figure}[h!]
   			\begin{minipage}{\textwidth}
   			\centering
       		\includegraphics[width=1.3\textwidth, angle=90, origin=c]{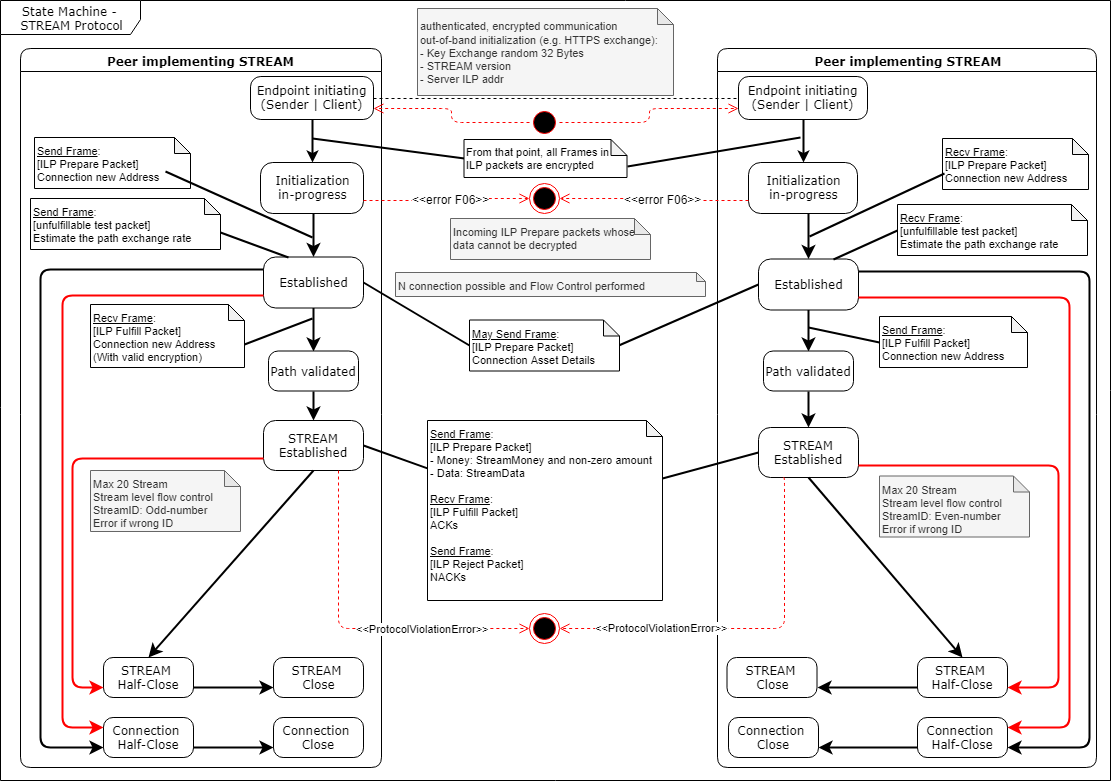}
			\caption[STREAM protocol: \acrshort{fsm} diagram]{STREAM protocol: \acrshort{fsm} diagram.}
            \label{fig: STREAM_FSM}
        	\end{minipage}
		\end{figure}

Further details on the STREAM protocol are illustrated in the finite state machine diagram of the STREAM protocol, presented in Figure \ref{fig: STREAM_FSM}.

\subsubsection{The Interledger Protocol}

The \acrfull{ilp}, currently at version 4, is the main protocol facilitating the Inter Ledger money transfers. It provides a solution to route payments across disconnected ledgers while minimizing the sender and receiver's risk of losing funds. What makes it different from previous versions is that it is optimized for sending many low value packets:

\textit{"We talked about the idea of streaming payments, where if you make payments so efficient that you could pay for like a milliliter of beer or a second of video. That's the way we think about efficiency of payments."}\cite{beer}

It is made for payment channels, which means faster and cheaper payments, while also accommodating any type of ledger. 

    \begin{figure}[h!]
   		\begin{minipage}{\textwidth}
   		\centering
        \includegraphics[width=0.8\textwidth]{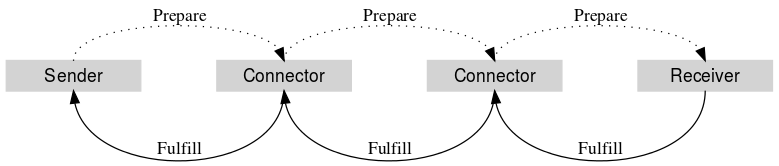}
		\caption[ILP packet flow]{ILP packet flow. \cite{ILParch}}
        \label{fig: packetflow}
        \end{minipage}
	\end{figure}

\acrshort{ilp}v4 involves \textit{Hashed Time Lock Agreements (HTLA)} \cite{ilpv4paper}, and makes use of three packet types:
     \begin{itemize}
         \item \textit{Prepare}, corresponding to request, with the following fields:
            \begin{itemize}
                \item \textit{destination} - \acrshort{ilp} address,
                \item \textit{amount} - UInt64, 
                \item \textit{condition} - UInt256,
                \item \textit{expiration} - timestamp,
                \item \textit{end-to-end (sender-receiver) data} - OCTET STRING. 
            \end{itemize}
            Example of an ilp-prepare packet:
            \begin{lstlisting}
{
amount= 69368000, 
executionCondition= fHII9adb3JY3D5drSNSoquLTIUJJhNLMeiiADnW4li0=, 
expiresAt= 2019-06-19T11:04:18.149Z, 
destination= g.conn1.ilsp_clients.mduni.local.viby9ZjztwCVMtptFjaueqsdlIxWSUba
        y7Jo3BxJyGc.elrqFEKZEc8BMcZ4PDUiPEAF, 
data= t6lmRiiFZecXhltYNsnyPYSgPld+Itmn+NefM5ytnFJiFDuMieyF9b2vB
        o2HPiNm34GpCBlU/HoGaCAsOQ==
}
            \end{lstlisting}
         \item \textit{Fulfill,} corresponding to response, and carrying the following fields:
            \begin{itemize}
                \item \textit{execution condition fulfillment} - UInt256, \\
                This is the proof that the receiver has been paid, so the fulfill packets are relayed back by the connectors from the receiver to the sender. It consists of a simple pre-image of a hash, and only the receiver can know this information.
                \item \textit{end-to-end (sender-receiver) data} - OCTET STRING.
            \end{itemize}
            The components of the prepare and fulfill packets concerning HTLA are:\\
            - \textit{amount, time} (expiration), and \textit{condition} for Prepare, and \\
            - the \textit{execution condition fulfillment (the hash)} for the Fulfill packet, which must be received before \textit{expiration}. This implies that the machines involved in the process should be time-synchronized. This is not an absolute enforcement, but any time offsets will packet rejection chances.
         \item \textit{Reject,} corresponding generally to error messages. They can be returned either by the receiver or the connectors in specific conditions and consist of:
            \begin{itemize}
                \item \textit{a standardized error code}, 
                \item \textit{triggered by:} - \acrshort{ilp} address; \\ 
                is the identifier of the participant that originally generated the error,
                \item \textit{user-readable error message} - UTF8String,
                \item \textit{machine readable error data} - OCTET STRING.
            \end{itemize}
     \end{itemize}

The connectors forward the \textit{prepare} packets from the \textit{sender} to the \textit{receiver}, and relay back the response or the reject from the \textit{receiver} to the \textit{sender,} as shown in Figure \ref{fig: packetflow}. As such, \acrshort{ilp} v4 uses a chaining of HTLAs to achieve an end-to-end transfer \cite{ilpv4paper}. In \acrshort{ilp} v4, HTLAs are mainly supported over Simple Payment Channels. Simple Payment Channels are generally supported by today's major blockchains like BTC, ETH, \gls{xrp},.. \cite{htlapaychan,htlaspc}.

Concerning Figure \ref{fig: packetflow} and \acrshort{ilp} v4: even if the original \acrshort{ilp} packet is prepared by the \textit{Sender} and addressed for the \textit{Receiver} (end-to-end), the transfer from the \textit{Sender} to the \textit{Receiver} will be in fact a chaining of transfers between the directly connected (and trusted) peers. Each pair of directly connected peers generally uses a dedicated, separate Payment Channel to settle their obligations \cite{ilpv4paper,htlapaychan,htlaspc}. Other means are possible \cite{htlaspc}, but not really used or supported \cite{htlapaychan}. \\

        \begin{figure}[h!]
   			\begin{minipage}{\textwidth}
   			\centering
       		\includegraphics[width=1\textwidth]{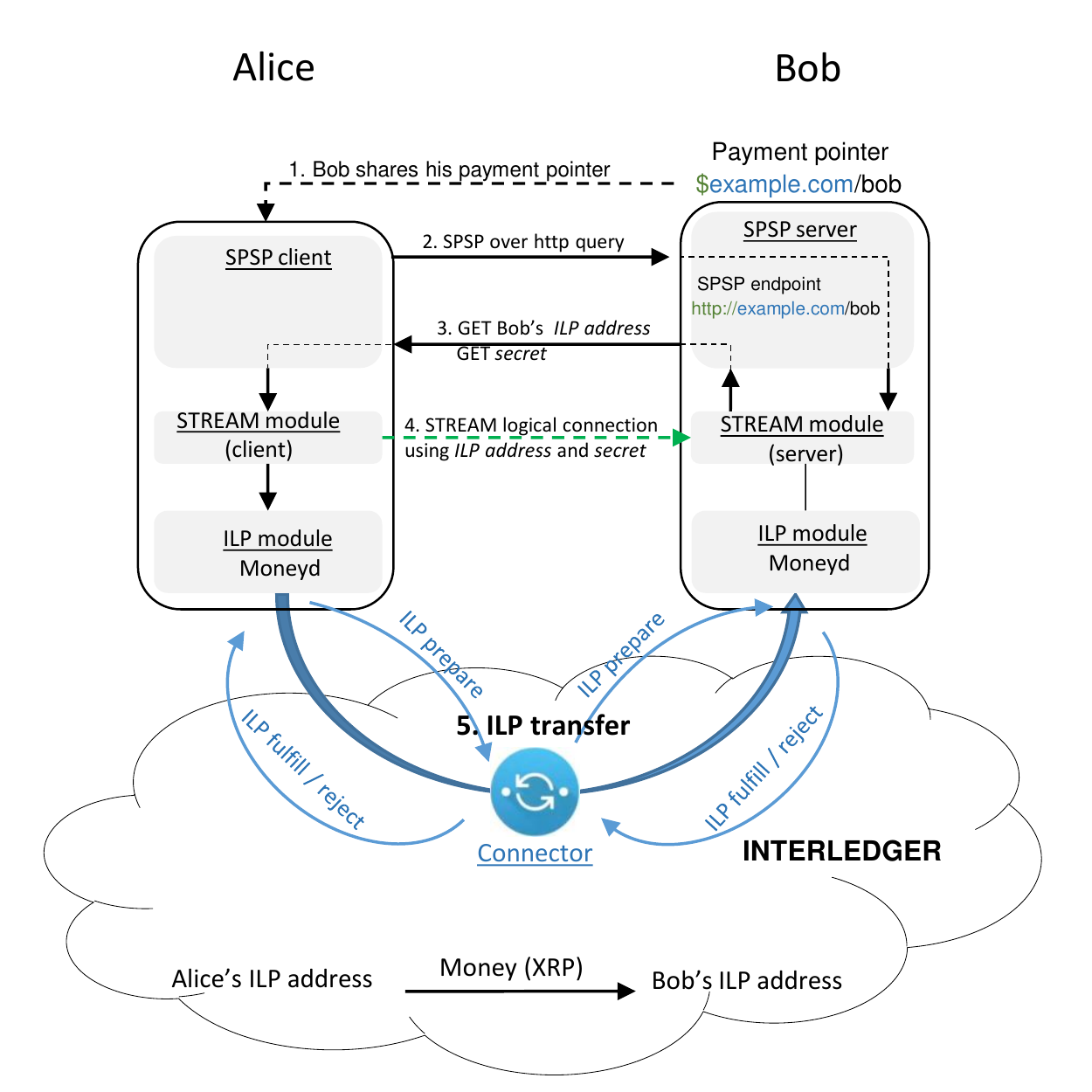}
			\caption[Example \ref{exmp_ilp}: ILP]{Example \ref{exmp_ilp}: ILP.}
            \label{fig: Ex3}
        	\end{minipage}
		\end{figure}
		
	\begin{figure}[h!]
   		\begin{minipage}{\textwidth}
   		\centering
        \includegraphics[width=1.3\textwidth,angle=90,origin=c]{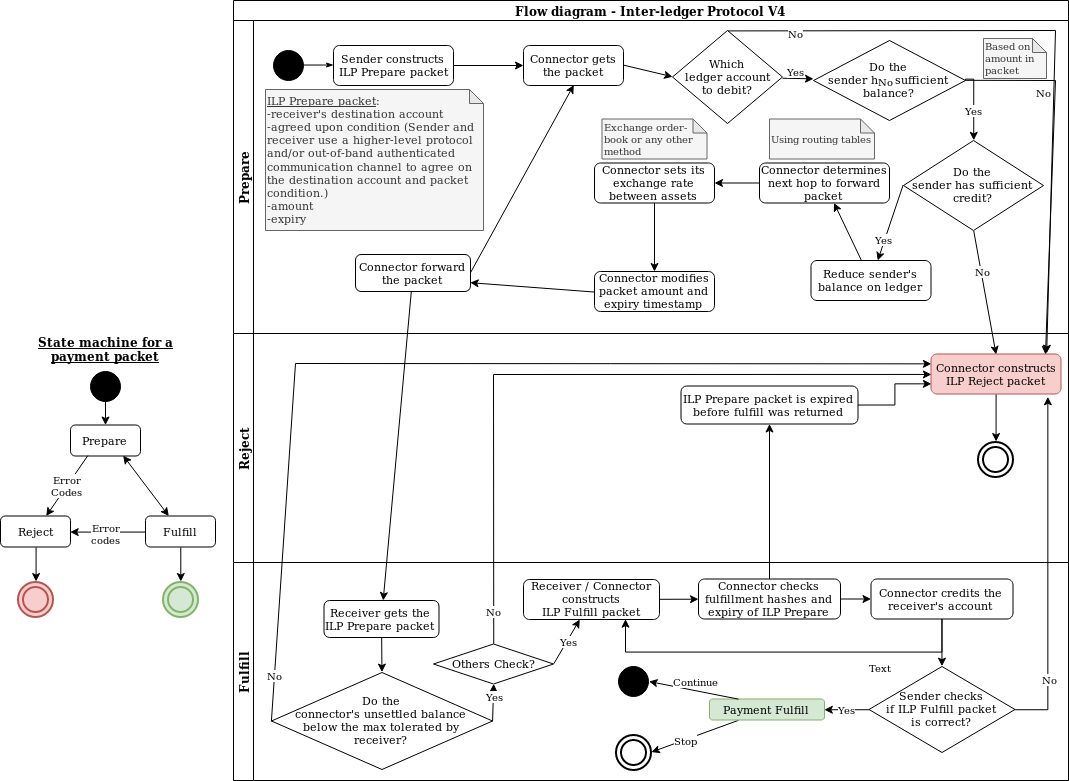}
		\caption[ILPv4 flow diagram]{ILPv4 flow diagram.}
        \label{fig: ILPv4diag}
        \end{minipage}
	\end{figure}
		
\begin{exmp}
\label{exmp_ilp}
 We will further expand on the Examples \ref{exmp_spsp} and \ref{exmp_stream}, using the Figure \ref{fig: Ex3}. 

\begin{itemize}
    \item Alice's \acrshort{spsp} client:
        \begin{itemize}
            \item resolves the \textit{payment pointer "\textcolor[HTML]{008000}{\$}\textcolor[HTML]{0087BD}{example.com}/bob"} to \textit{\textcolor[HTML]{008000}{https://}\textcolor[HTML]{0087BD}{example.com}/bob}
            \item connects over HTTP to Bob's \acrshort{spsp} server at address \textit{\textcolor[HTML]{008000}{https://}\textcolor[HTML]{0087BD}{example.com}/bob} (2)
            \item queries the \acrshort{spsp} server for Bob's \textit{\acrshort{ilp} address} and a \textit{unique secret} (2). The \acrshort{spsp} server forwards the request to the STREAM server module and fetches the answer
        \end{itemize}
    \item Bob's \acrshort{spsp} server sends Bob's \textit{\acrshort{ilp} address} and the \textit{secret} to Alice's \acrshort{spsp} client (3)
    \item Alice's \acrshort{spsp} client passes the credentials to the STREAM client module which initiates a logical STREAM connection over \acrshort{ilp}, using the \acrshort{ilp} module, in our case, \gls{moneyd} (4)
    \item The \acrshort{ilp} module, \gls{moneyd}, sends the \acrshort{ilp} packets corresponding to the STREAM virtual connection towards its upstream parent connector, which further routes them to its child, Bob's \gls{moneyd} module (5).
\end{itemize}
\end{exmp}

The STREAM module is able to break the payment into multiple packets, which would be sent over \acrshort{ilp} using \textit{prepare-fulfill-error} packets. The STREAM module at the \textit{receiver's} end will finally reassemble the payment.

Wrapping-up \acrshort{ilp}, Figure \ref{fig: ILPv4diag} presents the finite state machine diagram of an \acrshort{ilp} packet and the protocol flow chart. 

\subsubsection{The Bilateral Transfer Protocol} 
The \acrfull{btp} emerged as a necessity, due to a combination of \acrshort{ilp} goals (fast and cheap transactions) and the realities of some ledgers (expensive and/or slow settlements). With \acrshort{btp}, two parties can send funds directly to each other, up to a maximum amount they are willing to trust before settlement. \acrshort{btp} is used between connectors (\gls{moneyd} included) for transferring \acrshort{ilp} packets and messages necessary to exchange payments, settlement, configuration and routing information. 

        \begin{figure}[h!]
   			\begin{minipage}{\textwidth}
   			\centering
       		\includegraphics[width=0.6\textwidth]{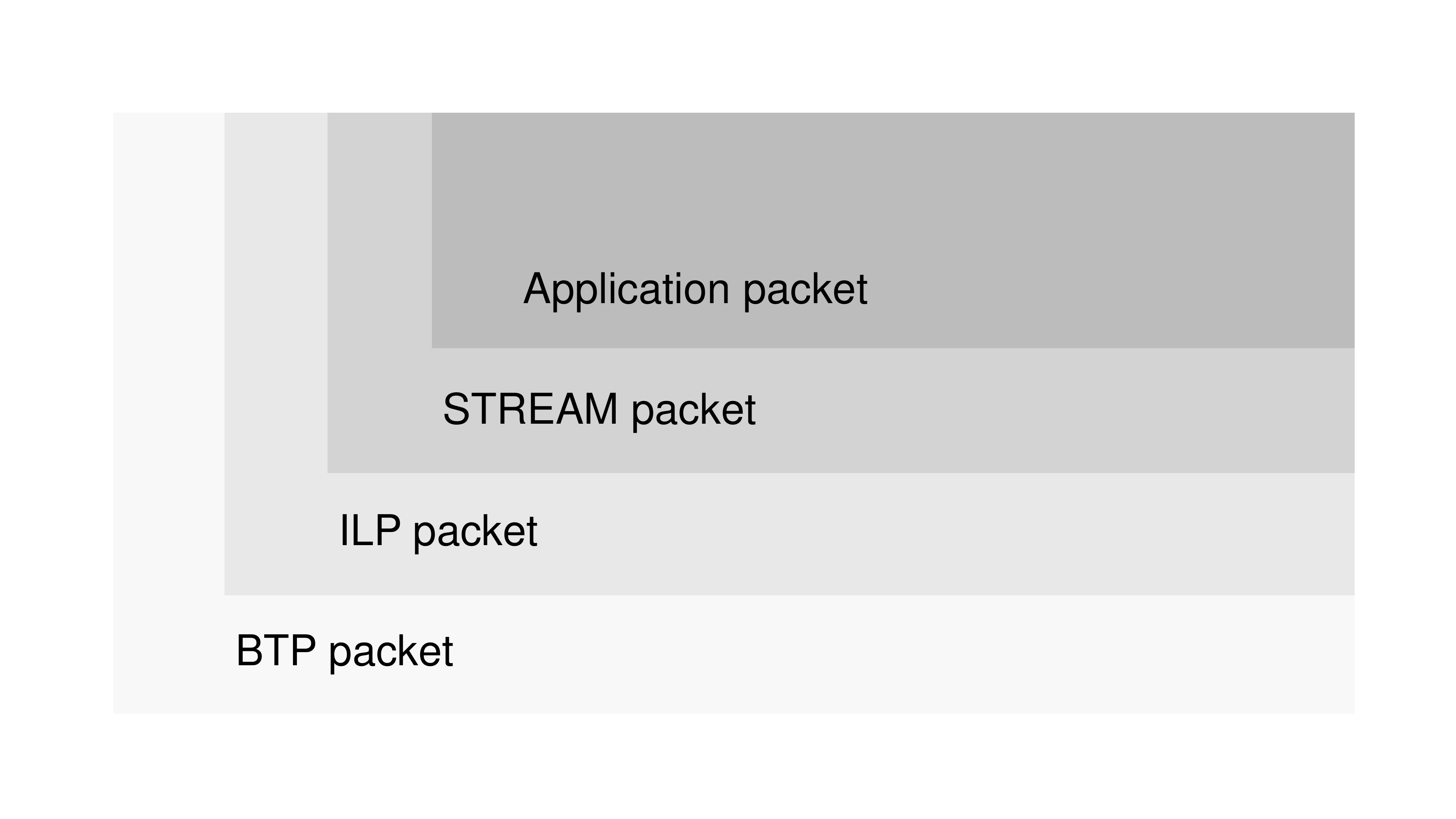}
			\caption[Packet data structure]{Packet data structure. \cite{protorel,ILPadvdiag}}
            \label{fig: packets}
        	\end{minipage}
		\end{figure}

\begin{figure}[h!]
   			\begin{minipage}{\textwidth}
   			\centering
       		\includegraphics[width=0.4\textwidth]{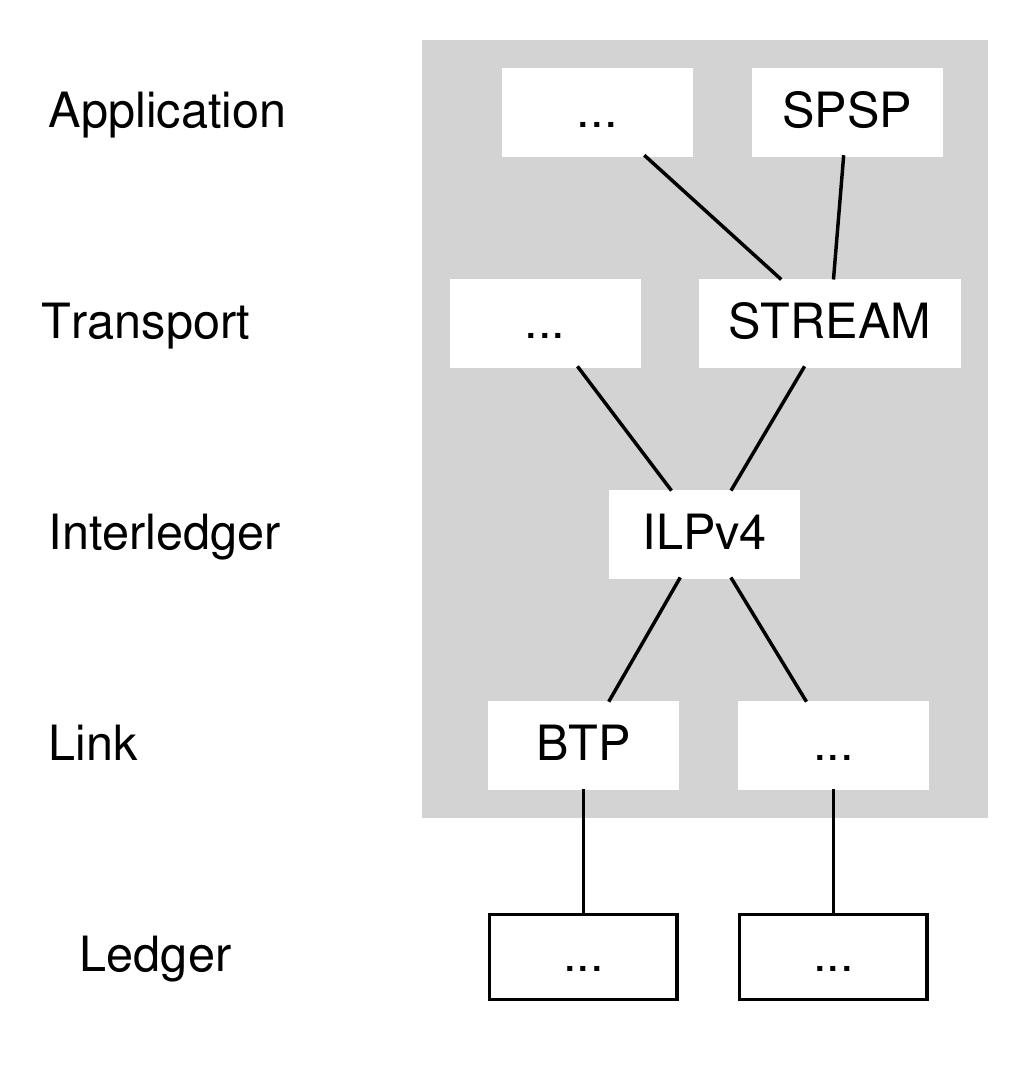}
			\caption[The protocol suite]{The protocol suite. \cite{ILParch}}
            \label{fig: protosuite}
        	\end{minipage}
		\end{figure}

As shown in Figure \ref{fig: packets}, \acrshort{btp} is a "carrier" for \acrshort{ilp} packets and as such, for other protocols like STREAM for example. \acrshort{btp} establishes the "link" between connectors, on top of which the \acrshort{ilp} packets are being sent. When setting-up the connector plugins, one also generally sets-up a \acrshort{btp} connection. The data is sent over web socket connections. One of the peers acts as a server while the other is connected as a client. It implements a Bilateral Ledger, where the two peers keep track of their (yet) un-settled accounts and balances. The Bilateral Ledger, a micro-ledger kept by the two peers in-between them, is not to be confused with the Underlying Ledger - the main ledger where all accounts and transactions are stored, e.g. the Ripple ledger. With regards to Figure \ref{fig: protosuite}, it is to be noted that \acrshort{ilp} can still work without \acrshort{btp} \cite{BTP}.

With regards to the current state of BTP, the following post by Ewan Schwartz is worth mentioning: \textit{"BTP is a binary request/response protocol implemented over WebSockets. It originally included message types for Prepare, Fulfill, Reject and Transfer, but BTP 2.0, which is used today, stripped out nearly everything except request/response semantics, authentication, and "sub-protocol" naming".} \cite{scalingconnc} \\

\begin{exmp}
\label{exmp_btp}
\indent We can now complete our diagrams presented in Examples \ref{exmp_spsp}, \ref{exmp_stream} and \ref{exmp_ilp} with the \acrshort{btp} protocol, which is illustrated in Figure \ref{fig: exmpBTP}. 

In order to connect to Interledger, each Alice and Bob's \acrshort{ilp} modules establish a \acrshort{btp} connection over wss with the parent connector. As long as they are connected to Interledger, this connection will be live. The \acrshort{ilp} packets will travel over \acrshort{btp}. While opening the \acrshort{btp} connection, both of them also negotiate a unique paychan with their direct peer, the connector. 

It is to be noted that while we made this choice for clarity, the order we presented \acrshort{spsp}, STREAM, \acrshort{ilp} and \acrshort{btp} is not necessarily the real temporal order of events. This will be further explained in Examples \ref{exmp_xrpxrp}, \ref{exmp_xrpeth}, and \ref{exmp_xrpetc2c}.

        \begin{figure}[h!]
   			\begin{minipage}{\textwidth}
   			\centering
       		\includegraphics[width=0.8\textwidth]{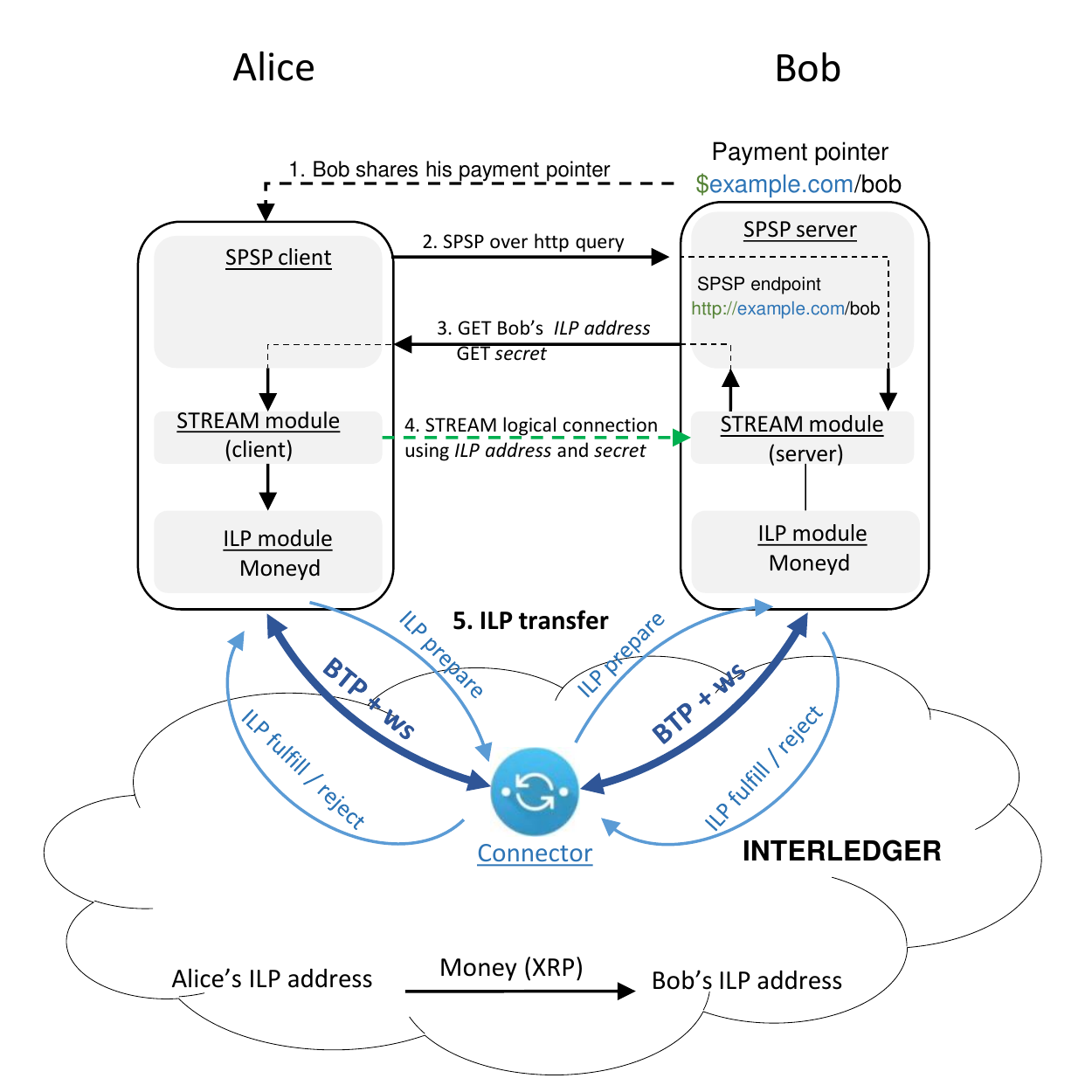}
			\caption[Example \ref{exmp_btp}: BTP ]{Example \ref{exmp_btp}: BTP. \cite{protorel,ILPadvdiag}}
            \label{fig: exmpBTP}
        	\end{minipage}
		\end{figure}

\end{exmp}

	\begin{figure}[h!]
   		\begin{minipage}{\textwidth}
   		\centering
        \includegraphics[width=1.3\textwidth,angle=90,origin=c]{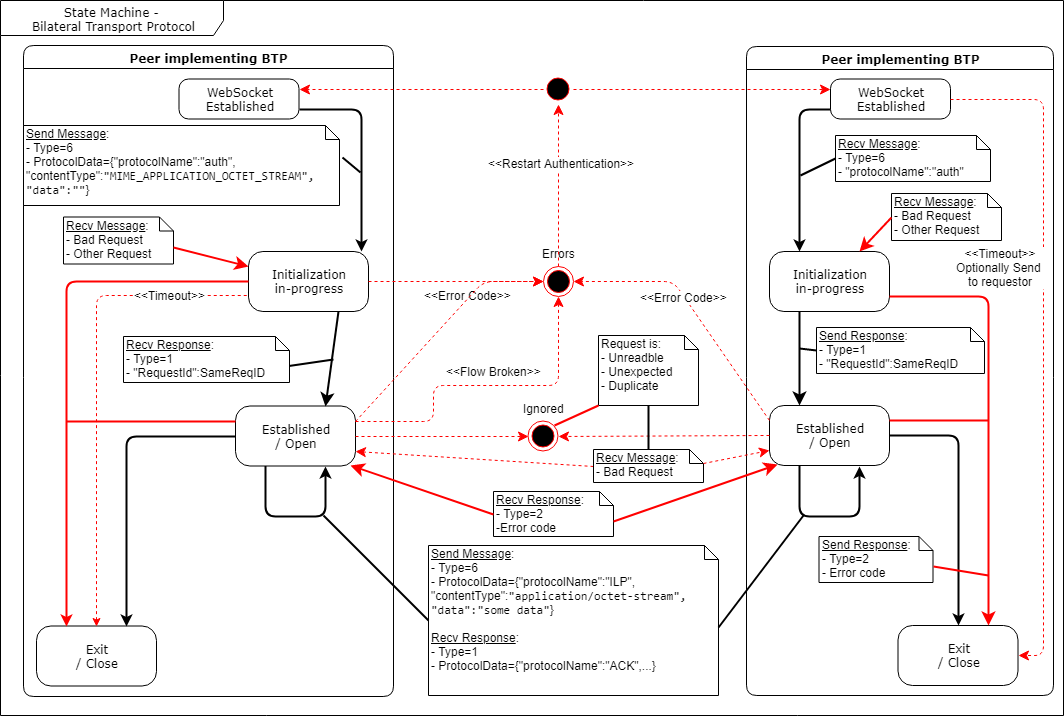}
		\caption[BTP: the finite state machine diagram]{BTP: the finite state machine diagram.}
        \label{fig: BTPdiag}
        \end{minipage}
	\end{figure}

The finite state machine of the \acrshort{btp} protocol is presented in Figure \ref{fig: BTPdiag}. 

\begin{figure}[p]
            \vspace{-3cm}
            \hspace{-1cm}
       		\includegraphics[width=18cm]{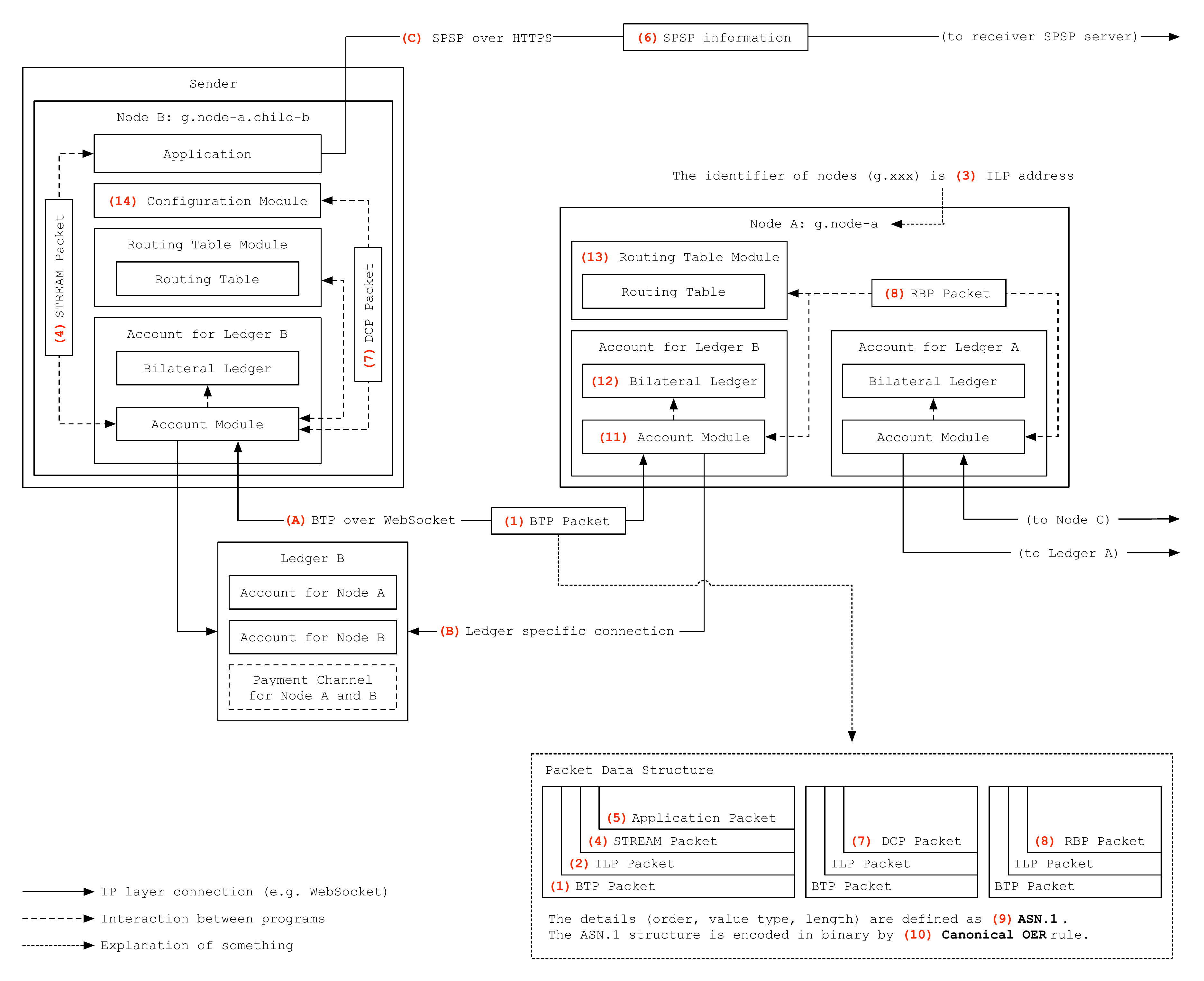}
			\caption[Protocols and details]{Protocols and details. Advanced diagram. \cite{protorel,ILPadvdiag}}
            \label{fig: protoadv}					
		\end{figure}

 
The relationship between protocols, and especially the STREAM protocol, can be best understood by referring to Figure \ref{fig: protoadv} at the end and reading the thorough explanations provided by \cite{protorel,ILPadvdiag}.\\

\indent Other protocols examples are the \textbf{Interledger Dynamic Configuration Protocol} (ILDCP), or the \textbf{Route Broadcasting Protocol} (RBP). DCP is built over \acrshort{ilp} and used to exchange node information such as \acrshort{ilp} address, while RBP is used to transfer routing information. Both use the data field in the \acrshort{ilp} packets \cite{protorel}.\\

In the examples that follow, we are going to make use of \acrshort{btp}, \acrshort{ilp}, STREAM and \acrshort{spsp}, a protocol suite also depicted in Figure \ref{fig: protosuite}.

\section{Customer apps for money transfer}
\label{section: custapps}
    
    \subsubsection[Moneyd, Moneyd-GUI and SPSP]{A customer environment comprising Moneyd, Moneyd-GUI and SPSP client/server.}
    \label{section: mdspsp}
        \textbf{\gls{moneyd}:} \textit{'\gls{moneyd}} \textit{provides a quick on-ramp as the "home router" of the Interledger, giving apps on your computer access to send and receive money. Although \gls{moneyd} can just as easily connect to the production Interledger, it is not currently designed for heavy production use, so it lacks features like budgeting that would let you give apps different spending limits and permissions levels.'} \cite{mdexpl}
        
        \gls{moneyd} connects to the Interledger and sends and receives \acrshort{ilp} packets for you. It is a simplified, end-user version of a connector, and as such, it will not send or receive routes as a regular connector does. Apps running on your machine that require access to \acrshort{ilp} will connect to the Interledger through \gls{moneyd}.
        
        When you fire-up \gls{moneyd}:
        \begin{itemize}
            \item \gls{moneyd} loads \textit{\textasciitilde/.moneyd.json} and instantiates an \acrshort{ilp} Connector
            \item The connector (i.e. "\gls{moneyd}") opens a web socket connection to its parent connector (configured in \textit{\textasciitilde/.moneyd.json}) and creates a payment channel, which can be in \gls{xrp}, Ethereum,.. depending on the uplink connection
            \item The connector (i.e. "\gls{moneyd}") listens on the local port 7768 to process \acrshort{ilp} payments. All your local apps (like \acrshort{spsp}) will connect here for \acrshort{ilp}.
        \end{itemize}
        
        To install \gls{moneyd} for \gls{xrp}\footnote{\href{https://github.com/interledgerjs/moneyd}{https://github.com/interledgerjs/moneyd}, accessed June 2019}, just type the following into the terminal: \\
        
        \textit{npm install -g moneyd moneyd-uplink-xrp}\\
        
        The best way to understand it is to configure it in advanced mode and start in DEBUG mode.
        
        Configure it using: \textit{'moneyd xrp:configure --advanced'}. A configuration file will be created in the user root folder, as \textit{ \textasciitilde/.moneyd.json}. If needed, this file can be inspected and manually updated afterwards.
        
        Start in DEBUG mode using: \textit{'DEBUG=* moneyd xrp:start --admin-api-port 7769'}. The option \textit{'--admin-api-port 7769'} will open the port 7769 as \gls{moneyd} admin port, such that Moneyd-\acrshort{gui} can connect to it and administration can be performed in a web browser. \\
        
        \textbf{Moneyd-\acrshort{gui}} \textit{is a frontend for \gls{moneyd} (or the reference js connector), which displays statistics, sends and receives money, and helps with troubleshooting} \footnote{\href{https://medium.com/interledger-blog/use-interledger-with-moneyd-gui-21dee0dc8ba0}{https://medium.com/interledger-blog/use-interledger-with-moneyd-gui-21dee0dc8ba0}, accessed June 2019}. Running on the same machine with \gls{moneyd} or with the js reference Connector, it can be installed with\footnote{\href{https://github.com/interledgerjs/moneyd-gui}{https://github.com/interledgerjs/moneyd-gui}, accessed June 2019}:\\ 
        \textit{npm install -g moneyd-gui.} \\
        It will connect automatically to the port above, after being started with: \\
        \textit{'npm start'} in \textit{'/home/user/moneyd-gui'}. \\
        In a web browser, it can be accessed by default at \href{http://127.0.0.1/7770}{http://127.0.0.1/7770}. \\
        
\begin{figure}[h!]
   			\begin{minipage}{\textwidth}
   			\centering
       		\includegraphics[width=1\textwidth]{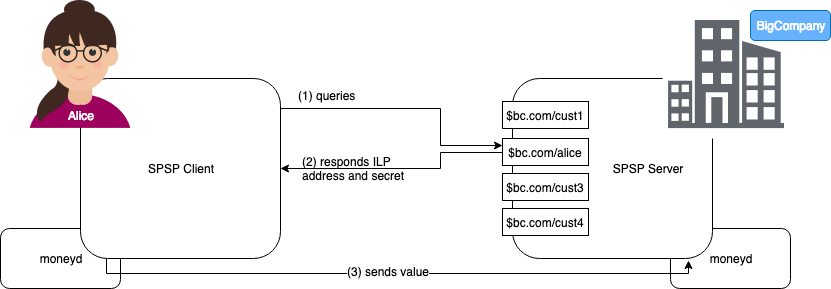}
			\caption[SPSP and Moneyd]{SPSP and Moneyd  \cite{spspmd}. Between the two Moneyd instances there are usually multiple connectors providing the \acrshort{ilp} service.}
            \label{fig: spspMd}
        	\end{minipage}
		\end{figure}

        \textbf{\acrshort{spsp}:} Another business case illustrating the way \acrshort{spsp} and \gls{moneyd}/\acrshort{ilp} are working together (because \gls{moneyd} is implementing \acrshort{ilp} for settlement) was well illustrated in Figure \ref{fig: spspMd}. Both Alice's and BigCompany's \acrshort{spsp} client and server are connected to \acrshort{ilp} through \gls{moneyd}. Alice's \acrshort{spsp} client connects to BigCompany's \acrshort{spsp} server which exposes an HTTPS endpoint abstracted as a payment pointer \cite{spspmd}.
        
        The protocol follows like below \cite{spspmd}:
        \begin{itemize}
            \item Alice queries her customer payment pointer (1), and
            \item receives the \acrshort{ilp} \textit{destination\_account} and a \textit{shared\_secret} (2)
            \item Using the data obtained, Alice's \acrshort{spsp} client starts a STREAM connection and sends the money to BigCompany (3).
        \end{itemize}   

        \indent  \textit{'The \acrshort{spsp} module calls the Interledger module with the address and other parameters in the Interledger packet to send a payment. 
        The Interledger module would send a transfer to the next connector or destination account along with the Interledger packet and according to the parameters given. The transfer and Interledger packet would be received by the next host's Interledger module and handled by each successive connector and finally the destination's \acrshort{spsp} module'} \cite{ilpccp}.
        
        \indent As a side note, \acrshort{spsp} first started with PSK and was later upgraded to STREAM. \newline
        
        Using the below procedure, one can send and receive money by writing a single line in the terminal. An \acrshort{spsp} server\footnote{\href{https://github.com/interledgerjs/ilp-spsp-server}{https://github.com/interledgerjs/ilp-spsp-server}, accessed June 2019} and a client\footnote{\href{https://github.com/interledgerjs/ilp-spsp}{https://github.com/interledgerjs/ilp-spsp}, accessed June 2019} should be installed on the recipient's and sender's machines, respectively. They work in pair and have as pre-requisite \gls{moneyd} but can also work alongside the reference connector. \\
        \indent The following lines are provided for the case of a private independent network using IPs as payment address identifiers. For real-life situations where addresses in the form of \textit{'\$sharafian.com'} can be provided, the original github guidelines can be used.
        
        In order to transfer money, we should first have a receiver address, i.e. \textbf{start the server} which listens for an incoming payment. Using 'DEBUG' will provide more info on what happens behind. \\ \newline
        \textbf{\acrshort{spsp} 'server' listening for an incoming transfer} \\
        Install the server with:\\
        \textit{'DEBUG=* ilp-spsp-server --localtunnel false --port 6000'} \\
        \textit{'--localtunnel false'}: forces the use of IP as the address instead of creating a tunnel and obtaining an address like \textit{'\$mysubdomain.localtunnel.me'} (analogue to \textit{"\$example.com/bob"}). \\
        \textit{'--port 6000'}: the server will listen for incoming payment from an \acrshort{spsp} 'client' on this port. \\ \newline
        \textbf{An \acrshort{spsp} 'client' sending money to the server} \\
        The money can be then \textbf{sent} by another user/machine having a similar setup, with: \\
        \textit{'DEBUG=ilp* ilp-spsp send --receiver http://192.168.1.116:6000 --amount 100'}. \\
        Where \textit{'http://192.168.1.116:6000'} is the receiving server's IP and port. \newline
        
To conclude this section we provide the following very nice explanation on \acrshort{spsp}, \acrshort{ilp}, \gls{moneyd} by Ben Sharafian on the \acrshort{ilp} forum, which we feel helps consolidate and clarify the larger picture:
        
\textit{In Interledger, the sender and receiver don't have any settlement relationship nor do they have any trust relationship.}

\textit{The only relationship is to your direct peer. If you're connecting to Interledger through \gls{moneyd}, we would call this peer your "upstream connector" or "uplink". When you send packets through your upstream connector, \gls{moneyd} keeps track of the amounts and settles them.}

\textit{\acrshort{spsp} isn't doing any of the tracking for settlements, that all lies at the Interledger layer. This lets you save a lot of headache when implementing high level protocols: settlement is always abstracted away}.

\textit{\gls{moneyd} does settle on-ledger. The configuration that \gls{moneyd} uses to connect to its upstream connector determines the currency that this happens in. (for \gls{xrp}, moneyd-uplink-xrp).} 

\textit{The currency that \acrshort{spsp} pays in depends on what currency is used by the \gls{moneyd} it connects to. If you use a \gls{moneyd} uplink in \gls{xrp}, then the currency will be \gls{xrp}} \cite{sharafianspsp}. \newline

\begin{exmp}
\label{exmp_xrpxrp}
\textbf{\gls{xrp} payment using \gls{moneyd}, \acrshort{spsp} and a connector.}\\

        We consider Figure \ref{fig: XRPpay} a good starting example because while involving a relatively simple structure, it still illustrates the main structural components of the Interledger ecosystem: 
        
        \begin{itemize}
            \item A ledger, i.e. the Ripple ledger
            \item A connector
            \item Customers:
            \begin{itemize}
                 \item Alice, operating Machine A with \gls{moneyd}-\gls{xrp} and \acrshort{spsp}
            \item Bob, operating Machine B with \gls{moneyd}-\gls{xrp} and \acrshort{spsp}
            \end{itemize}
            \item The main protocols: \acrshort{btp}, \acrshort{ilp}, STREAM and \acrshort{spsp}.
        \end{itemize}
        
        \begin{figure}[h!]
   			\begin{minipage}{\textwidth}
       		\includegraphics[width=1\textwidth]{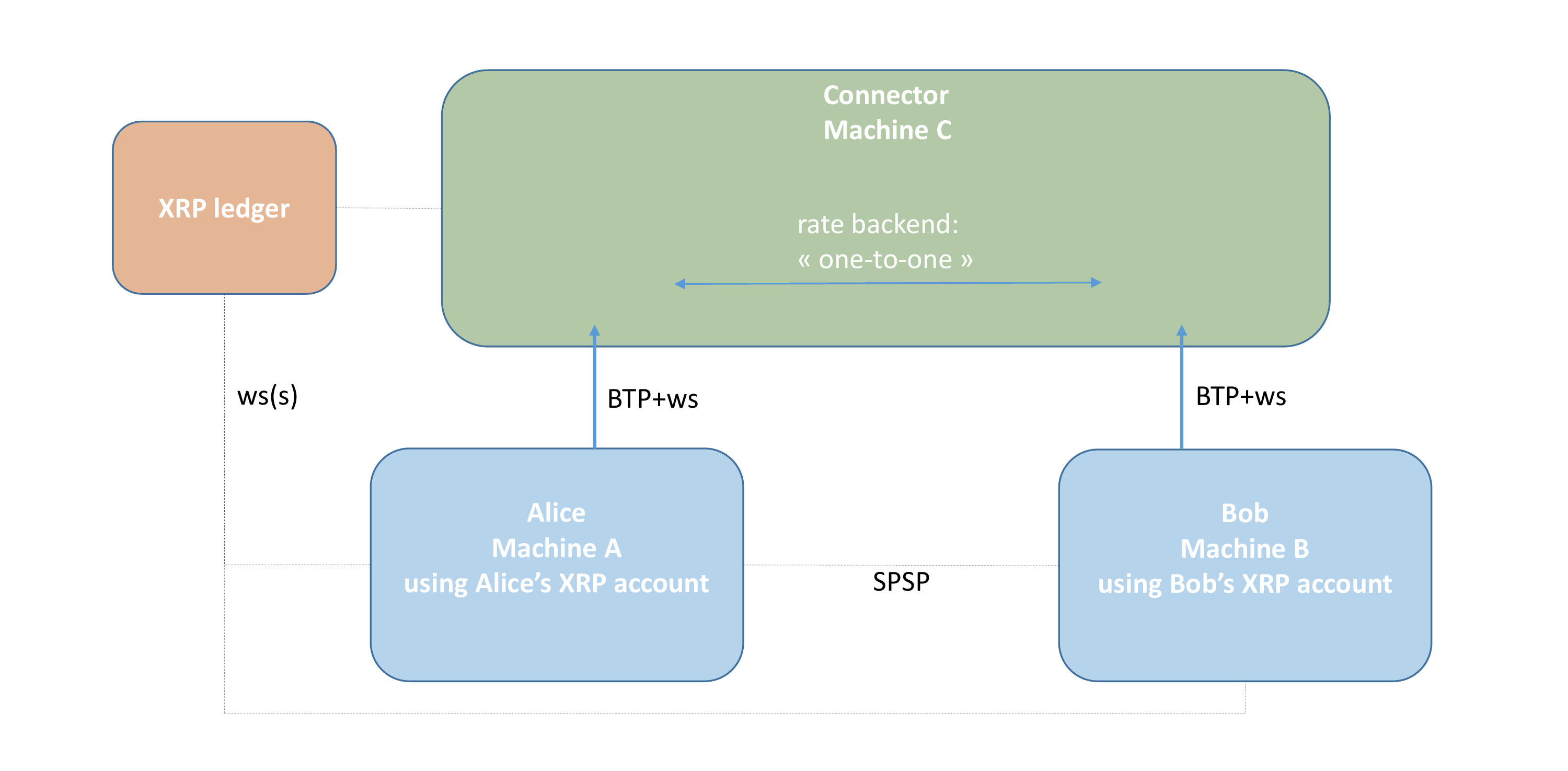}
			\caption[Example \ref{exmp_xrpxrp}: \gls{xrp} payment]{Example \ref{exmp_xrpxrp}: \gls{xrp} payment.}
            \label{fig: XRPpay}
        	\end{minipage}
		\end{figure}
        
        Alice, Bob and the Connector are connecting to the ledger over a web socket connection to settle their obligations and exchange ledger-related data with the ledger. In order to access \acrshort{ilp}, Alice and Bob are also connecting to the Connector through \acrshort{btp}+ws connections. The Connector works like an \acrshort{ilp} and payment bridge between them, and because everything happens in \gls{xrp}, the exchange rate applied is 1, and the backend used by the connector is "one-to-one". Further, Alice and Bob can initiate exchanges of value, i.e. \gls{xrp} payments, in this case by using \acrshort{spsp}.
		
        The case is detailed in Figure \ref{fig: XRPpayadv}, where we can identify:
        
        \begin{itemize}
            \item The \acrfull{ilsp} machine, containing:
                \begin{itemize}
                    \item The connector, where we can identify:
                        \begin{itemize}
                            \item The connector core
                            \item Plugins
                            \item The rate backend, set as "one-to-one"
                        \end{itemize}
                    \item Moneyd-\acrshort{gui}
                    \item A web browser connecting locally to Moneyd-\acrshort{gui} as a visual admin interface on http://127.0.0.1:7770
                \end{itemize}
            \item The \gls{xrp} ledger, accessible over a web socket connection, usually at port 51233
            \item Alice's machine, comprising:
                \begin{itemize}
                    \item \gls{moneyd}-\gls{xrp}:
                        \begin{itemize}
                            \item \gls{moneyd}-core
                            \item \gls{xrp} plugin
                            \item \gls{xrp} uplink
                        \end{itemize}
                    \item Visual admin interface:
                        \begin{itemize}
                            \item Moneyd-\acrshort{gui}
                            \item Web browser
                        \end{itemize}
                    \item \acrshort{spsp} client and server
                \end{itemize}
            \item Bob's machine, with a similar setup.
        \end{itemize}
        
        \begin{figure}[h!]
   			\begin{minipage}{\textwidth}
       		\includegraphics[width=1\textwidth]{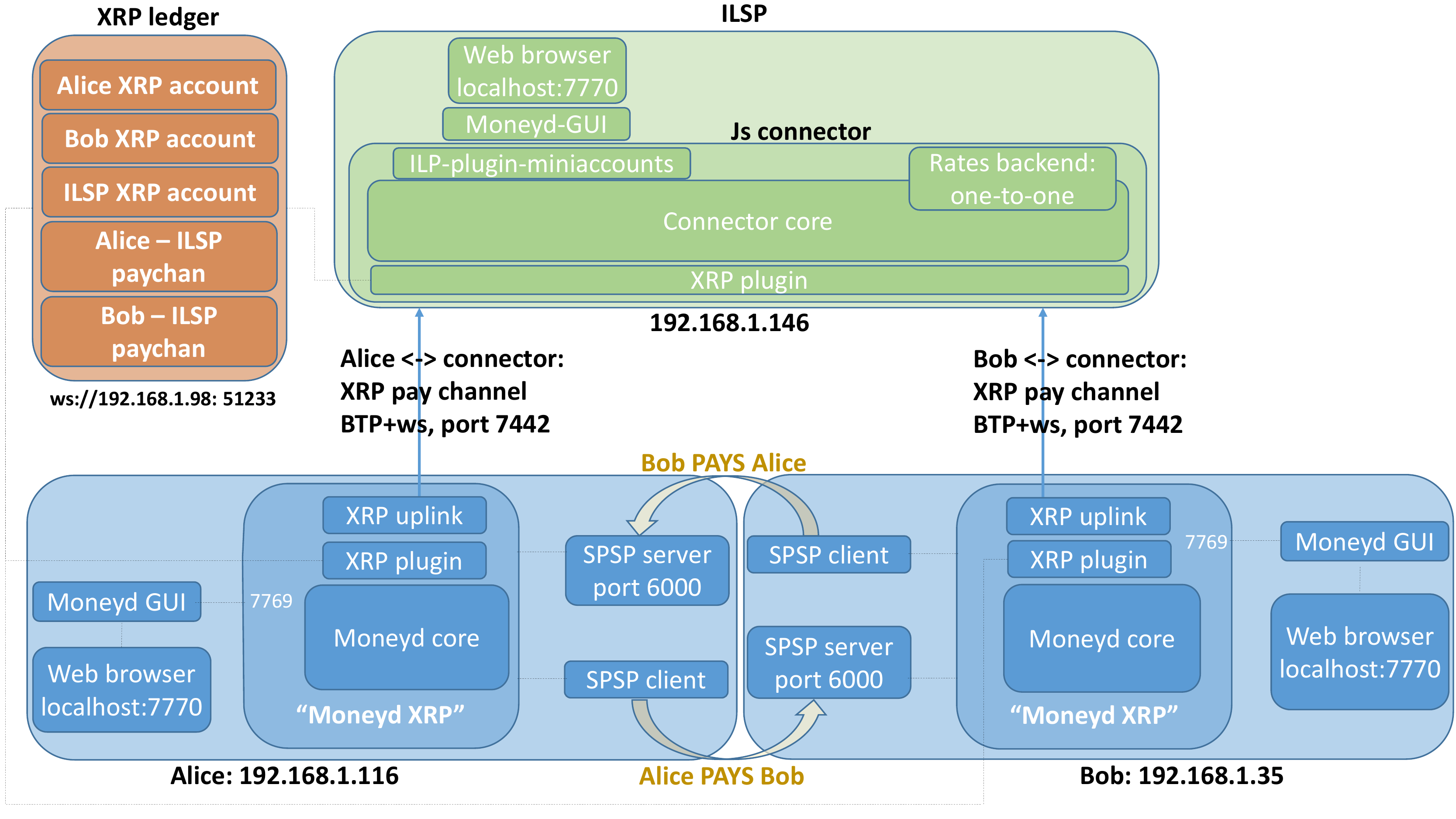}
			\caption[Example \ref{exmp_xrpxrp}: \gls{xrp} payment, advanced]{Example \ref{exmp_xrpxrp}: \gls{xrp} payment, advanced.}
            \label{fig: XRPpayadv}
        	\end{minipage}
		\end{figure}
		
		Using \gls{moneyd}-\gls{xrp}, Alice "dials-up" to, and opens a paychan with the Connector, thus enabling \acrshort{ilp} access on her machine. The connection with the Connector is made over \acrshort{btp}+ws. Between other functions, \acrshort{btp} acts as a "carrier" for \acrshort{ilp}. The paychan is recorded on the ledger. Alice is able to administer \gls{moneyd} through a web browser, using Moneyd-\acrshort{gui}. The \acrshort{spsp} app running on her machine will get \acrshort{ilp} access through \gls{moneyd}. Bob's situation is similar. \\
		\indent The ledger holds Alice's, Bob's and \acrshort{ilsp}'s \gls{xrp} accounts, and the paychans corresponding to the pairs Alice - \acrshort{ilsp} (Connector) and Bob - \acrshort{ilsp} (Connector). \\
		\indent When Alice sends a payment addressed to Bob, what happens is that the connector pays Bob on behalf of Alice (if Bob is able to provide the payment condition), and this transaction is recorded between the Connector and Bob. Further, the Connector presents to Alice the payment condition he just got from Bob, and Alice pays the Connector in exchange. This transaction is also recorded between Alice and the Connector. The value has moved from Alice to Bob, and in turn, two transactions have been recorded: Alice to Connector, and the Connector to Bob. Further, it is up to these pairs (Alice-Connector and Connector-Bob) to settle these transactions according to conditions agreed between each other (using the pay channels). \newline
		
		 \begin{figure}[h!]
   			\begin{minipage}{\textwidth}
       		\includegraphics[width=1\textwidth]{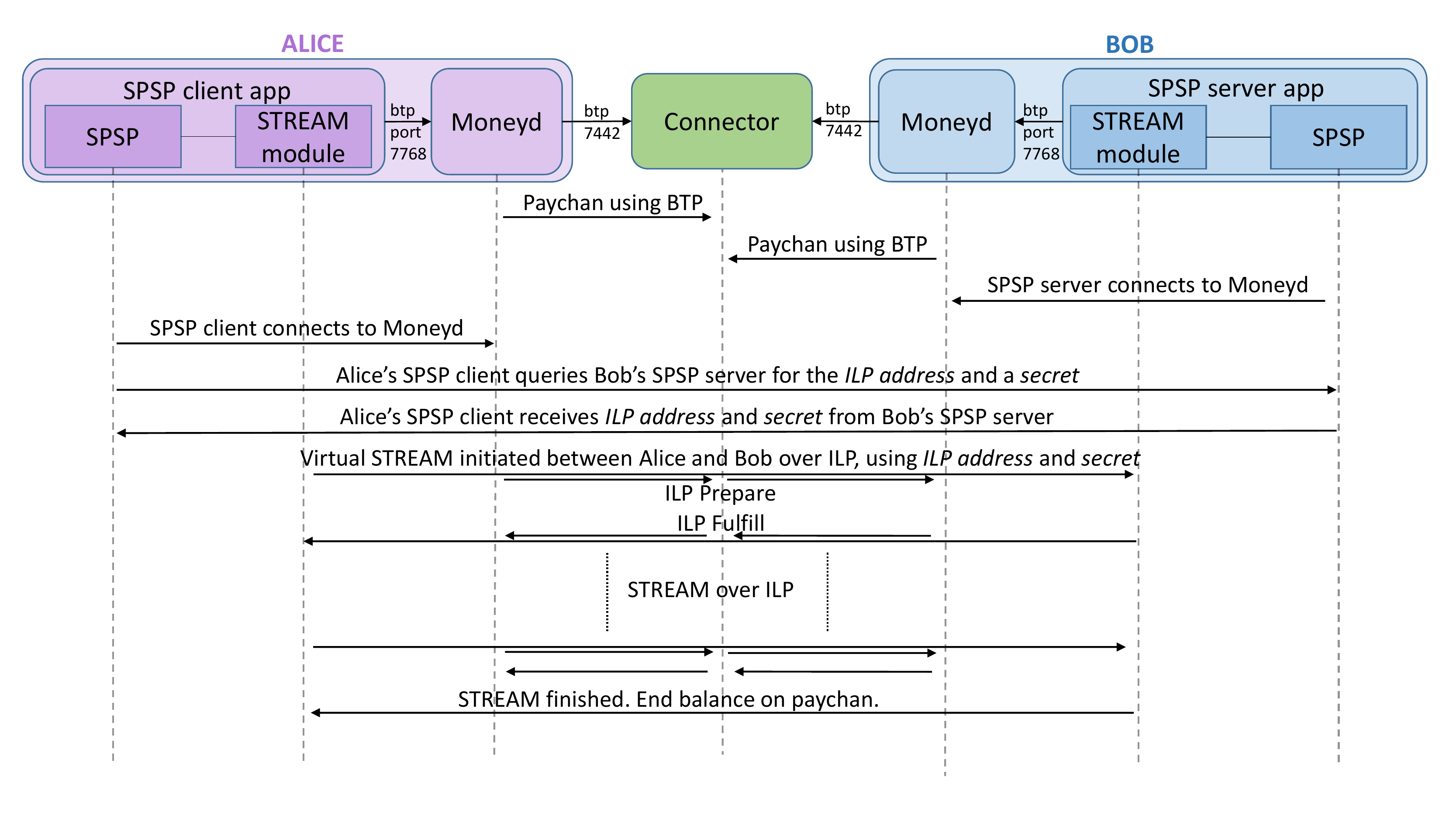}
			\caption[Example \ref{exmp_xrpxrp}: Protocol sequence]{Example \ref{exmp_xrpxrp}: Protocol sequence. \acrshort{btp}, \acrshort{ilp}, STREAM and \acrshort{spsp} are being used.}
            \label{fig: sequence}
        	\end{minipage}
		\end{figure}
		
		From a technical point of view, the sequence is illustrated in Figure \ref{fig: sequence}, and follows like below:
		
		\begin{itemize}
		    \item Alice and Bod start-up and connect their \gls{moneyd} instances to the \acrshort{ilp} connector. Each of them opens dedicated payment channels with the connector. Their transactions will be performed on the payment channels through \gls{moneyd} and \acrshort{ilp}. \\
		    
		    This is an example of a \acrshort{btp} packet sent between \gls{moneyd} and the Connector in order to open a paychan:
		    \begin{lstlisting}
btpPacket: { type: 6,
   requestId: 1890145753,
   data: { protocolData: [ [Object], [Object], [Object] ] } } ; data: { protocolData:
    [ { protocolName: 'channel',
        contentType: 0,
        data:
         <Buffer 15 8b 1c 3d a9 97 e5 b6 af 10 2d 90 51 b2 cd 3d 8c 58 96 55 02 ec dc 3a 07 02 f9 2c b7 88 61 1b> },
      { protocolName: 'channel_signature',
        contentType: 0,
        data:
         <Buffer 62 68 63 4c af b5 72 0f 2f 38 58 4a 81 03 cc 68 5a 62 ef 9a 43 91 80 73 40 31 03 8f 49 f8 f5 ae 88 64 fd 7f 14 39 39 d7 5b c3 a2 17 e1 7f c7 d2 49 a4 ... > },
      { protocolName: 'fund_channel',
        contentType: 1,
        data:
         <Buffer 72 70 4e 33 55 50 6a 59 61 45 72 74 34 52 57 34 67 41 69 71 75 63 4d 43 66 4c 35 6e 4a 4a 43 34 59 7a> } ] } ;
		    \end{lstlisting}
		    
		    \item Bob starts his \acrshort{spsp} server application, which connects to his local \gls{moneyd} instance and starts listening for incoming connections from an \acrshort{spsp} client.
		    \item Using her \acrshort{spsp} client, Alice queries the HTTP payment pointer presented by Bob.
		    \item Alice receives an \acrshort{ilp} address to use as \textit{destination account} for the transaction and a \textit{shared secret}.
		    \item Using this information, Alice pays Bob by starting a STREAM payment over \acrshort{ilp}. The \acrshort{ilp} service is provided by \gls{moneyd}. The \textit{shared secret} is used by STREAM to authenticate and encrypt multiple packets, as well as to generate the conditions and fulfillments.
		    \item STREAM divides the larger payments into packets and reassembles them. The packets are encapsulated inside, and sent along with, \acrshort{ilp} packets in the data field, and will be retrieved by the STREAM endpoint. During the STREAM connection, one of the parties acts as a client (the initiator) and the other acts as a server (the one accepting the connection). As STREAM is implemented by the \acrshort{spsp} application, the STREAM server will be already listening for a STREAM client connection. \\
		    
		    \indent Below is an example of a STREAM prepare packet carried by an \acrshort{ilp} packet sent over \acrshort{btp}:
		    \begin{lstlisting}
ILP-PLUGIN-BTP: handleIncomigWsMessage: binaryMessage:
<Buffer 06 1f 9d 97 68 81 e9 01 01 03 69 6c 70 00 81 e0 0c 81 dd 00 00 00 00 95 02 f9 00 32 30 31 39 30 36 31 39 30 39 34 33 30 31 35 30 39 45 04 2b e1 cd 98 ... >
ILP-PLUGIN-BTP: handle incoming packet from:  ; btpPacket: { type: 6,
requestId: 530421608,
data: { protocolData: [ [Object] ] } } ; data: { protocolData:
    [ { protocolName: 'ilp',
        contentType: 0,
        data:
         <Buffer 0c 81 dd 00 00 00 00 95 02 f9 00 32 30 31 39 30 36 31 39 30 39 34 33 30 31 35 30 39 45 04 2b e1 cd 98 68 71 95 50 c5 de 4a f2 1c f1 eb 4e 79 6e 95 cb ... > } ] } ;
ilp-plugin-xrp-paychan: received btp packet. type=TYPE_MESSAGE requestId=530421608 info=ilp-prepare
ILP_PACKET: binary: <Buffer 0c 81 dd 00 00 00 00 95 02 f9 00 32 30 31 39 30 36 31 39 30 39 34 33 30 31 35 30 39 45 04 2b e1 cd 98 68 71 95 50 c5 de 4a f2 1c f1 eb 4e 79 6e 95 cb ... >
ILP_PACKET: type: 12
ILP_PACKET: contents: <Buffer 00 00 00 00 95 02 f9 00 32 30 31 39 30 36 31 39 30 39 34 33 30 31 35 30 39 45 04 2b e1 cd 98 68 71 95 50 c5 de 4a f2 1c f1 eb 4e 79 6e 95 cb d8 f6 5a ... >
ILP_PACKET: deserializeIlpPrepare:
{
amount= 2500000000 , 
executionCondition= RQQr4c2YaHGVUMXeSvIc8etOeW6Vy9j2WlDZYKIZUbM= ,
expiresAt= 2019-06-19T09:43:01.509Z , 
destination= g.conn1.ilsp_clients.mduni.local.NL8f2khL-VmasfzfA-
    w_ds5F15J063Tn4oxDwoXTjGw.gHvuhB1r5GN0UQikoCGahPsj , 
data= YFwVZXQYK7pDTrprLcFOYbyt9qGQm+0APnOaBw5w5iUvvEggyB4Le0J8Bjbav7FKGyJ6Ih95xT8
    lss4BCQ==
}
\end{lstlisting}

The first byte "06" of the \acrshort{btp} packet is the \acrshort{btp} packet type, 6 in this case. \\
The next 4 bytes "1f 9d 97 68", are the request\_id in hex - 530421608 in this case. \\
We can also infer from the ILP-packet header that "0c" is the packet type - 12 in this case, and from the body, that "95 02 f9 00" represents the \textit{amount} = 2500000000. \\
This confirms the packet type as STREAM, because it is type 12 \cite{streamrfc, BTP}.

Below is a second example of a \acrshort{btp} packet carrying a STREAM fulfill:
\begin{lstlisting}
{ type: 1,
   requestId: 1054375881,
   data: { protocolData: [ [Object] ] } } ; data: { protocolData:
    [ { protocolName: 'ilp',
        contentType: 0,
        data:
         <Buffer 0d 5e 78 d3 d3 3e 33 27 b9 44 a1 45 92 f8 d8 98 28 8c 96 e2 20 00 af 8f bd eb 0d a3 24 04 79 0f 9b 75 3d 12 ef 89 a6 79 c1 a5 cc 53 ef c6 0f c1 60 8a ... > } ] } ;
\end{lstlisting}
The first byte "0d" is the packet type - 13, meaning fulfill.
The packet structure is like below:
\begin{lstlisting}
{ 
type: 13,
   typeString: 'ilp_fulfill',
   data:
    { fulfillment:
       <Buffer 78 d3 d3 3e 33 27 b9 44 a1 45 92 f8 d8 98 28 8c 96 e2 20 00 af 8f bd eb 0d a3 24 04 79 0f 9b 75>,
      data:
       <Buffer 12 ef 89 a6 79 c1 a5 cc 53 ef c6 0f c1 60 8a 71 38 b5 72 70 a8 f7 54 16 1c 30 65 f5 f1 9e fd 8f ee a6 d0 63 85 36 49 fd ab 5e 18 a6 d9 40 04 d4 5a 61 ... > } 
}
\end{lstlisting}

		    \item when the transfer is finished, the STREAM connection is closed.
		    \item at this moment the balances are updated on the payment channels opened between the parties involved, accordingly. The value can be redeemed out of each payment channel by each participant.
		    \item after claiming the funds, the payment channel can be either closed or used for other transactions.
		\end{itemize}
		
		The Interledger balance between Alice and Bob is continuously updated. If for some reason the STREAM connection is interrupted before the total amount is transferred, the amounts already transferred are not lost. \\
		\indent \textit{"Once peered, the two connectors both track the Interledger account balance and adjust it for every \acrshort{ilp} Packet successfully routed between them." \cite{peecleset}}
		
		Regarding the setup and configuration, for this example, we are going to consider the \gls{xrp} Ledger and the connector black boxes, and provide instructions for \gls{moneyd} and \acrshort{spsp}.
		
	\begin{itemize}	
	\item \indent On both Alice's and Bob's machines:
		\begin{itemize}
		\item If not already installed, install \textit{node.js}
		\item Install \gls{moneyd}, Moneyd-\acrshort{gui}, \acrshort{spsp} server and \acrshort{spsp} client apps:
    		\begin{itemize}
    		    \item \textit{"npm install -g moneyd moneyd-uplink-xrp"}
    		    \item \textit{"npm install -g moneyd-gui"}
    		    \item \textit{"npm install -g ilp-spsp-server ilp-spsp"}
    		\end{itemize}
	    \item Configure \gls{moneyd}:
    		\begin{itemize}
    		    \item \textit{"moneyd  xrp:configure  --advanced"}\\
    		    There will be four questions:
    		    \begin{itemize}
    		        \item \textit{? \acrshort{btp} host of parent connector:} \\ 
    		        We are going to use the IP:port(7442) of the connector. 
    		        \item \textit{? Name to assign to this channel:} \\ Can keep the autogenerated proposal or enter custom name.
                    \item \textit{? \gls{xrp} secret:} \\ Alice's/Bob's \gls{xrp} account secret: "sXXXXXXXXXXXXXXXXXXXXXXXX"
                    \item \textit{? Rippled server:}\\
                    The IP:port of the local Ripple server (ws://192.168.1.98:51235) or for example, "wss://s1.ripple.com".
    		    \end{itemize}
    		\end{itemize}
    	\item Start \gls{moneyd}, Moneyd-\acrshort{gui} and the browser interface:
    	    \begin{itemize}
    		    \item Start \gls{moneyd}: \\
    		    \textit{"DEBUG=*  moneyd  xrp:start  --admin-api-port  7769"}
    		    \item Start Moneyd-\acrshort{gui} by issuing: \\
    		    \textit{"npm start"} in /home/user/moneyd-gui
    		    \item Start a web browser and go to:\\
    		    \textit{http://127.0.0.1:7770}\\
    		    In order for all Moneyd-\acrshort{gui}'s graphical interface elements to load, you should also have internet access.
    		\end{itemize}
    	\end{itemize}
    	
    \item \indent On Bob's machine, start the \acrshort{spsp} server:\\
    	\indent \textit{DEBUG=* ilp-spsp-server --localtunnel false --port 6000} \\
    	
    \item \indent From Alice's machine initiate the \acrshort{spsp} transfer: \\
    	\indent \textit{DEBUG=ilp* ilp-spsp send --receiver http://192.168.1.116:6000 --amount 100} \\
    	Obviously, the above IP is just an example and you should use here Bob's machine's IP, according to your network setup.
\end{itemize}

Additional information and useful commands for \gls{moneyd} can be found on \gls{moneyd}'s github page. We provide the Table \ref{tab: moneydcomm} below, for general reference.

    \begin{table}[h]
        \centering
		\caption[Useful Moneyd commands]{Useful Moneyd commands.}
		\label{tab: moneydcomm}
        \setlength{\tabcolsep}{0.7em}
			\begin{tabular}{l|l}
                command         & effect \\
                \hline
                moneyd  xrp:configure  --advanced & configure Moneyd in advanced mode\\
                moneyd  xrp:start  --admin-api-port  7769            & enable the admin api port for Moneyd-GUI use \\
                moneyd start --unsafe-allow-extensions         & allow web browser payments \\
                moneyd xrp:info    & XRP account balance and outstanding paychans \\
                moneyd xrp:cleanup   & close paychans (get the money back from paychans) \\
                moneyd help           & list of Moneyd flags\\
                moneyd help $<$command$>$  & info on a specific command
			\end{tabular}
    \end{table}

For reference, we also provide an example of a \gls{moneyd}-\gls{xrp} configuration file, \textit{.moneyd.json}:

\begin{lstlisting}
{
  "version": 1,
  "uplinks": {
    "xrp": {
      "relation": "parent",
      "plugin": "/home/user/.nvm/versions/node/v10.15.3/lib/node_modules/moneyd-uplink-xrp
            /node_modules/ilp-plugin-xrp-asym-client/index.js",
      "assetCode": "XRP",
      "assetScale": 9,
      "balance": {
        "minimum": "-Infinity",
        "maximum": "20000000000",
        "settleThreshold": "50000000",
        "settleTo": "1000000"
      },
      "sendRoutes": false,
      "receiveRoutes": false,
      "options": {
        "currencyScale": 6,
        "server": "btp+ws://mduni:85941fd308ac69bfe7a4f6b9726430ea9ee6e6e654bb19b40a419c7a029b6fa7
            @192.168.1.146:7443",
        "secret": "ssVe1jBi2SUU5HQX6YPoTpRdRDG69",
        "address": "rpN3UPjYaErt4RW4gAiqucMCfL5nJJC4Yz",
        "xrpServer": "ws://192.168.1.98:51233"
      }
    }
  }
} 
\end{lstlisting}
\end{exmp}

\begin{exmp}
\label{exmp_xrpeth}
\textbf{\gls{xrp}-ETH \acrshort{ilp} payment using \gls{moneyd}, \acrshort{spsp} and a connector} \\
        
        We will discuss the configuration presented in Figure \ref{fig: ILPpaya}. It is comprised of:
        \begin{itemize}
            \item The \textit{\gls{xrp} ledger}, or the \gls{xrp} network, made up of servers running the "Rippled" software. Mainly, the ledger holds the account balances for all users and validates the transactions performed in-between users.
            \item The \textit{ETH ledger}, with a similar function.
            \item \textit{Alice}, holding an account on the \gls{xrp} ledger, operating Machine A, and running a user-level \acrshort{ilp} \gls{xrp} app, in this case \gls{moneyd}-\gls{xrp} and \acrshort{spsp}.
            \item Bob, holding an account on the ETH ledger, operating Machine B, and running a user-level \acrshort{ilp} Ethereum app, in this case \gls{moneyd}-ETH and \acrshort{spsp}.
            \item A \textit{Connector}, having 2 accounts - one on each ledger. The connector will act as a facilitator - an intermediary between the two users. It will accept \gls{xrp} from Alice and will forward the corresponding value, denominated in ETH, applying its exchange rate, to Bob.
        \end{itemize}
        
        \begin{figure}[h!]
   			\begin{minipage}{\textwidth}
       		\includegraphics[width=1\textwidth]{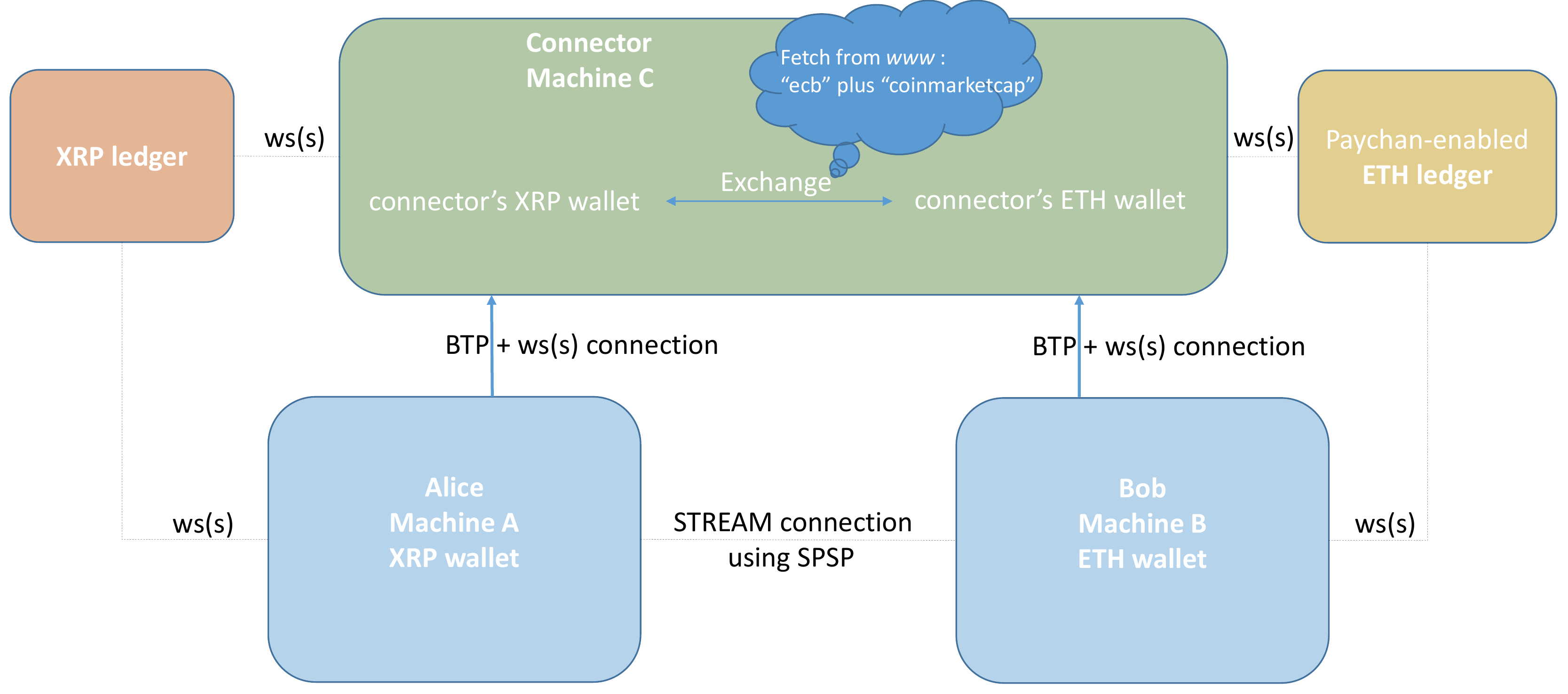}
			\caption[Example \ref{exmp_xrpeth}: Interledger payment]{Example \ref{exmp_xrpeth}: Interledger payment.}
            \label{fig: ILPpaya}
        	\end{minipage}
		\end{figure}
        
        A more advanced representation of the same setup is provided in Figure \ref{fig: ILPpay} and explained below. In order to be able to settle the payments in ETH, \textit{Machinomy} smart contract has to be deployed on the \textit{ETH ledger}.
        
        \begin{figure}[h!]
   			\begin{minipage}{\textwidth}
       		\includegraphics[width=1\textwidth]{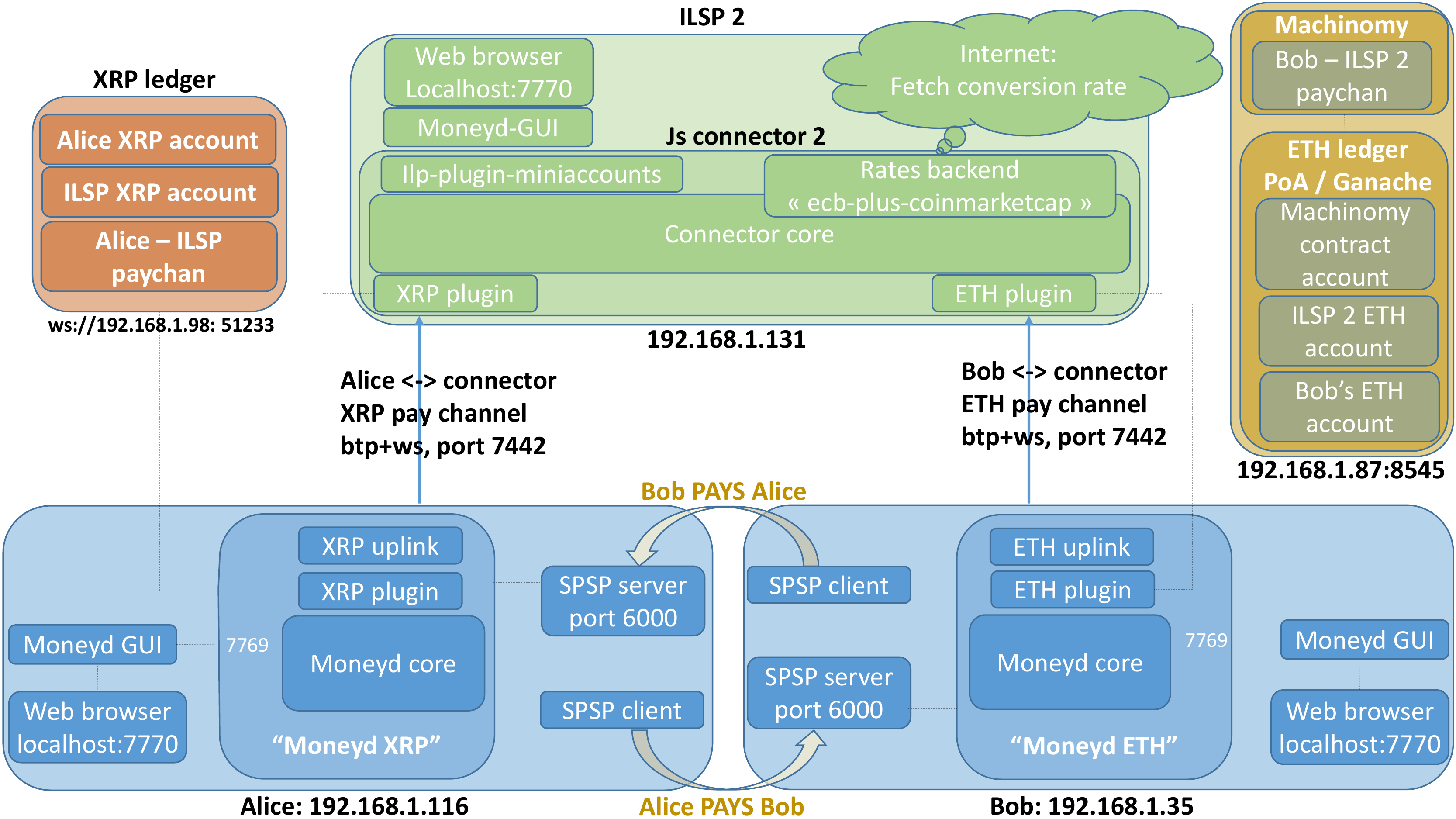}
			\caption[Example \ref{exmp_xrpeth}: Inter-Ledger payment, advanced]{Example \ref{exmp_xrpeth}: Interledger payment, advanced.}
            \label{fig: ILPpay}
        	\end{minipage}
		\end{figure}
        
        \begin{itemize}
            \item \textit{Alice} negotiates and opens a paychan denominated in \gls{xrp} with the \textit{connector}
            \item \textit{Bob} negotiates and opens a paychan denominated in ETH with the \textit{connector}
            \item Alice and Bob's machines comprise the following:
                \begin{itemize}
                    \item \gls{moneyd}-\gls{xrp} (Alice) or ETH (Bob), comprising of:
                        \begin{itemize}
                            \item \gls{moneyd}-core
                            \item \gls{xrp}/ETH plugin, providing the settlement means 
                            \item \gls{xrp}/ETH uplink, providing the uplink to the connector
                        \end{itemize}
                    \item \gls{moneyd}-\acrshort{gui}, providing a visual admin interface
                    \item \acrshort{spsp} modules:
                        \begin{itemize}
                            \item \acrshort{spsp} server: listens for connections from \acrshort{spsp} clients and receives  payments
                            \item \acrshort{spsp} client: connects to \acrshort{spsp} servers and sends  payments
                        \end{itemize}
                \end{itemize}
            \item The connector, comprising of:
                \begin{itemize}
                    \item Connector core
                    \item Different plugins:
                        \begin{itemize}
                            \item \gls{xrp} plugin
                            \item ETH plugin
                            \item Possibly, "ilp-plugin-mini-accounts" - to make use of Moneyd-\acrshort{gui} as a visual admin interface
                            \item Possibly other plugins
                        \end{itemize}
                    \item The rates backend, which fetches the exchange rates from the internet. We will be using "ecb-plus-coinmarketcap". Other possibilities are: ecb, ecb-plus-xrp, , one-to-one. "One-to-one" applies an exchange rate of 1 to everything and is used by connectors operating in a single currency environment.
                \end{itemize}
        \end{itemize}

	\begin{figure}[h!]
   			\begin{minipage}{\textwidth}
       		\includegraphics[width=1\textwidth]{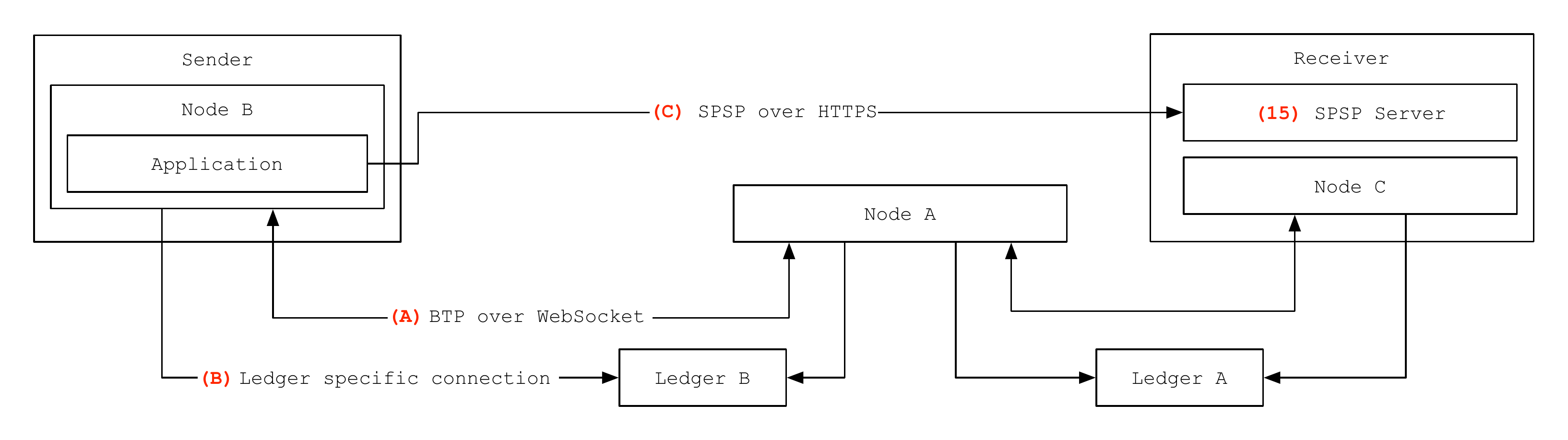}
			\caption[Perspective: connections]{Perspective: connections. \cite{protorel}}
            \label{fig: prspprt}
        	\end{minipage}
		\end{figure}
	Into perspective, the situation can be represented as in Figure \ref{fig: prspprt}, where:		
	\begin{itemize}
	    \item \textit{Sender} and \textit{Receiver} are Alice and Bob
	    \item \textit{Node B, Node C} are \gls{moneyd}
	    \item \textit{Node A} is the connector
	    \item \textit{Application} is the \acrshort{spsp} client
	    \item Ledger A and B are the \gls{xrp} and ETH ledgers
	    \item \gls{moneyd} connects to the connector over \textit{\acrshort{btp} (A)} and also has a \textit{ledger specific connection (B)} for settlement. 
	\end{itemize}
		
	The protocol interactions are the same as in Example 1, and also Alice's machine is the same as in Example 1. The main differences are that on the ETH ledger, for settlement, Machinomy smart contract must be deployed, and that the paychan between Bob and the \acrshort{ilsp} (Connector) is recorded there instead of the \gls{xrp} ledger. As such, on the \gls{xrp} ledger we find:
	\begin{itemize}
	    \item Alice's \gls{xrp} account
	    \item \acrshort{ilsp} (the connector) \gls{xrp} account
	    \item Alice - \acrshort{ilsp} paychan,
	\end{itemize}
	
	while on the ETH side:
	
	\begin{itemize}
	    \item Bob's ETH account
	    \item \acrshort{ilsp} (connector) ETH account
	    \item Bob - \acrshort{ilsp} paychan
	    \item The Machinomy smart contract account, deployed in order to help manage the paychans and settlements on Ethereum.
	\end{itemize}
    
    An advanced diagram of connections and protocols interactions is provided by Ripple in \cite{ILPadvdiag}. The explanations are extensive and beyond the scope of this paper, but they can be retrieved by the interested readers by following the link to the reference.
    
    For orientation, we provide as example a \gls{moneyd}-ETH configuration file:
    
\begin{lstlisting}
    {
  "version": 1,
  "uplinks": {
    "eth": {
      "relation": "parent",
      "plugin": "/home/user/node_modules/ilp-plugin-ethereum/index.js",
      "assetCode": "ETH",
      "assetScale": 9,
      "sendRoutes": false,
      "receiveRoutes": false,
      "options": {
        "role": "client",
        "ethereumPrivateKey": "0x72f3b5a36a6719492913f6480b8b5036bf5cc5f312152351886c8e216fc63288",
        "ethereumProvider": "kovan",
        "outgoingChannelAmount": "50000000",
        "balance": {
          "maximum": "1000000",
          "settleTo": "0",
          "settleThreshold": "300000"
        },
        "server": "btp+ws://ASDG:294a4788a4b0a7a048332c7d2390e6ce06bcd63e59585493f50e8738650
        a948a@192.168.1.131:7442"
      }
    }
  }
}
\end{lstlisting}
\end{exmp} 
    \subsubsection{@Kava-Labs: Switch API}
    \label{switch}
        \gls{switchapi}\footnote{\href{https://github.com/Kava-Labs/ilp-sdk/blob/master/README.md}{https://github.com/Kava-Labs/ilp-sdk/blob/master/README.md}, accessed June 2019} has been built mostly for cryptocurrencies trading like from \gls{xrp} to ETH or Lightning. This means that the accounts involved in the currency swap belong to the same user. It 'streams money', meaning that for example, a 20 units transfer would be split into small chunks and each of these chunks would be separately sent on the paychan until the whole amount is sent \cite{switchapi1,switchapi2}.
        
        Switch API handles multiple uplinks, with dedicated plugins for each currency - \gls{xrp}, ETH and Lightning. We investigated \gls{xrp} and ETH. For communicating with the \gls{xrp} connector we set up a dedicated \gls{xrp} plugin\footnote{\href{https://github.com/Kava-Labs/ilp-plugin-xrp-paychan}{https://github.com/Kava-Labs/ilp-plugin-xrp-paychan}, accessed June 2019}, while for the Ethereum uplink we use a dedicated Ethereum plugin\footnote{\href{https://github.com/interledgerjs/ilp-plugin-ethereum}{https://github.com/interledgerjs/ilp-plugin-ethereum}, accessed June 2019}.\\
         \indent To handle the ETH settlement, Machinomy contracts have to be deployed on the ETH network, as explained in Section \ref{ETHMach}. We have tested a stream payment between \gls{xrp} and ETH using Ganache\footnote{\href{https://truffleframework.com/ganache}{https://truffleframework.com/ganache}, accessed June 2019} as ETH provider. 
         
         Some particular aspects of running \gls{switchapi} - as of May 2019:
         \begin{itemize}
            \item The modules "ethers" and "ilp-plugin-ethereum" must be updated to the last version on the connector machine and \gls{switchapi} machine.
            \item  When setting up \gls{switchapi} the credentials must be lowercase.
            \item After each run, it creates a config file in \textit{/home/user/.switch/config}. If run with the same credentials (for tests), this file must be manually deleted or would output the warning "can not create duplicate uplink".
           
            \item When using a private ETH network, Ganache included, the network ID and the Machinomy contract address should be set. We have used the Kovan network ID, 42, which we have set in Ganache, while in the following files, we have changed the address to the Machinomy contract address deployed on the ETH network (Ganache):
            
            \begin{itemize}
                \item  \textit{'/home/user/node\_modules/ilp-plugin-ethereum/build/utils/channel.js'}  on the \textbf{reference node.js connector handling the ETH uplink} - the \acrshort{ilsp} 2 connector in Figure \ref{fig: netovw}, AND 
                \item in the same file \textbf{on the machine running the \gls{switchapi}} app\\
            \begin{lstlisting}
42: {
        unidirectional: {
            abi: Unidirectional_testnet_json_1.default,
            address: '0xa711d0a8b93faacd0f0f1897c11a1d7286d29720'
        }
    }
            \end{lstlisting}
            
    The Machinomy contract address is the "Unidirectional contract" address deployed by Machinomy.
            \end{itemize}
           
            \item Settings regarding settlement, \textbf{on the machine running \gls{switchapi}}, when running \gls{switchapi} on private \gls{xrp} and ETH networks:
            \begin{itemize}
                \item File: \textit{'switch-api/build/settlement/machinomy.js'}: \\
                \begin{lstlisting}
        remoteConnectors: {
            local: {
                'Kava Labs': (token) => `btp+ws://:${token}@192.168.1.131:7442`   // Reference ETH connector IP:port. ILSP 2 in Figure 26.
            },
            testnet: {
                'Kava Labs': (token) => `btp+ws://:${token}@192.168.1.131:7442`   // Reference ETH connector IP:port. ILSP 2 in Figure 26.
                 },
            mainnet: {
                  'Kava Labs': (token) => `btp+ws://:${token}@192.168.1.131:7442`
                  }
        }
                \end{lstlisting}
                \item File: \textit{'switch-api/build/settlement/xrp-paychan.js'}: \\
                \begin{lstlisting}
const getXrpServerWebsocketUri = (ledgerEnv) => ledgerEnv === 'mainnet'
      ? 'ws://192.168.1.98:51233'   // XRP validator IP
      : 'ws://192.168.1.98:51233';  // XRP validator IP
   . . . . .. . . . .. . . .. . . .. . . .. . . .
   remoteConnectors: {              //XRP parent connector
                      local: {
                              'Kava Labs': (token) => `btp+ws://:${token}@192.168.1.146:7444` // XRP referenceConnector - ILSP 1 in Figure 26.
                      },
                      testnet: {
                              'Kava Labs': (token) => `btp+ws://:${token}@192.168.1.146:7444` // XRP referenceConnector - ILSP 1 in Figure 26.
                      },
                      mainnet: {
                              'Kava Labs': (token) => `btp+ws://:${token}@192.168.1.146:7444` // XRP referenceConnector - ILSP 1 in Figure 26.
                      }
              }[ledgerEnv],         
                \end{lstlisting}
            \end{itemize}
    \item Additional settings \textbf{on the machine running \gls{switchapi}}. The "Ethers" module provides support for setting up different providers \footnote{\href{https://docs.ethers.io/ethers.js/html/api-providers.html}{https://docs.ethers.io/ethers.js/html/api-providers.html}, accessed June 2019}. 
        \begin{itemize}
            \item file: /home/user/node\_modules/ethers/utils/networks.js: \\
            \begin{lstlisting}
       kovan: {
              chainId: 42,
              name: 'kovan',
              _defaultProvider: etcDefaultProvider('http://192.168.1.87:8545') // set ETH provider IP and port (Ganache)
        }
            \end{lstlisting}
            \item in file /home/user/node\_modules/ethers/ethers.js, set network to kovan:
            \begin{lstlisting}
    function getDefaultProvider(network) {
        console.log('ETHERS.js get default provider (network): network:', network);
        if (network == null) {
              network = 'kovan'; //set kovan
        }
            \end{lstlisting}
            \item in file /home/user/node\_modules/ilp-plugin-ethereum/build/index.js, set provider to kovan:
            \begin{lstlisting}
class EthereumPlugin extends eventemitter2_1.EventEmitter2 {
     constructor({ role = 'client', ethereumPrivateKey, ethereumProvider = 'kovan', getGasPrice, outgoingChanne
            \end{lstlisting}
        \end{itemize}
    \item Additional setting \textbf{on the machine running the connector providing the ETH link}:
    \begin{itemize}
        \item in file  /home/user/node\_modules/ethers/utils/networks.js:
        \begin{lstlisting}
kovan: {
        chainId: 42,
        name: 'kovan', _defaultProvider: etcDefaultProvider('http://192.168.1.87:8545') //ETH provider IP:port (Ganache)
        }
        \end{lstlisting}
    \end{itemize}
    \item Example script which can be used for streaming \gls{xrp}-ETH using \gls{switchapi} \cite{switchapi1}:

        \begin{lstlisting}
const { connect } = require('@kava-labs/switch-api')
const BigNumber = require('bignumber.js')

async function run() {
  // Connect the API
  console.log('*** example-js ***: adding API')
  const api = await connect()

   //Add new uplink with an account
     console.log('**** example-js ****: addING uplink machinomy')
       const ethUplink = await api.add({
       settlerType: 'machinomy',
       privateKey: '6da09c0a78255932210aaf5b9f61046a00e9e3ab389c7357e388c4b35682342e'
       }) //switch Api wallet ETH
     console.log('*** example-js ***: addED uplink eth')

  // Add new uplink with an XRP testnet credential
     console.log('*** example-js ***: addING uplink XRP')
       const xrpUplink = await api.add({
       settlerType: 'xrp-paychan',
       secret: 'sasa3hrRUndoxAMoXEc3MMyZHNL3W' //switch API wallet XRP
       })
     console.log('*** example-js ***: addED uplink XRP')

  // Display the amount in client custody, in real-time
  xrpUplink.balance$.subscribe(amount => {
    console.log('XRP Interledger balance:', amount.toString())
  })
  ethUplink.balance$.subscribe(amount => {
    console.log('ETH Interledger balance:', amount.toString())
  })

  // Deposit 20 XRP into a payment channel
  console.log('EXAMPLE.js: start depositing 20XRP')
  await api.deposit({
    uplink: xrpUplink,
    amount: new BigNumber(20)
  })
  console.log('EXAMPLE.js: depositED 20xrp')


  // Deposit 0.05 ETH into a payment channel
  console.log('EXAMPLE.js: start depositing 0.05ETH')
  await api.deposit({
    uplink: ethUplink,
    amount: new BigNumber(0.05)
  })
  console.log('EXAMPLE.js: depositED 0.05ETH')

  // Stream 10 XRP to ETH, prefunding only $0.05 at a time
  // If the connector cheats or the exchange rate is too low, your funds are safe!
  await api.streamMoney({
    amount: new BigNumber(10),
    source: xrpUplink,
    dest: ethUplink
  })

  await api.disconnect()
}

run().catch(err => console.error(err))

        \end{lstlisting}

\end{itemize}

This file can be placed in \gls{switchapi} home directory and run with: \\
\textit{DEBUG=* node --inspect ./file-name.js} \newline

On top of this, Kava Labs has built an app for swapping the BTC, ETH and \gls{xrp} cryptocurrencies just in a matter of seconds\footnote{\href{https://github.com/Kava-Labs/switch}{https://github.com/Kava-Labs/switch}, accessed June 2019}. 
    
\section{The connectors}
\label{sec:connectors}
\begin{figure}[h!]
   			\begin{minipage}{\textwidth}
   			\centering
       		\includegraphics[width=0.6\textwidth]{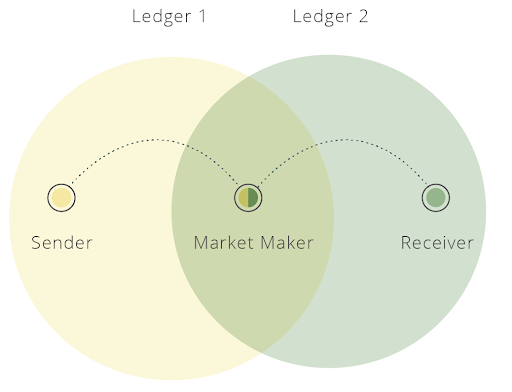}
			\caption[A connector holding two wallets on two different networks]{A connector (money maker) holding two wallets on two different networks.}
            \label{fig: 2wallets}
        	\end{minipage}
		\end{figure}

    Connectors are transaction 'intermediaries' lying in-between the payer and the payee, connecting them and facilitating the transaction. They are the 'market makers' or 'liquidity providers', and their role is especially evident when the sender's and receiver's wallets hold different currencies, as depicted in Figure \ref{fig: 2wallets}. A connector would take the sender's money in the sender's currency and pay the receiver with the receiver's currency while charging a small fee for the service. In order to be able to do this, a connector owns two wallets, one on each currency involved in the transaction.
    
\begin{figure}[p]
            \vspace{-3cm}
       		\includegraphics[width=24cm,angle=90,origin=c]{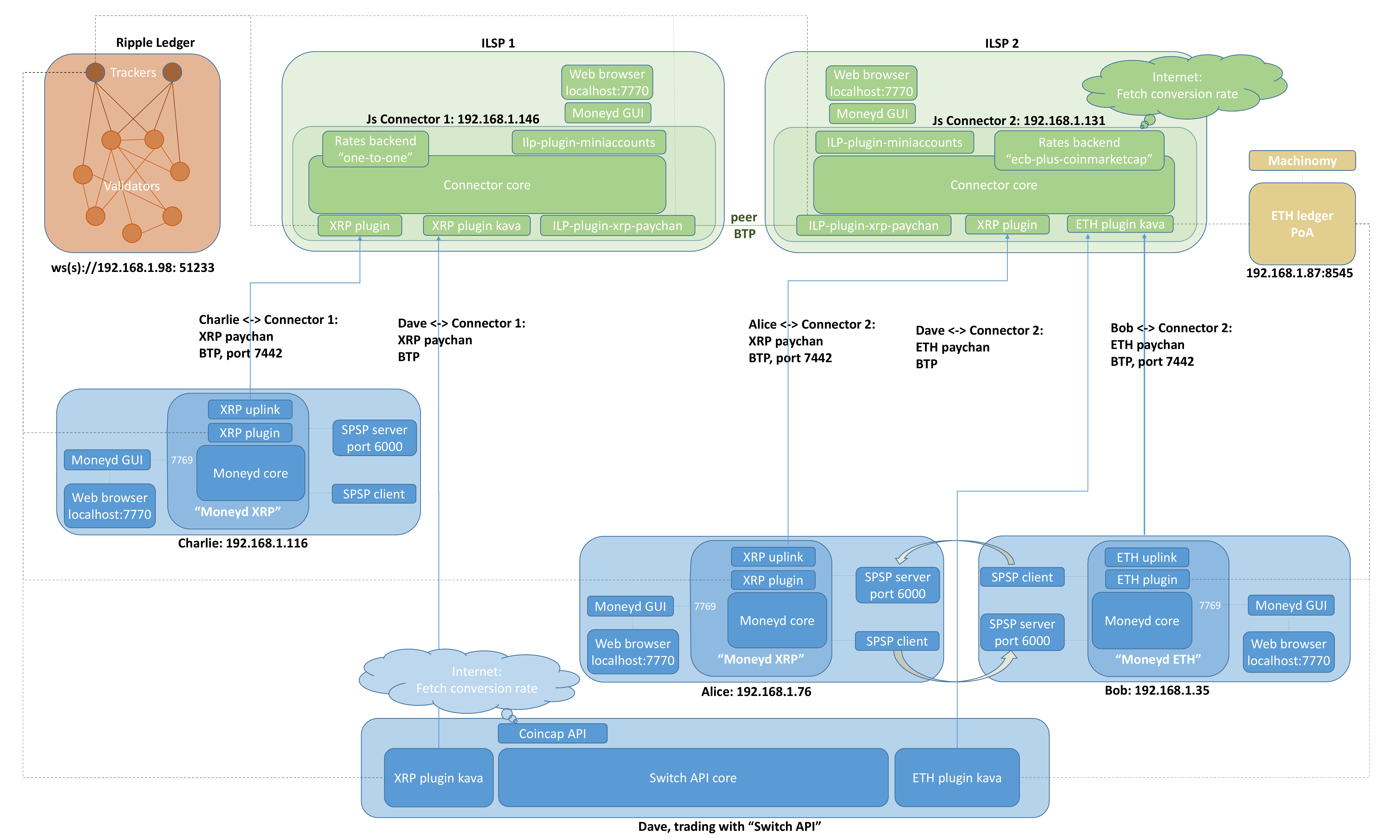}
			\caption[Ecosystem overview]{Ecosystem overview. The machines involved are time-synchronized using time servers. The ETH gas price and the currency rates are fetched from online. To keep the diagram readable we didn't illustrate all plugin connections to the ledgers; each plugin provides for connection to the appropriate ledger using wss or ws.}
            \label{fig: netovw}					
		\end{figure}

    \textit{'A connector is a host holding a balance on two or more ledgers. Connectors trade a debit against their balance on one ledger for a credit against their balance on another as a means of facilitating the payment between the two ledgers.'} \cite{connbailieswartzthomas} \\
    \indent While providing their services, the connectors act as an internet service provider would. In Interledger, the entities running the connectors are known as \textit{Interledger Service Providers} or \textit{\acrshort{ilsp}s}. For example, the \acrshort{ilp} reference connector written in js\footnote{\href{https://github.com/interledgerjs/ilp-connector}{https://github.com/interledgerjs/ilp-connector}, accessed June 2019} can be run by an \acrshort{ilsp}. An \acrshort{ilsp} can run one or more connectors. \gls{moneyd} is a stripped version of a connector which is not sending or receiving routes, and is used as a \textit{"home router"}, by end-users or customers in order to dial-up and connect to the \acrshort{ilsp}s running a connector.
    As such, \gls{moneyd}\footnote{\href{https://github.com/interledgerjs/moneyd}{https://github.com/interledgerjs/moneyd}, accessed June 2019} or \gls{switchapi}\footnote{\href{https://github.com/Kava-Labs/ilp-sdk/blob/master/README.md}{https://github.com/Kava-Labs/ilp-sdk/blob/master/README.md}, accessed June 2019}  are examples of 'customer' apps connecting to their preferred \acrshort{ilsp} and sending requests, as shown in Figure \ref{fig: netovw}. \\
    \indent \textit{'Connectors implement the Interledger protocol to forward payments between ledgers and relay errors back along the path. Connectors implement (or include a module that implements) the ledger protocol of the ledgers on which they hold accounts. Connectors also implement the Connector to Connector Protocol (CCP) to coordinate routing and other Interledger control information.'}\cite{ilpccp}

\begin{figure}[h!]
   			\begin{minipage}{\textwidth}
       		\includegraphics[width=1\textwidth]{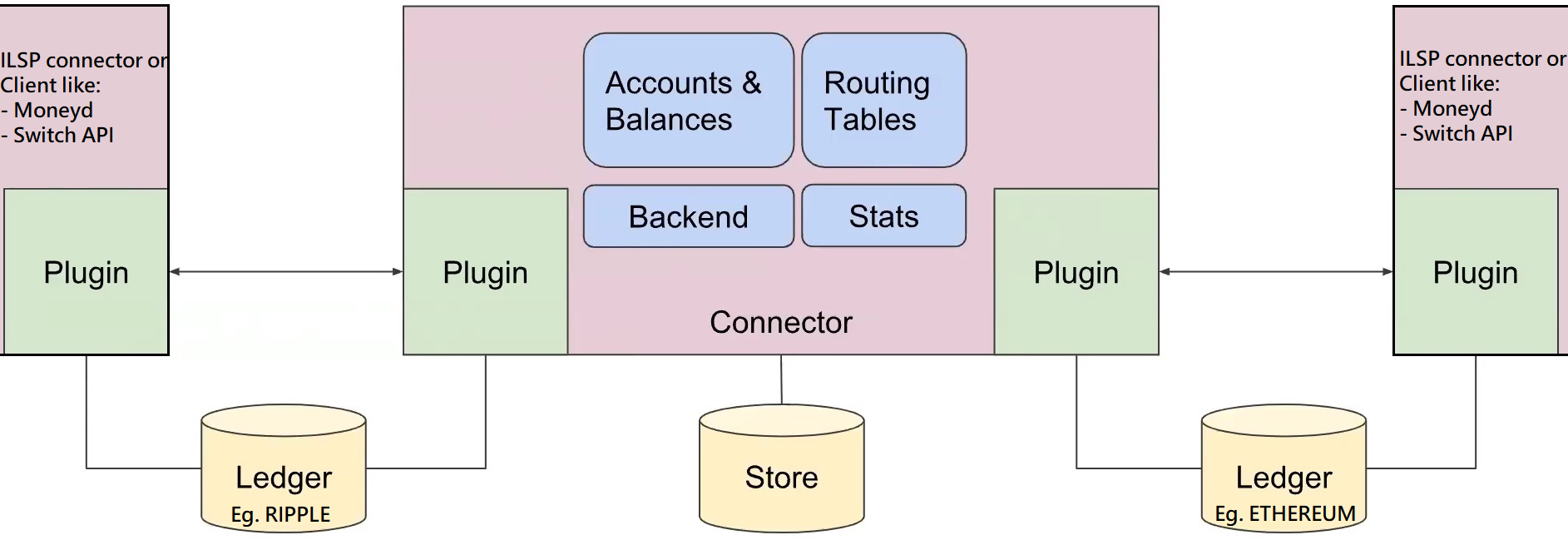}
			\caption[Architecture overview]{Architecture overview. \cite{bailiecall}}
            \label{fig: 2}
        	\end{minipage}
		\end{figure}

    Currently, the connectors rely on plugins to \textit{settle} the transactions (new architectures are currently being considered or implemented, e.g. the Rafiki connector). \\
    \indent Making the analogy to a real-life example, swiping a credit card at a cashier's desk is considered a \textit{payment}. The \textit{settlement} occurs when the money is debited from the card holder's bank and credited to the merchant's bank. When you swipe the credit card and introduce your PIN you create and sign an irrevocable obligation for payment. On an Interledger paychan, this signed obligation for payment is known as a \textit{"claim"}. A \textit{redeemed claim} would translate to a bank transaction which has been "cleared" or "went through" (the money completely left the payer's bank account, and are visible and available in the payee's account; or analogously, the money completely left the sender's wallet/ledger and have shown up on the receiver's wallet, ledger and currency). \\
    \indent Concerning the plugins, they are installed on the same machine with the connector and configured according to purpose. 
    This architecture is illustrated in Figure \ref{fig: 2}. \newline
    
    Some plugin examples would be:
         \begin{itemize}[noitemsep, leftmargin=*]
    	\item {\textbf{ilp-plugin-xrp-paychan}}\footnote{\href{https://github.com/interledgerjs/ilp-plugin-xrp-paychan}{https://github.com/interledgerjs/ilp-plugin-xrp-paychan}, accessed June 2019} creates a direct peer relation with other connectors. It is an Unconditional Payment Channel plugin, where one has to trust his peer for the in-flight XRP amounts.
    	\item {\textbf{ilp-plugin-xrp-asym server}}\footnote{\href{https://github.com/interledgerjs/ilp-plugin-xrp-asym-server}{https://github.com/interledgerjs/ilp-plugin-xrp-asym-server}, accessed June 2019} enables the \acrshort{ilsp} server to accept new client connections and creates an internal ILP account for each of them. It is the plugin appropriate for provider - customer relationships. This service will be exposed publicly and 'customers' will connect to it.
    \item {\textbf{ilp-plugin-xrp-asym-client}}\footnote{\href{https://github.com/interledgerjs/ilp-plugin-xrp-asym-client/}{https://github.com/interledgerjs/ilp-plugin-xrp-asym-client/}, accessed June 2019} will be used by a 'customer' to connect to his provider's plugin, i.e the \textit{ilp-plugin-xrp-asym-server} above. 
    	\item {\textbf{moneyd's uplink-xrp plugin}}\footnote{\href{https://github.com/interledgerjs/moneyd-uplink-xrp}{https://github.com/interledgerjs/moneyd-uplink-xrp}, accessed June 2019} makes use of \textit{ilp-plugin-xrp-asym-client}.
    	\item {\textbf{ilp-plugin-mini-accounts}}\footnote{\href{https://github.com/interledgerjs/ilp-plugin-mini-accounts}{https://github.com/interledgerjs/ilp-plugin-mini-accounts}, accessed June 2019} can be used to connect Moneyd-\acrshort{gui} to \gls{moneyd} or to the reference \acrshort{ilsp} connector, for example. 
    	\item Kava Labs has been involved in the development of  {\textbf{@kava-labs/ilp-plugin-xrp-paychan}}\footnote{\href{https://github.com/Kava-Labs/ilp-plugin-xrp-paychan}{https://github.com/Kava-Labs/ilp-plugin-xrp-paychan}, accessed June 2019} and {\textbf{ilp-plugin-ethereum}}\footnote{\href{https://github.com/interledgerjs/ilp-plugin-ethereum}{https://github.com/interledgerjs/ilp-plugin-ethereum}, accessed June 2019}. Both can be used in conjunction with \gls{switchapi}. Ilp-plugin-ethereum settles Interledger payments with ether and is powered by Machinomy smart contracts for unidirectional payment channel.
    \end{itemize}
    
\begin{figure}[h!]
    \centering
    \includegraphics[width=0.8\textwidth]{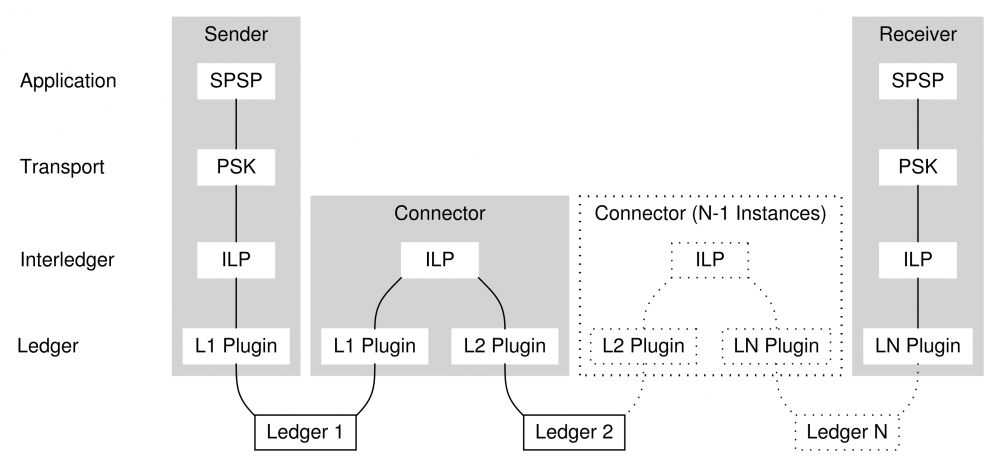}
    \caption[The protocol stack in the payment chain]{The protocol stack in the payment chain.}
    \label{fig: paychain}
\end{figure}

    Another way to illustrate a payment chain forwarding the payment through the connectors with the use of \acrshort{spsp}, the \acrshort{ilp} protocol, some of the plugins above and the servers from Section \ref{XRPledger} which form the \gls{xrp} ledger for example, is provided in Figure \ref{fig: paychain}, with some other ledger examples being the ETH or Lightning server networks. It is worth being noted that PSK was upgraded to STREAM. As such, the sender can pay in \gls{xrp} and the receiver can get his money in BTC. 
    
	\indent Below we reproduce a nice explanation on payment channels that we found worth adding: \\
	\textit{'In order to avoid having to go through the consensus process for each and every transaction, only the summary of several transactions is validated on the blockchain. The intermediate transactions are conducted outside of the Ripple Ledger, off-chain. Ripple Labs has said that with Payment Channels (introduced in summer 2017), several thousand transactions per second can be processed. This number is approaching the transaction capacity of the VISA network. Because of the increased efficiency, Payment Channels are a feasible micro-payment alternative.'}  \cite{paychanexpl}\\ 
	\indent Also, payment channels are worth being used with expensive or slow ledgers. The two parties' transactions are being performed on the paychan under the limits established. Sometimes they offset each other. When the conditions for settlement are met, the settlement can occur. This lowers the cost and time involved by the overall process. \newline
	
\begin{exmp}
\label{exmp_xrpetc2c}
\textbf{An advanced \gls{xrp}-ETH setup using two connectors} \\

    Alice, having an \gls{xrp} account, connects to the \acrshort{ilsp} 1 connector and establishes a paychan using the \gls{xrp} plugin and uplink. Bob, having an Ethereum account, connects to the \acrshort{ilsp} 2 Ethreum enabled connector using an Ethereum plugin and uplink. The two connectors are peered over \gls{xrp} and establish a paychan with the \textit{ilp-plugin-xrp-paychan}. When Alice sends money to Bob, the payment goes through \acrshort{ilsp} 1 and \acrshort{ilsp} 2. Alice will settle her balance with \acrshort{ilsp} 1; \acrshort{ilsp} 1 will settle its balance with \acrshort{ilsp} 2, and \acrshort{ilsp} 2 will settle with Bob. \\
    \indent Because \acrshort{ilsp} 1 works exclusively in \gls{xrp}, its exchange rate will always be 1, so the backend used will be "one-to-one". On the other hand, \acrshort{ilsp} 2 works on two ledgers and performs "the currency exchange" so the backend will make use of the "ecb-plus-coinmarketcap" option, and the rate will be fetched from the internet. \newline
    \indent On the \gls{xrp} ledger we find Alice's, \acrshort{ilsp} 1 and \acrshort{ilsp} 2's accounts, together with the \gls{xrp} paychans established between the pairs Alice - \acrshort{ilsp}1 and \acrshort{ilsp}1 - \acrshort{ilsp}2. Accordingly, on the ETH network we find \acrshort{ilsp}2's and Bob's accounts along with the paychan Bob - \acrshort{ilsp}2.
    
        \begin{figure}[h!]
            \hspace{-1cm}
   			\begin{minipage}{\textwidth}\centering
       		\includegraphics[width=1.1\textwidth]{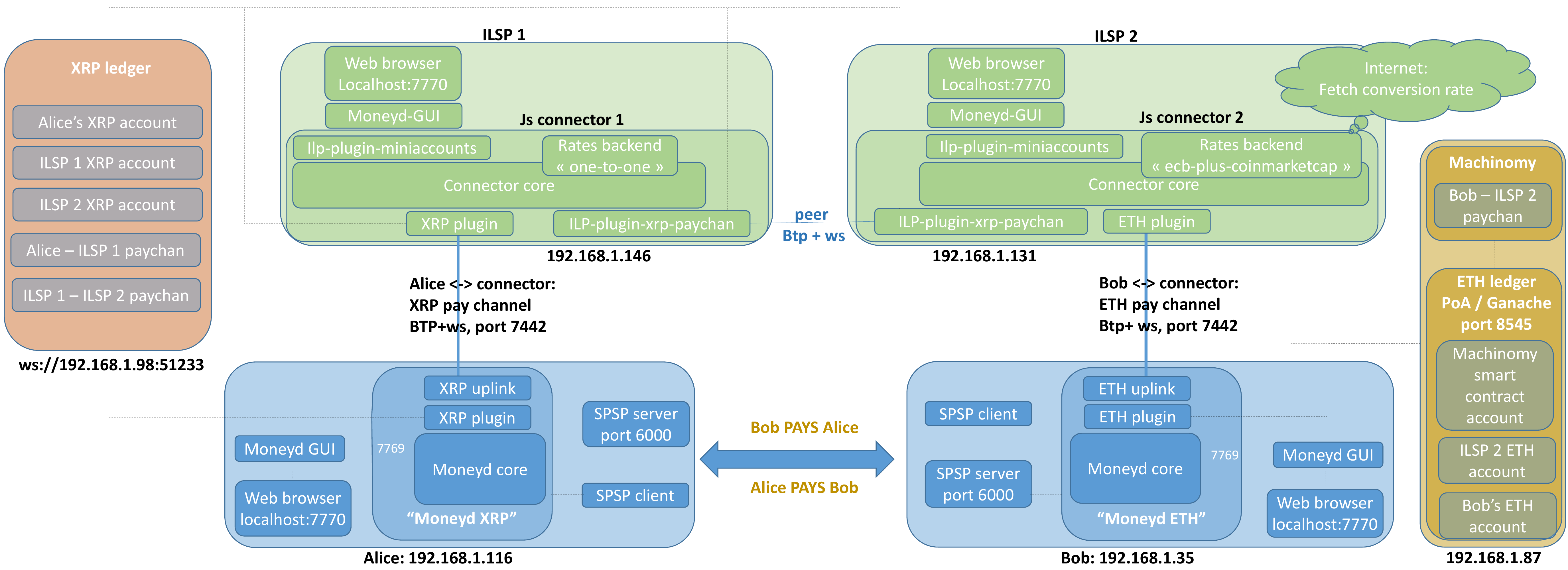}
			\caption[Example \ref{exmp_xrpetc2c}:  advanced Interledger payment]{Example \ref{exmp_xrpetc2c}:  advanced Interledger payment.}
            \label{fig: Example3}
        	\end{minipage}
		\end{figure}
		
    For this configuration, we will assume the ledgers are already up and running on the same Local Area Network hardwired or Wi-Fi. Also, internet access is available. Then, the deployment sequence is:
    
    \begin{itemize}
        \item Start the connectors, one by one, and wait for them to establish a payment channel between them. This will be shown in the connector logs. If using \textit{pm2}, logs can be enabled with \textit{"pm2 logs connector"}.
        \item Start the \gls{moneyd} instances for Alice and Bob and wait for each to establish paychans with the corresponding connector.
        \item Start Bob's \acrshort{spsp} server.
        \item From Alice's terminal, use the \acrshort{spsp} client to initiate an \acrshort{spsp} transaction.
        \item Optionally, Moneyd-\acrshort{gui} and a web browser can be used to administer \gls{moneyd} and the connectors.
    \end{itemize}
    
    The diagram presented in Figure \ref{fig: ex3interact} further illustrates how the nodes and protocols interact.
    
        \begin{figure}[h!]
        \hspace{-1cm}
   			\begin{minipage}{\textwidth}\centering
       		\includegraphics[width=1.1\textwidth]{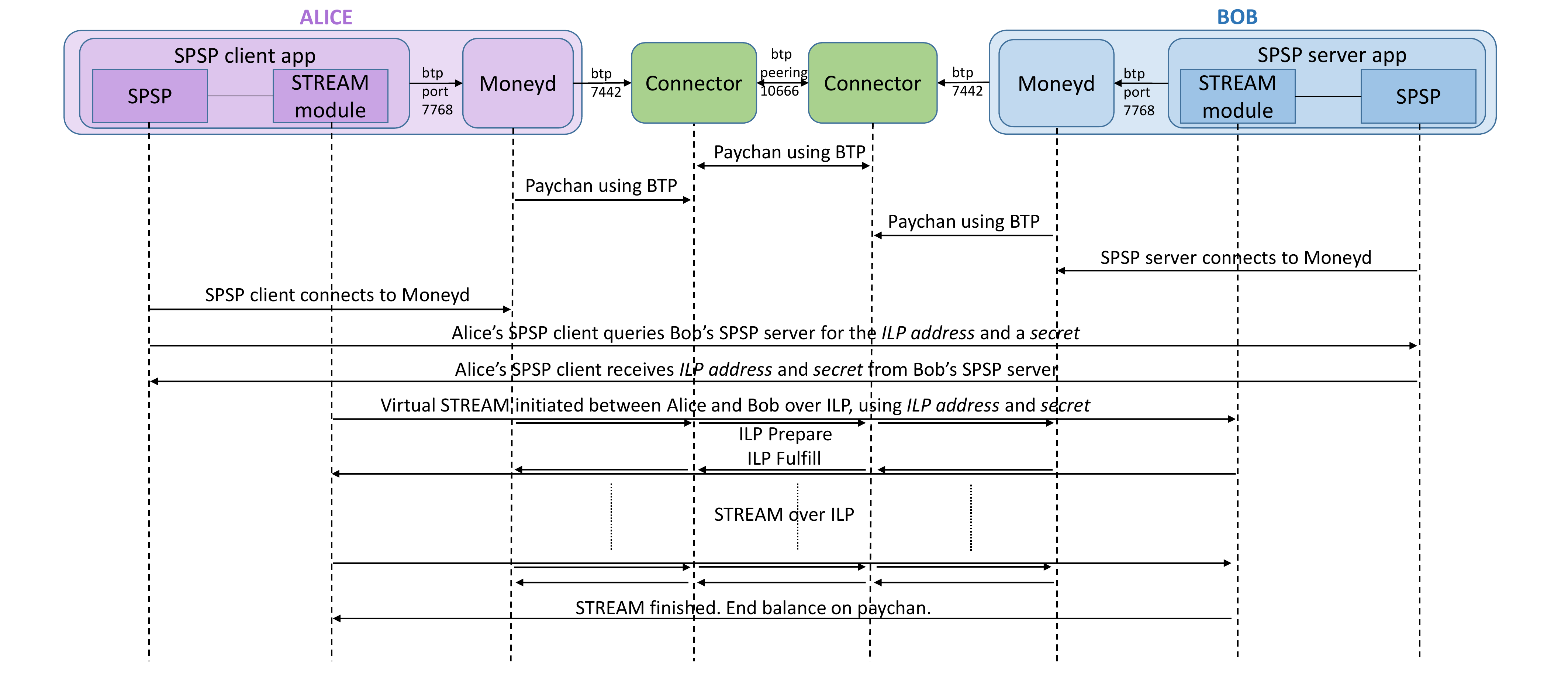}
			\caption[Example \ref{exmp_xrpetc2c}: interaction diagram]{Example \ref{exmp_xrpetc2c}: interaction diagram.}
            \label{fig: ex3interact}
        	\end{minipage}
		\end{figure}
		
    The setup for Alice and Bob's machines is similar to Examples 1 and 2. \\
    \indent In regards to installation and set-up of connectors, a new guide\footnote{\href{https://www.stratalabs.io/mainnet}{https://www.stratalabs.io/mainnet}, accessed June 2019}\textsuperscript{,}\footnote{\href{https://github.com/d1no007/easy-connector-bundle}{https://github.com/d1no007/easy-connector-bundle}, accessed June 2019} from Strata Labs has just been released, so you can try the bundle they propose if you want to hit the ground running. \\
    \indent The tutorial\footnote{\href{https://medium.com/interledger-blog/running-your-own-ilp-connector-c296a6dcf39a}{https://medium.com/interledger-blog/running-your-own-ilp-connector-c296a6dcf39a}, accessed June 2019} provided by Adrian Hope-Bailie is a very good and thorough step-by-step guide. Inspired from his guide on installing the reference connector, a faster and easier minimal set-up and a few tips are provided below. This procedure will miss some features in the original guide. For the advanced set-up, the original post can be followed. For convenience, the Step \# has been kept the same as in the original tutorial.
    \begin{itemize}
        \item Step 5: install node
            \begin{lstlisting}
    - curl -o- https://raw.githubusercontent.com/creationix/nvm/v0.34.0/install.sh | bash
    - Restart terminal
    - nvm install v10.15.3
            \end{lstlisting}
        \item Step 7: install "redis"\footnote{\href{https://www.digitalocean.com/community/tutorials/how-to-install-and-secure-redis-on-ubuntu-18-04}{https://www.digitalocean.com/community/tutorials/how-to-install-and-secure-redis-on-ubuntu-18-04}, accessed June 2019}
        \item Step 8: install pm2
        \item Step 10: get and fund an \gls{xrp} Ledger address. Alternative to \textit{'ripple-wallet-cli'}, it is also possible to generate a wallet directly, using the \textit{'wallet\_propose'} method: \\
        \textit{'user@saintmalo:~/rippled/ccabuild\$ ./rippled  --conf /home/user/rippled/cfg/rippled-example.cfg wallet\_propose'}.
        \item Step 12: pick an \acrshort{ilp} Address. The format should be \textit{'g.somethingunique'}. For an independent private network, we also used the 'production' settings and 'g' as address prefix. Any of the others (private, local, ..) did not seem to work right.
        \item Step 13: create your config file using \textit{'pm2 init'}. A file named \textit{'ecosystem.config.js'} will be created, possibly in the folder \textit{'home/user'}. Check it, update as needed, and move it to \textit{'/home/user/ilp-connector/'}.
        \item Step 14: start it with: \\
        \textit{'cd /home/user/ilp-connector: \$ pm2 start ecosystem.config.js'} \\
        Use \textit{'pm2 stop ecosystem.config.js'} to stop it, \textit{'pm2 restart ecosystem.config.js --update-env'} for restart, and \textit{'pm2 logs connector'} to see the logs. The log files are located in '/home/user/.pm2/logs/connector-out.log'.
    \end{itemize}

\noindent How to setup a connector: \\ 
First of all the following example is deployed in Ubuntu bionic (kernel 4.15 but not important). \\

\textit{sudo apt-cache madison npm} - at the time of writing 3.5.2.

\textit{sudo apt-cache madison nodejs} - at the time of writing version 8.10.

\textit{sudo apt-get install nodejs npm build-essential}

\textit{npm config get prefix} This will return the path for packages installed with -g. In our case, we will install the packages locally in the folder \textit{jsilp}.

\textit{npm install memdown}. We will use memdown for this example but for production, you should use another type of databases.

\textit{pm2 restart launch.config.js} \newline

    The connector will be much easier to admin and understand using an interface, at this moment Moneyd-\acrshort{gui}. Installed on the same machine as the connector, it will provide UI access in a browser at http://127.0.0.1/7770. For the graphical interface, Moneyd-\acrshort{gui} loads some resources from online.
    
    Below we provide an example configuration for the \acrshort{ilsp}1 Connector from Figure \ref{fig: netovw}. For our use-case we used ws, but in a real scenario wss is used.
    
    \begin{lstlisting}
'use strict'
const path = require('path')

const address = 'rMqUT7uGs6Sz1m9vFr7o85XJ3WDAvgzWmj' // <YOUR RIPPLE ADDRESS>
const secret = 'shjZQ2E3mYzxHf1VzYBJCQHqLvt7Y'       // <YOUR RIPPLE SECRET>

const peer1 = {
    relation: 'peer',                  // establish a 'peer' relationship.
    plugin: 'ilp-plugin-xrp-paychan',  //peer with another ILSP connector over XRP
    assetCode: 'XRP',
    assetScale: 9,     //"Interledger amounts are integers, but most currencies are typically represented as fractional units, e.g. cents. This property defines how many Interledger units make up one regular units. For dollars, this would usually be set to 9, so that Interledger amounts are expressed in nanodollars."
    balance: {
        maximum: '1000000000',
        settleThreshold: '-5000000000',
        settleTo: '0'
    },
    options: {
        listener: { //If you want your peer to connect to you as a ws client (which doesn't change the nature of the liquidity relationship) set the `listener` argument in the constructor.
        port: 10666, //this ws server listens for ws clients on port 10666
        secret: '2afe5e6cece84ed0027f9a2463edfa6358901bd6f1c9f3e1b0e43c13ff1ae2eb'  // this is the token that your peer must authenticate with.
        },
        //server: 'btp+ws://:its_a_secret@192.168.1.146:10666', //this connector would be a ws client connecting to its peer ws server at port 10666
        //It should be possible to use it without credentials like this, also: server: 'btp+ws://:@192.168.1.146:10666'
        // You may specify both the server and client options; in that case it is not deterministic which peer will end up as the ws client.
        rippledServer: 'ws://192.168.1.98:51233',               //the server that you submit XRP transactions to //MAINNET - wss://s2.ripple.com
        peerAddress: 'rLR52VSZG3wqSrkcpfkSnaKnYoYyPoJJgy',      //<PEER RIPPLE ADDRESS>
        address,
        secret
     }
}

const ilspServer = {                                  //MoneyD XRP clients
    relation: 'child',                                //Moneyd apps will be 'children'
    plugin: 'ilp-plugin-xrp-asym-server',             //plugin that exposes the ILSP server to downstream clients
    assetCode: 'XRP',
    assetScale: 6,
    options: {
        port: 7443,                                   //port on which to listen to client apps
        xrpServer: 'ws://192.168.1.98:51233',         //MAINNET - wss://s2.ripple.com
        address,
        secret
    }
}

const SwitchAPIServer = {
      relation: 'child',                              //Switch API connects as a 'child'
      plugin: '@kava-labs/ilp-plugin-xrp-paychan',
      assetCode: 'XRP',
      assetScale: 6,
      options: {
          role: 'server',
          port: 7444,                                 //Switch API will connect on this port
          xrpSecret: 'shjZQ2E3mYzxHf1VzYBJCQHqLvt7Y', //this connector's secret
          xrpServer: 'ws://192.168.1.98:51233',
          // Very asymmetric... you fund a channel for $0.50 in XRP, we'll open one to you for $10!
          outgoingChannelAmount: '32658000',          // ~= 10$ in XRP drops
          minIncomingChannelAmount: '1632900',        // ~= 0.5$ in XRP drops
          // Use plugin maxPacketAmount (and not connector middleware) so F08s occur before T04s
          maxPacketAmount: '653200'                   // ~= 0.2$ in XRP drops
      }
}

const moneydGui = {                         //MoneyD GUI for this connector
  relation: 'child',
  plugin: 'ilp-plugin-mini-accounts',
  assetCode: 'XRP',
  assetScale: 6,
  options: {
    port: 7768                             //MoneyD GUI will connect on this port
  }
}

const connectorApp = {
    name: 'connector',
    env: {
        DEBUG: 'ilp*,connector*',
        CONNECTOR_ENV: 'production',
        CONNECTOR_ADMIN_API: true,
        CONNECTOR_ADMIN_API_PORT: 7769,              //this should not conflict with moneydGUI, set here on 7768
        CONNECTOR_ILP_ADDRESS: 'g.conn1',            //<YOUR ILP ADDRESS>
        CONNECTOR_BACKEND: 'one-to-one',
        CONNECTOR_SPREAD: '0',
        CONNECTOR_STORE: 'memdown',                  //comment this if using the store below
        //CONNECTOR_STORE: 'ilp-store-redis',        //if using a store
        //CONNECTOR_STORE_CONFIG: JSON.stringify({
        //    prefix: 'connector',
        //    port: 6379
        //}),
        CONNECTOR_ACCOUNTS: JSON.stringify({
            conn2: peer1,                            //arbitrary names easy to remember
            ilsp_clients: ilspServer,
            moneyd_GUI: moneydGui,
            switchXRP: SwitchAPIServer
        })
    },
    script: path.resolve(__dirname, 'src/index.js')
}

module.exports = { apps: [ connectorApp ] }
    \end{lstlisting}
  
  Further, we reproduce an example configuration for the \acrshort{ilsp}2 Connector in Figure \ref{fig: netovw}, which is peered with the \acrshort{ilsp}1 Connector. This connector has two wallets, one in \gls{xrp} and one in ETH, so it is able to provide cross payments between \gls{xrp} and ETH. This use case fits the architecture illustrated in Figure \ref{fig: 2}. It is also equipped with Moneyd-\acrshort{gui} for easier administration through a Chrome browser (recommended), and can as well perform \acrshort{spsp} payments given \acrshort{spsp} is installed. More on \acrshort{spsp} in Section \ref{section: mdspsp}.
  
  \begin{lstlisting}
'use strict'
const path = require('path')

const address = 'rLR52VSZG3wqSrkcpfkSnaKnYoYyPoJJgy'  // <YOUR RIPPLE ADDRESS>
const secret = 'ssrnzXKsJKWDh9cFpmZSLWHN3D5HM'        // <YOUR RIPPLE SECRET>

//to get the gas price
const { convert, usd, gwei } = require('@kava-labs/crypto-rate-utils')
const axios = require('axios')

const getGasPrice = async () => {
  const { data } = await axios.get(
    'https://ethgasstation.info/json/ethgasAPI.json'
  )

  return convert(gwei(data.fast / 10), wei())
}
//
const peer1 = {
    relation: 'peer',
    plugin: 'ilp-plugin-xrp-paychan', //peer with other connector/node over XRP
    assetCode: 'XRP',
    assetScale: 9,
    balance: {
        maximum: '1000000000',
        settleThreshold: '-5000000000',
        settleTo: '0'
    },
    options: {
        //listener: {       //this connector would be a server listening on port 10666
        //port: 10666,
        //secret: '2afe5e6cece84ed0027f9a2463edfa6358901bd6f1c9f3e1b0e43c13ff1ae2ea'    // this is the token that your peer must authenticate with.
        //},
        server: 'btp+ws://yourcustomsequence:2afe5e6cece84ed0027f9a2463edfa6358901bd6f1c9f3e1b0e43c13ff1ae2eb@192.168.1.146:10666',       //this connector is a ws client connecting to its ws server at port 10666
        rippledServer: 'ws://192.168.1.98:51233',        //PORT?    //wss://s2.ripple.com    // ?Specify the server that you submit XRP transactions to?
        peerAddress: 'rMqUT7uGs6Sz1m9vFr7o85XJ3WDAvgzWmj',    //<PEER RIPPLE ADDRESS>
        address,
        secret
     }
}

const peerETH = {
      relation: 'child',
      plugin: 'ilp-plugin-ethereum',
      assetCode: 'ETH',
      assetScale: 9,
      options: {
          role: 'server',
          port: 7442,
          ethereumPrivateKey: '0x43c50a578883922df30a33eb74418fb568c0081c40256e4675df02dcc28b6ef6',   //this connector's ETH address; different from machinomy contract address
          ethereumProvider: 'kovan', //goes to ETH plugin as identifier
          getGasPrice: getGasPrice, //'20000000000',
          outgoingChannelAmount: '71440000',       //10 usd
          minIncomingChannelAmount: '3570000',  // 0.5usd
          // In plugin (and not connector middleware) so F08s occur before T04s
          maxPacketAmount: '1430000'    // 0.2USD 
      }
}

const ilspServer = {
    relation: 'child',
    plugin: 'ilp-plugin-xrp-asym-server',   // ILSP server for downstream clients
    assetCode: 'XRP',
    assetScale: 6,
    options: {
        port: 7443,                         //port on which to listen to client apps
        xrpServer: 'ws://192.168.1.98:51233',  //MAINNET wss://s2.ripple.com
        address,
        secret
    }
}

const moneydGui = {
  relation: 'child',
  plugin: 'ilp-plugin-mini-accounts',
  assetCode: 'XRP',
  assetScale: 6,
  options: {
    port: 7768
  }
}

const connectorApp = {
    name: 'connector',
    env: {
        DEBUG: 'ilp*,connector*',
        CONNECTOR_ENV: 'production',
        CONNECTOR_ADMIN_API: true,
        CONNECTOR_ADMIN_API_PORT: 7769,
        CONNECTOR_ILP_ADDRESS: 'g.conn2',    //<YOUR ILP ADDRESS>
        CONNECTOR_BACKEND: 'one-to-one',
        CONNECTOR_SPREAD: '0',
        CONNECTOR_STORE: 'memdown',
        //CONNECTOR_STORE: 'ilp-store-redis',
        //CONNECTOR_STORE_CONFIG: JSON.stringify({
        //    prefix: 'connector',
        //    port: 6379
        //}),
        CONNECTOR_ACCOUNTS: JSON.stringify({
            conn1: peer1,                 
            ilsp_clients: ilspServer,
            moneyd_GUI: moneydGui,
            peer_ETH: peerETH
        })
    },
    script: path.resolve(__dirname, 'src/index.js')
}

module.exports = { apps: [ connectorApp ] }
  \end{lstlisting}
  
  This connector supports \acrshort{spsp} client-server. This is explained in Section \ref{section: mdspsp}. 
  
\end{exmp}
  Other connectors, functional but still in development, are the \textit{Rafiki}\footnote{\href{https://github.com/interledgerjs/rafiki}{https://github.com/interledgerjs/rafiki}, accessed June 2019} connector and the \textit{Rust}\footnote{\href{https://github.com/emschwartz/interledger-rs/tree/master}{https://github.com/emschwartz/interledger-rs/tree/master}, accessed June 2019} connector. A basis for a Java connector is also in the works, as \textit{Quilt}\footnote{\href{https://www.hyperledger.org/projects/quilt}{https://www.hyperledger.org/projects/quilt}, accessed July 2019}. \newline
\indent Rafiki is a modular connector which is meant to improve on the "reference" js connector in regards to practical aspects like cloud deployment, more manageable and 'hot' changes of configuration, etc \cite{introrafiki}. \newline 
\indent The Rust connector is meant to be a faster connector for high traffic. One important update is the new concept of "Settlement engine" and the elimination of the plugins. Official information on the Settlement Engines' architecture can be found in the new RFC on the Interledger website\footnote{\href{https://interledger.org/rfcs/0038-settlement-engines/}{https://interledger.org/rfcs/0038-settlement-engines/}, accessed January 2020}. Also because of the modular architecture, at least in the case of Rust and at the present moment, in order to run a connector, the user needs to separately start 3 processes: the connector itself, the settlement engine, and in some cases the Redis database. Separate configuration need to be provided\footnote{\href{https://github.com/interledger-rs/interledger-rs/tree/master/examples}{https://github.com/interledger-rs/interledger-rs/tree/master/examples}, accessed January 2020}. 

\section{The ledgers} 
\label{section: ledgers}

\subsection{The Ripple ledger}
\label{XRPledger}

The Rippled XRP ledger is made up of two types of servers: \textit{"trackers"} (or \textit{stock servers}) and \textit{"validators"}. They run the same piece of software \cite{installripple}, just with a different configuration. Although they can answer user queries, validators should ideally just process the transactions they receive from the trackers. \\ \newline
\textit{'Ideally, validating nodes are clustered with at least two stock nodes, to prevent DoS attacks and to preserve availability while updating the stock nodes. This configuration enables the validating node to be cut off from the internet, except for messages to/from other trusted nodes in the cluster and SSH connections via a LAN connection. Using two stock nodes provides redundant communication to the validating node, which is useful in case one of the stock nodes crashes or goes offline. However, this means a validating node has 3x the cost, 3x the monitoring, and 3x the time commitment of a stock node.  Production validating nodes should have at least 32 GB of memory as well as a 50 GB+ solid state drive.  I encourage operator to refrain from making \acrshort{api} calls (monitoring excepted) on validating nodes'.} \cite{stockserver}\\ \newline
Validators participate in the consensus process and vote on fees and amendments. Trackers are meant to be placed in-between validators and the rest of the network, pick-up traffic and forward it to the validators. They work as relays, protecting the validators. They also can hold the full history of the ledger and answer queries about old ledgers. On the other hand, the validators can work with minimal stored history. \newline
    
Below is the procedure to build and cold-start an independent local validators cluster. One aspect to keep in mind is that \textit{'there is no rippled setting that defines which network it uses. Instead, it uses the consensus of validators it trusts to know which ledger to accept as the truth. When different consensus groups of rippled instances only trust other members of the same group, each group continues as a parallel network. Even if malicious or misbehaving computers connect to both networks, the consensus process overrides the confusion as long as the members of each network are not configured to trust members of another network in excess of their quorum settings.'} \cite{privatenet}

\subsubsection{Preparation}
To build a parallel Rippled servers (validators, trackers) network, which, in its entirety is also called the \textit{"Rippled Ledger"}, the minimal required hardware resources\footnote{\href{https://developers.ripple.com/system-requirements.html}{https://developers.ripple.com/system-requirements.html}, accessed June 2019} need to be planned in advance. Depending on the available resources, each Rippled server can be deployed on a different physical machine or not. However, for high traffic use-cases, in order to streamline I/O, each server could have its own physical SSD.
    
To install the pre-packaged Rippled server, the instructions on the Ripple developer portal should be followed\footnote{\href{https://developers.ripple.com/install-rippled.html}{https://developers.ripple.com/install-rippled.html}, accessed June 2019}. Then, another guide is disseminated on the developer portal to install Rippled from the source code\footnote{\href{https://developers.ripple.com/build-run-rippled-ubuntu.html}{https://developers.ripple.com/build-run-rippled-ubuntu.html}, accessed June 2019}. After that, the following steps must be followed:
    \begin{itemize}[noitemsep, leftmargin=*]
    	\item Build the validators keys and tokens, using the published documentation\footnote{\href{https://developers.ripple.com/run-rippled-as-a-validator.html\#enable-validation-on-your-rippled-server}{https://developers.ripple.com/run-rippled-as-a-validator.html\#enable-validation-on-your-rippled-server}, accessed June 2019} and code\footnote{\href{https://github.com/ripple/validator-keys-tool}{https://github.com/ripple/validator-keys-tool}, accessed June 2019}.
    	    \begin{itemize}
    	        \item Create keys: 
    	            \begin{lstlisting} 
~/validator-keys-tool/build/gcc.debug$ ./validator-keys create_keys 
    	            \end{lstlisting}
    	        \item Create tokens:
    	            \begin{lstlisting}
~/validator-keys-tool/build/gcc.debug$ ./validator-keys create_token --keyfile /home/user/.ripple/validator-keys.json
                    \end{lstlisting}
    	    \end{itemize}
        \item Add generated $[validator\_token]$ to 'rippled.cfg'.
        \item Add generated $[validators]$ public keys to 'validators.txt'. Comment the rest of 'validators.txt'.
        \item In 'rippled.cfg', add the peer validators' IPs in the field $[ips\_fixed]$ in the form of IP:port (51235).
        \item Check that 'validator.txt' file name is the same with the name referenced by 'rippled.cfg'.
        \item Configure clustering as per the Ripple documentation\footnote{\href{https://developers.ripple.com/cluster-rippled-servers.html}{https://developers.ripple.com/cluster-rippled-servers.html}, accessed June 2019}, using the validation\_create\footnote{\href{https://developers.ripple.com/validation\_create.html}{https://developers.ripple.com/validation\_create.html}, accessed June 2019} method:
            \begin{lstlisting} 
~/rippled/ccabuild$  ./rippled --conf /home/user/rippled/cfg/rippled-example.cfg validation_create
    	    \end{lstlisting}
    \end{itemize}
    
\subsubsection{Start up}

In the case when Docker images have been used, after creating the Docker image of the Rippled server, this can be loaded and started on each physical/virtual server machine with the following:
    \begin{lstlisting}
    - 'sudo docker load -i /path/to/your_docker_image.tar'
    - 'sudo docker images' - to check the image name
    - 'sudo docker run -ti -u root --network host --name <container_name> <image_name>'
    - 'sudo docker exec -ti -u root <container_name> bash'. To open a second terminal to the container, just run the command again into a fresh terminal window.
    \end{lstlisting}
    With the docker images loaded on each Rippled server machine, the actual Rippled validators servers network can be cold-started as follows:
    \begin{itemize}	    
    	\item Start the first Rippled server with 'quorum 1' and wait a few minutes for it to stabilize:
    	    \begin{lstlisting}
./rippled --conf /home/user/rippled/cfg/rippled-example.cfg --quorum 1
            \end{lstlisting}
        \item Start the remaining servers with the same command, waiting for each to stabilize, first.
        \item Restart the servers in the same order, waiting a few minutes for each to stabilize before starting the next, with 'quorum 2':
            \begin{lstlisting}
./rippled --conf /home/user/rippled/cfg/rippled-example.cfg --quorum 2
            \end{lstlisting}
        In a new terminal window handling the Docker container, use the "stop" command in Table \ref{tab: ripplecomm} to gracefully stop the servers before restart. 
    \end{itemize}
    
In this minimal set-up though, if any of the servers is restarted, it will lose previously kept ledger history - even with full history enabled. This won't stop it working after restart, as validators do not need full history to work properly. To be able to access previous ledger history, tracking servers should be also set up. Data \acrshort{api}\footnote{\href{https://developers.ripple.com/data-api.html}{https://developers.ripple.com/data-api.html}, accessed June 2019} is a useful history tool which could also be set-up if desired, although setting it up on a private network seems not too obvious.
    
'Ripple \acrshort{api}'\footnote{\href{https://developers.ripple.com/rippleapi-reference.html}{https://developers.ripple.com/rippleapi-reference.html}, accessed June 2019} provides the means to interact with the server. For example, in Ripple, all the money are created in the beginning, and stored in an account with a hard-coded address, called the "Genesis account'. One can check the 'Genesis account' with:

            \begin{lstlisting}
./rippled  --conf /home/user/rippled/cfg/rippled-example.cfg account_info rHb9CJAWyB4rj91VRWn96DkukG4bwdtyTh
            \end{lstlisting}
    
 Some other useful server commands are provided in Table \ref{tab: ripplecomm}. These can be entered from a separate terminal window handling the docker container.\newline
 
    \begin{table}[h]
        \centering
		\caption[Useful Rippled server commands]{Useful Rippled server commands.}
		\label{tab: ripplecomm}
        \setlength{\tabcolsep}{0.7em}
			\begin{tabular}{l|l}
                command         & effect \\
                \hline
                wallet\_propose & create a new wallet with random seed credentials (inactive until funded)\\
                stop            & gracefully stop the server \\
                restart         & restart the server \\
                server\_info    & various easy-to-read info about the server \\
                server\_state   & almost same info as above, but easier-to-process instead of easy-to-read \\
                peers           & info on peer validators: connected? ledger sequences available? ...\\
			\end{tabular}
    \end{table}
    
Immediately after creating and starting the validators cluster network (which form the \gls{xrp} ledger), one can open a few accounts with the 'wallet\_propose' command above, and fund them using for example the following simple procedure. Regarding the wallets, they can be sometimes classified as 'hot' or 'cold' wallets. The difference is that 'hot' wallets are connected to the internet, while 'cold' wallets are not. 'Hot' wallets provide the advantage of quick access but lower security, like anything connected to the internet. 'Cold' wallets are slower to access (need to connect) but more secure due to generally not being online. It is generally recommended to hold only the amounts necessary for daily operation in the 'hot' wallet, while the bulk of the money would be kept offline.
    
     \begin{itemize}[noitemsep, leftmargin=*]
    	\item Install Ripple-API for javascript\footnote{\href{https://developers.ripple.com/get-started-with-rippleapi-for-javascript.html}{https://developers.ripple.com/get-started-with-rippleapi-for-javascript.html}, accessed June 2019}.
    	\item Place the two example scripts in the app folder: \textit{'/home/user/ripple\_api/get-account-info.js'}.
    	\item Run them with \textit{'./node\_modules/.bin/babel-node get-account-info.js'}. The code should run on one of the Ripple servers.
    \end{itemize}

Example script - get account info:
\begin{lstlisting}
//GET ACCOUNT INFO
'use strict';
const RippleAPI = require('ripple-lib').RippleAPI;

const api = new RippleAPI({
  server: 'ws://localhost:6006'
});
api.connect().then(() => {
  /* begin custom code ------------------------------------ */
  const testAddress = 'rHb9CJAWyB4rj91VRWn96DkukG4bwdtyTh';

  console.log('getting account info for', testAddress);
  return api.getAccountInfo(testAddress);

}).then(info => {
  console.log(info);
  console.log('getAccountInfo done');

  /* end custom code -------------------------------------- */
}).then(() => {
  return api.disconnect();
}).then(() => {
  console.log('done and disconnected.');
}).catch(console.error);  
\end{lstlisting}

Example script - fund an account:
 
 \begin{lstlisting} 
 //Account funding
 const RippleAPI = require('ripple-lib').RippleAPI

// SENDER - ADDRESS 1
const ADDRESS_1 = "rHb9CJAWyB4rj91VRWn96DkukG4bwdtyTh"
const SECRET_1 = "snoPBrXtMeMyMHUVTgbuqAfg1SUTb"

// RECEIVER - ADDRESS 2
const ADDRESS_2 = "rMqUT7uGs6Sz1m9vFr7o85XJ3WDAvgzWmj"

const instructions = {maxLedgerVersionOffset: 5}
const currency = 'XRP'
const amount = '20000000' // this is not 'drops' but XRP

const payment = {
  source: {
    address: ADDRESS_1,
    maxAmount: {
      value: amount,
      currency: currency
    }
  },
  destination: {
    address: ADDRESS_2,
    amount: {
      value: amount,
      currency: currency
    }
  }
}

const api = new RippleAPI({
  //server: 'wss://s1.ripple.com'                      //MAINNET
  //server: 'wss://s.altnet.rippletest.net:51233'   // TESTNET
  server: 'ws://localhost:6006'                        // Localhost
})

api.connect().then(() => {
  console.log('Connected...')
  api.preparePayment(ADDRESS_1, payment, instructions).then(prepared => {
    const {signedTransaction, id} = api.sign(prepared.txJSON, SECRET_1)
    console.log(id)
    api.submit(signedTransaction).then(result => {
      console.log(JSON.stringify(result, null, 2))
      api.disconnect()
    })
  })
}).catch(console.error)
\end{lstlisting}   
    
The known amendments seem not to be automatically enabled after cold starting a private network. In order to force them, we added the \textit{[features]} stanza in each validator's config file. Otherwise, the validators would apparently work, but, when trying to open for example a paychan, would throw the error "logic not enabled' - because the paychan amendment is not enabled. According to documentation\footnote{\href{https://developers.ripple.com/amendments.html}{https://developers.ripple.com/amendments.html}, accessed June 2019}, for an amendment to become enabled, it needs the support of 80\% of validators' votes for two weeks. If it loses this support, the amendment is temporarily disabled, and it can be re-enabled after it re-gains this support.  \\
    
\noindent [features] \\
    PayChan \\
    Escrow \\
    CryptoConditions \\
    fix1528 \\
    ....... \\
    
    Below is an example config file for a private network cluster of 3 validators. The file is located in \textit{'home/user/rippled/cfg'}. We used a docker container with a compiled version of Rippled.
    \begin{lstlisting}

[server]
port_rpc_admin_local
port_peer
port_ws_admin_local
port_ws_public
port_public

[port_rpc_admin_local]
port = 5005
ip = 127.0.0.1
admin = 127.0.0.1
protocol = http

[port_peer]  //talk to other validators
port = 51235
ip = 0.0.0.0
protocol = peer

[port_ws_admin_local]
port = 6006
ip = 127.0.0.1
admin = 127.0.0.1
protocol = ws

[port_ws_public]
port = 6005
ip = 127.0.0.1
protocol = wss

[port_public] //connectors, moneyd, switch API will connect here
ip = 0.0.0.0
port = 51233
protocol = ws

[node_size] //required for full history
huge

# This is primary persistent datastore for Rippled.  This includes transaction
# metadata, account states, and ledger headers.  Helpful information can be
# found here: https://ripple.com/wiki/NodeBackEnd
# delete old ledgers while maintaining at least 2000. Do not require an
# external administrative command to initiate deletion.
[node_db]   //NuDB type required for full history
type=NuDB
path=/var/lib/rippled/db/nudb
#open_files=2000 //these are not needed for NuDB
#filter_bits=12
#cache_mb=256
#file_size_mb=8
#file_size_mult=2
#online_delete=2000
#advisory_delete=0

[ledger_history] //although enabled, full history seems not to work
                 //correctly for validators, will need trackers for this.
full

[database_path]
/var/lib/rippled/db

# This needs to be an absolute directory reference, not a relative one.
# Modify this value as required.
[debug_logfile]
/var/log/rippled/debug.log

[sntp_servers] // servers for time sync
time.windows.com
time.apple.com
time.nist.gov
pool.ntp.org

# File containing trusted validator keys or validator list publishers.
# Unless an absolute path is specified, it will be considered relative to the
# folder in which the rippled.cfg file is located.
[validators_file]
validators-example.txt

# Turn down default logging to save disk space in the long run.
# Valid values here are trace, debug, info, warning, error, and fatal
[rpc_startup]
{ "command": "log_level", "severity": "trace" } //verbose logging

# If ssl_verify is 1, certificates will be validated.
# To allow the use of self-signed certificates for development or internal use,
# set to ssl_verify to 0.
[ssl_verify]
0

[ips_fixed]
192.168.1.97 51235 //IPs and ports of the other 2 peer validators
192.168.1.132 51235

[peer_private]
1

[node_seed]
shEm9dGAs2aq6MMe9XsXYXKrPmqft

[cluster_nodes]
n9LPJFoTLxVbTtdWADZzPpCwACwC3aLAYGhFcNNR61fD9DTc2w5L ripdbg1
n9KUMms9ZrDgHU7rN9pRTRGMKEWy5Ghk3qj53aCPAbJRur2sTqwp ripdbg3

[validator-token]
eyJtYW5pZmVzdCI6IkpBQUFBQUZ4SWUwYkVlUVp1bGNsKzRadk44cGhXUWJNNWhlV3RKY0hN
YUVKcUpadWVRWm9jWE1oQXYvVWY3MmlaQ0VQZndPZTd0TjNaY0V1UnFDd2Q3U2JkU3hPTnJq
TXlsNWlka2N3UlFJaEFKc3IzL3g2U0RiRGprOHc0Mks2eU91M1FPbW4vNjVIeTM4bkxjbnJa
c1ROQWlBSnRlRTRpdjVqSjRJMytvS0VseEFjTmFUL3VoQnRlSVFyK29RdmVoemJESEFTUU53
RnpLN21kV3lUaTZoTWY4SUJTRUxmZHI1cjhuMFdIeE5BSGNHSXJURDV1N09BK3FKZWZLMzkw
Smx3aE5ydGVLL09LWS8rQldDUHo0ejQ4VXptaHd3PSIsInZhbGlkYXRpb25fc2VjcmV0X2tl
eSI6IkJGMTcyRjJBMzNGQTZDOTdBQ0JBODhBNTA0NThGQzZFRURENzBCNjEwMzdEMjcwNjgz
RTQ3MzRBNUY2OURGRkMifQ==

[features]
PayChan
Escrow
CryptoConditions
fix1528
DepositPreauth
FeeEscalation
fix1373
MultiSign
TickSize
fix1623
fix1515
TrustSetAuth
fix1513
fix1512
fix1571
Flow
fix1201
fix1523
fix1543
SortedDirectories
EnforceInvariants
fix1368
DepositAuth
fix1578

    \end{lstlisting}
    
\subsection[The Ethereum ledger]{The ETH ledger. Connecting the \gls{xrp} and ETH ledgers through 'Machinomy'}
\label{ETHMach}
    For the scope of this work, we will assimilate the ETH network to a black-box holding the ETH wallet accounts, executing commands and providing immediate response. For testing purposes, such a friendly 'black-box' can be 'Ganache'\footnote{\href{https://truffleframework.com/docs/ganache/quickstart}{https://truffleframework.com/docs/ganache/quickstart, accessed June 2019}, accessed June 2019}, previously called 'TestRPC'. After download, Ganache can be started directly:
    \begin{lstlisting}
    cd /Downloads
    ./ganache-1.3.1-x86_64.AppImage
    \end{lstlisting}
    
    'Machinomy'\footnote{\href{https://machinomy.com/}{https://machinomy.com/}, accessed June 2019} is used to connect the \gls{xrp} and ETH ledgers. It achieves this by deploying a specific contract on the ETH ledger. One contract manages all the channels for Ether micropayments (all the sender-receiver pairs). Thus, Machinomy creates the settlement capability when \acrshort{ilp} payment interacts with the ETH ledger.
    
    \textit{'Machinomy is a micropayments SDK for Ethereum platform. State channels is a design pattern for instant blockchain transactions. It moves most of the transactions off-chain. As transactions do not touch the blockchain, fees and waiting times are eliminated, in a secure way.'} \cite{machinomy}
    
    Machinomy should be installed\footnote{\href{https://github.com/machinomy/machinomy}{https://github.com/machinomy/machinomy}, accessed June 2019} on the same machine with the ETH provider, in this case, Ganache. After installing Machinomy, a contract can be deployed on the ETH network using the following:
    \begin{lstlisting}
        cd machinomy/node_modules/@machinomy/contracts 
        yarn truffle migrate --reset
    \end{lstlisting}
    Checking back in Ganache after Machinomy contract deployment, you will notice that a small amount of ETH has been subtracted from the first account, and in the \textit{Transactions} tab, the contract has been deployed.
    
        After a Ganache restart, the \textit{'--reset'} option has to be used because Ganache is not persistent. The contract will be deployed on the first Ganache account. The other accounts can be used by ETH client wallets.
        
        After deploying the Machinomy contract on the ETH network, apps like \gls{switchapi} can be used to exchange \gls{xrp} and ETH back and forth.
        The plugins should be set to access Ganache using \textit{http://ganache\_IP:ganache\_port}. A detailed explanation on \gls{switchapi} is provided in Section \ref{switch}.\\ \newline
        
\section{Evaluation and discussion}
\label{section:eval}

In this paper, we have provided the details on how to set-up a private \acrshort{ilp} network comprising of two ledgers - \gls{xrp} and ETH, \acrshort{ilp} service providers (connectors), and customer apps (\gls{moneyd}, \acrshort{spsp}, \gls{switchapi}). The payments can be streamed from  one ledger to the other with the \gls{switchapi} app, making use of Machinomy smart contracts deployed on the ETH ledger. Time (the time on the machines must be synchronised), conversion rates and gas price are fetched from the internet.\\

In our opinion, at the present moment, as one moves from the core - the Rippled servers making-up the ledger, to the periphery - the customer apps, the support and availability of apps decreases. The most information to be found concerns the Rippled servers, while in regards to customer apps we have tried so far, at present only \gls{moneyd}-\gls{xrp} seems fairly supported. Some of the plugins are undergoing changes (e.g. ETH plugin), and with the advent of connectors like Rafiki, they may be, at least partially, replaced with new approaches like the "settlement engine". Intuitively this is the way the ecosystem should be built and we are confident the future will bring many improvements.
 
\section{Conclusions and future work}
\label{section:conclusions}

 Sometimes abstract concepts are explained separately from the actual implementation, making it difficult to make the connections. This work fills a hole in the documentation regarding a lack of a comprehensive high level view of the ecosystem and how the different pieces are joined together.

We are currently studying the all-new @Coil/Rafiki which is still in beta and will soon provide the results.
\section*{Acknowledgements}

This  work  was  supported  by  the  Luxembourg  National  Research Fund (FNR) through grant PRIDE15/10621687/SPsquared. In addition, we thankfully acknowledge the support from the RIPPLE University Blockchain Research Initiative (UBRI) framework for our research.

\clearpage
 
\printglossary[type=\acronymtype]
 
\printglossary
\begin{footnotesize}


\bibliographystyle{unsrt}
\bibliography{References}

\end{footnotesize}

\end{document}